\documentclass[12pt]{article}

\usepackage[letterpaper,margin=0.6in]{geometry}
\usepackage{newtxtext,newtxmath}
\usepackage{setspace}

\usepackage{amsmath,bm}

\usepackage{graphicx,epstopdf,psfrag,wrapfig}
\usepackage{caption,subcaption}
\usepackage{booktabs,array,multirow,makecell,hhline,adjustbox,dcolumn,longtable}
\usepackage{siunitx,l3draw}

\usepackage{float}
\usepackage[section]{placeins}
\usepackage[round,authoryear]{natbib}
\usepackage[colorlinks=true,linkcolor=blue,citecolor=blue,urlcolor=blue]{hyperref}

\usepackage{xcolor,soul,comment,todonotes,pifont}
\usepackage[normalem]{ulem}

\newcommand{\vk}[1]{{\color{black}#1}}

\title{On the scaling of bubble interactions in dynamic turbulence: theoretical, numerical, and experimental study}

\author{
Vivek Kumar$^{1,2,3}$, Prasoon Suchandra$^{1,2}$, Shivam Prajapati$^{1,2}$,\\
Suhas S. Jain$^{4}$, and Cyrus K. Aidun$^{1,2}$\\[1ex]
\small $^{1}$George W. Woodruff School of Mechanical Engineering, Georgia Institute of Technology, Atlanta, GA 30332, USA\\
\small $^{2}$Center for Multiphase Flow Research, Georgia Institute of Technology, Atlanta, GA 30332, USA\\
\small $^{3}$Renewable Bioproducts Institute, Georgia Institute of Technology, Atlanta, GA 30332, USA\\
\small $^{4}$Daniel Guggenheim School of Aerospace Engineering, Georgia Institute of Technology, Atlanta, GA 30332, USA\\[1ex]
\small Corresponding author: \href{mailto:cyrus.aidun@me.gatech.edu}{cyrus.aidun@me.gatech.edu}
}
\date{}

\begin{document}
\maketitle

\begin{abstract}
This study investigates high-Reynolds-number dilute bubbly decaying homogeneous isotropic turbulence (HIT), characterized by Taylor-scale Reynolds numbers $\mathrm{Re}_{\lambda}\sim\mathcal{O}(10^2\text{--}10^3)$ and void fractions $\langle \phi \rangle\sim\mathcal{O}(1\%)$. Results show that the turbulent kinetic energy and dissipation rate ($\varepsilon$) follow robust power-law decay in time $t$ with decay exponent $m$, with $\varepsilon\sim t^{-m}$, closely resembling the classical Saffman- and Loitsyanskii-type decay regimes of single-phase turbulence. The bubble population, however, reorganizes according to its size relative to the evolving Hinze scale, $d_H$. Since $d_H\sim \varepsilon^{-2/5}$, for $m>5/3$ the Hinze scale grows faster than the characteristic bubble size, $d$, so that $d/d_H\sim t^{(15-9m)/25}$ decreases with time in the early mixed regime. The population is therefore driven from super-Hinze toward sub-Hinze sizes, passing through a mixed regime in which coalescence dominates while breakup remains active, before entering a pure-coalescence regime. Counterintuitively, the characteristic bubble size grows faster in the mixed regime because residual breakup increases the number density of smaller bubbles and thereby enhances coalescence, whose rate scales quadratically with bubble number density. Direct numerical simulation (DNS) of dilute bubble-laden HIT at $\mathrm{Re}_\lambda\sim \mathcal{O}(10^2)$ shows that the dissipation remains close to single-phase HIT, with $m\simeq 2.2$--$2.4$, although modest deviations arise from interfacial energy exchange. Starting from an equilibrated bubble-size probability density function (PDF) spanning both sub-Hinze and super-Hinze bubbles, the PDF rapidly shifts toward smaller $d/d_H$ as $d_H$ grows faster than the characteristic bubble size. Before the transition, the PDF exhibits a dual power-law structure, with a $d^{-3/2}$ range associated with capillary effects and a steeper $d^{-10/3}$ range associated with inertial breakup. After the transition, it approaches a single $d^{-3/2}$ scaling, indicating that the early breakup pathway is suppressed. Theory and DNS show faster growth, $d(t)\sim t^{(15+m)/25}$, in the mixed regime, followed by the pure-coalescence scaling $d(t)\sim t^{(3-m)/2}$. In the pure-coalescence regime, a Smoluchowski-type balance gives number density [$n(t)\sim t^{-(9-3m)/2}$], interfacial area [$A(t)\sim t^{(m-3)/2}$], and the per-bubble coalescence hazard follows the universal law $r(t)\sim t^{-1}$, while the volumetric coalescence rate scales as $R(t)\sim t^{(3m-11)/2}$. The Hinze-scale drift mechanism and the associated transition from mixed breakup--coalescence dynamics to a sub-Hinze coalescence-dominated regime are further assessed in a practical, spatially developing pump-driven turbulent bubbly flow as an extension of~\citet{kumar2026bubble} (J. Fluid Mech., vol. 1033, 2026, A6), obtained at inlet $\mathrm{Re}_\lambda \sim O(10^3)$. In this spatially developing duct flow, confinement, inhomogeneity, and sustained wall production reduce the effective dissipation decay exponent, but the turbulence still follows an approximate power law with $m\simeq 1.8$--$2.0$ near the duct centre. Using this measured exponent, the theoretical and DNS-based scalings for bubble diameter, interfacial area, and bubble count align with the experimental observations. These results identify Hinze-scale drift as the organizing mechanism for bubble interactions in both idealized HIT and pump-driven decaying turbulence, providing source-term constraints for non-stationary population-balance and interfacial-area-transport closures with applications to nuclear cooling, multiphase web forming technologies, paper manufacturing, and chemical reactors.
\end{abstract}

\noindent\textbf{Keywords:} Bubbly flow, HIT decaying flow, Sub-Hinze regime, Coalescence regime, Scaling analysis 
\par\medskip

\section{Introduction}

Gas–liquid turbulent flows are ubiquitous in natural phenomena and industrial applications, including chemical reactors, biomedical systems, oil–gas pipelines, aeration and wastewater treatment units, flotation devices, and heat exchangers~\citep{garcia2009bioreactor,kantarci2005bubble,delnoij1997computational,kumar2024hydrostatic,kulkarni2007mass}. In these systems, dispersed bubbles strongly influence mass transfer, mixing, drag, particle separation, and reaction rates~\citep{clift2005bubbles,alameedy2025comprehensive,fabre1992modeling}. These effects become particularly pronounced in high-Reynolds-number flows, such as those generated in multiphase pumps, where intense turbulence is deliberately produced to disperse gas into the liquid phase. Turbulent stresses fragment larger bubbles into smaller ones, thereby establishing the initial bubble size distribution and governing the downstream evolution of the two-phase mixture~\citep{hinze1955fundamentals,Kolmogorov1949,zhang2018investigation}. Similar turbulence-driven bubble interactions arise in natural systems, including oceanic wave breaking~\citep{deane2002scale,chan2018bubble}, biomedical flows~\citep{kumar2023particle}, and volcanic eruptions~\citep{liao2009review,liao2010review}, underscoring their broad physical relevance. At smaller scales, bubbly turbulence is controlled by stochastic interfacial processes in which turbulent eddies deform bubbles, promote collisions, and thin the intervening liquid films, leading to topological events such as coalescence and breakup that continuously redistribute interfacial area and modify the bubble population~\citep{martinez1999breakup,coulaloglou1977description}. Quantifying the rates of breakup and coalescence and their dependence on evolving turbulence and interfacial physics is therefore central to understanding and modelling turbulent bubbly flows.

The evolution of the bubble-size distribution in turbulent flows is governed primarily by the competing processes of bubble breakup and coalescence. From a modelling perspective, this distribution is typically predicted by solving a population-balance equation coupled with the turbulent flow field, in which empirically closed breakup and coalescence kernels control the transfer of bubble number density across size classes in both space and time~\citep{lehr2002bubble}. These mechanisms have been investigated extensively using theoretical analysis and numerical simulations, revealing that turbulent dissipation, inertial stresses, and interfacial forces jointly determine the statistical evolution of bubble populations~\citep{Li2024,Yu2024}. Bubble breakup occurs when turbulent stresses overcome surface tension forces, leading to fragmentation once a critical Weber number ($\mathrm{We}_c$) is exceeded~\citep{Kolmogorov1949,hinze1955fundamentals,ni2024deformation}. Corresponding to the critical Weber number ($\mathrm{We}_c$), the characteristic bubble diameter is defined as the Hinze scale, $d_H$, given by
\begin{equation} \label{eq:hinze scale}
d_H = \mathrm{We}_c^{3/5}
\left(\frac{\gamma}{\rho_\ell}\right)^{3/5}
\varepsilon^{-2/5},
\end{equation}
where $\varepsilon$ denotes the turbulent kinetic energy dissipation rate, $\gamma$ the interfacial tension, and $\rho_\ell$ the density of the carrier liquid~\citep{hinze1955fundamentals,hesketh1987bubble,salibindla2020lift,ni2024deformation}. Building on this framework, classical breakup models such as those of ~\citet{coulaloglou1977description,luo1996breakup,lehr2002bubble} relate breakup rates to the interaction between turbulent eddies and bubble interfaces, either through stochastic energy arguments, eddy–bubble collision integrals, or force-balance criteria. These models introduce empirical constants and closure relations to account for unresolved physics, including energy transfer efficiency, fragmentation probability, and daughter-bubble-size distributions. Further experimental measurements revealed a non-monotonic breakup frequency, increasing near the Hinze scale and decreasing for substantially larger bubbles~\citep{martinez1999breakup}, while DNS studies have reported memoryless breakup dynamics and a finite probability of breakup even for sub-Hinze bubbles~\citep{vela2022memoryless}. These simulations also show enhanced deformation when the bubble diameter approaches the scale of interacting turbulent eddies~\citep{calado2024dynamics}. In highly turbulent flows, intensified shear and a broader inertial cascade promote higher breakup frequencies for bubbles larger than the Hinze scale~\citep{Chen2021,Sajjadi2013}, reinforcing the Weber number as a key parameter governing breakup dynamics~\citep{Nguyen2013}.

Despite their widespread use, these breakup models are fundamentally constrained by assumptions regarding turbulence structure, empirical closures, and flow configuration. Most formulations assume statistically stationary, homogeneous turbulence, and rely on a single representative dissipation rate, while embedding empirical constants calibrated for specific geometries such as bubble columns or stirred tanks~\citep{liao2009review,yao2023breakup}. As a result, these constants implicitly incorporate geometry-dependent effects such as anisotropy, confinement, and mean shear, limiting their general applicability. At very high Reynolds numbers, turbulence exhibits strong intermittency and a wide range of local dissipation and strain-rate fluctuations, so that breakup is governed by rare, intense events rather than mean-field quantities. Consequently, models based on average dissipation or a single effective eddy scale cannot accurately capture breakup frequency. Moreover, in spatially evolving or decaying turbulence, key quantities such as $\varepsilon$, eddy turnover times, and the Hinze scale vary in time and space, altering both the breakup threshold and the active eddy population. Under such conditions, empirical coefficients and closure relations should, in principle, depend on local turbulence intensity and flow evolution, which is not accounted for in existing models. Additionally, most formulations neglect finite deformation time and assume quasi-instantaneous response of the interface, an assumption that becomes invalid when deformation and turbulent time scales are comparable at high Taylor-scale Reynolds number, $\mathrm{Re}_\lambda$ ($\approx \mathcal{U}\lambda_T/\nu$), where $\mathcal{U}$ is r.m.s of velocity fluctuation, $\lambda_T$ is Taylor microscale, and $\nu$ is the kinematic viscosity of the continuous phase. While controlled laboratory and numerical studies of bubbly flows are typically limited to $\mathrm{Re}_\lambda \lesssim \mathcal{O}(10^2\text{--}10^3)$, much higher Reynolds numbers are encountered in natural and industrial flows, highlighting the lack of validation of breakup closures in strongly turbulent, non-stationary regimes \citep{ni2024deformation,Li2024}.

On the other hand, among coalescence closures, the most widely used models are those of~\citet{coulaloglou1977description,prince1990bubble,chesters1991modelling,luo1998coalescence,lehr2002bubble}. In these formulations, the coalescence kernel is expressed as the product of a collision frequency and a coalescence efficiency, but they differ in how the collision process and film drainage are modeled. The \citet{coulaloglou1977description} model is based on a film-drainage framework in which turbulence-induced collisions lead to coalescence only if the liquid film thins to a critical thickness within a finite contact time; this formulation introduces empirical constants in both the collision frequency and the exponential form of the efficiency. \citet{prince1990bubble} extend this approach by incorporating turbulent, buoyancy-driven, and laminar-shear collision mechanisms, requiring empirical prefactors and characteristic time-scale closures. \citet{chesters1991modelling} provides a mechanistic basis for film drainage through lubrication theory, distinguishing interface mobility conditions and deriving drainage times, but still relies on simplified assumptions for interface properties and critical film thickness. \citet{luo1998coalescence} reformulate coalescence efficiency using characteristic drainage and contact times linked to turbulence scales, introducing closure constants that relate eddy properties to collision duration, while \citet{lehr2002bubble} propose a critical approach velocity criterion, in which empirical thresholds determine whether colliding bubbles coalesce or rebound. A common feature across these models is the reliance on calibrated constants, assumed collision geometries, and closure relations based on a representative dissipation rate and idealized turbulence structure.

These assumptions inherently embed dependence on flow configuration and turbulence level, limiting their generality. Most coalescence kernels are calibrated using data from specific geometries such as bubble columns, stirred tanks, or pipe flows, and therefore implicitly incorporate geometry-dependent effects such as anisotropy, mean shear, and confinement \citep{liao2009review,liao2010review}. Furthermore, collision frequency models typically scale with \((\varepsilon d)^{1/3}\), assuming Kolmogorov-type turbulence and weak intermittency, while coalescence efficiency depends on quasi-steady drainage and contact times.

At very high Reynolds numbers, $\mathrm{Re}_\lambda \sim \mathcal{O}(10^3)$, turbulence becomes strongly intermittent, so local dissipation and strain rates can deviate substantially from their mean values. Such fluctuations are not fully captured by mean-field closures. Experimental and atmospheric flows can reach \(\mathrm{Re}_\lambda \sim 10^4\text{--}10^5\) regimes, while laboratory grid turbulence and controlled facilities typically achieve \(\mathrm{Re}_\lambda \sim 10^3\) \citep{dhruva1997transverse,sreenivasan1997phenomenology,bodenschatz2014vdtt}. Even at \(\mathrm{Re}_\lambda \approx 10^3\), small-scale isotropy assumptions begin to break down due to persistent anisotropy and intermittency, particularly in shear-driven flows. In addition, in spatially evolving or decaying turbulence, key quantities such as \(\varepsilon\), eddy-turnover times, and the Hinze scale vary in time and space, altering both collision rates and film-drainage dynamics. Consequently, the empirical constants and closure coefficients in these models should, in principle, depend on local turbulence intensity, geometry, and flow evolution. Since such dependence is not included in existing formulations, these models can significantly mispredict coalescence rates in high-intensity, non-stationary flows, particularly in regimes where turbulence–interface coupling becomes strongly nonlinear \citep{ni2024deformation}.

These limitations point to a fundamental issue: existing coalescence and breakup closures are intrinsically non-universal, with empirical coefficients that are strongly configuration-dependent and often vary widely across studies. More critically, their underlying assumption of stationary, homogeneous turbulence becomes untenable in high-$\mathrm{Re}_\lambda$, evolving flows, where bubble–turbulence interactions are governed by intermittent, multi-scale, and inherently time-dependent dynamics. As a result, a predictive, physics-based understanding of bubble population evolution in such regimes remains largely unresolved.  The present work addresses this gap by developing a unified scaling framework for multiphase dynamics in strongly evolving turbulent bubbly flows, where both the turbulence field and the bubble population co-evolve and remain dynamically coupled across scales. In particular, the bubble-size distribution does not develop to an equilibrium state, but is continuously reshaped by the evolving turbulent cascade, leading to a fundamentally different regime from classical stationary formulations. To capture this behaviour, a theoretical framework derived in ~\citet{vivek_prl} based on a Smoluchowski-type population-balance equation is formulated, from which closed-form scaling laws for key multiphase quantities are derived under time-dependent turbulence. The analysis is first established for general decaying turbulence and subsequently specialized to dilute bubbly flows, with void fractions of order $1\%$, where the turbulence decay closely follows canonical single-phase homogeneous isotropic turbulence (HIT), thereby enabling a direct connection with classical turbulence theory. The resulting predictions are rigorously tested using high-fidelity direct numerical simulations (DNS) of HIT, employing the accurate conservative diffuse-interface/phase-field (ACDI) method \citep{jain2022a} that fully resolves bubble-turbulence interactions across scales. Crucially, the framework is further examined in a realistic, strongly inhomogeneous experimental configuration, where high-intensity turbulence is generated by a pump which evolves downstream in a duct flow. Despite the added complexity of non-stationarity and geometric confinement, our experimental observations exhibit remarkable agreement with both our theoretical and DNS predictions. These findings demonstrate that the proposed scaling framework captures the essential physics of bubble population evolution in regimes far beyond the validity of existing closures, thereby providing a new route toward predictive modelling of turbulent multiphase flows at high Reynolds numbers.

The rest of the article is organized as follows. Section~\ref{sec:theory section} presents the theoretical framework for decaying homogeneous isotropic turbulence and extends it to dilute bubbly flows. Section~\ref{sec:exp and DNS setup} describes the DNS configuration, experimental setup, and measurement techniques for the decaying turbulent duct flow. Section~\ref{section:Results and Discussions} presents the DNS and experimental results on bubble-population evolution and compares them with the theoretical predictions. Finally, Section~\ref{sec:conclusion} summarizes the main findings and discusses their implications for modelling dilute bubbly flows in strongly decaying turbulence.

\section{Theoretical framework} \label{sec:theory section}

In decaying HIT, the temporal decay of the dissipation rate is governed by the large-scale invariant associated with the low-wavenumber spectrum. If the energy spectrum behaves as $E(\kappa)\sim \kappa^{p-1}$ for $\kappa\to0$, the corresponding invariant implies $k\,\ell^{p}=\text{constant}$ with $p\in\{3,5\}$, where $k$ is the turbulent kinetic energy, $\ell$ is the integral length scale, and $\kappa$ is the wavenumber~\citep{saffman1967large,batchelor1956large}. This relation yields $\ell \propto k^{-1/p}$. Assuming a finite dissipation coefficient, $\varepsilon \sim C_\varepsilon k^{3/2}/\ell$, so that $\varepsilon \propto k^{(3/2+1/p)}$~\citep{vassilicos2015dissipation}. Here, \(C_\varepsilon\) is an order-unity constant. Since the kinetic energy evolution with time ($t$) satisfies $dk/dt=-\varepsilon$, one obtains $dk/dt\propto -k^{\,m_k}$ with $m_k=\tfrac{3}{2}+\tfrac{1}{p}$. Integrating this gives $k(t)\propto t^{-\,2p/(p+2)}$, from which $\ell(t)\propto t^{\,2/(p+2)}$ and the dissipation results in
\begin{equation}
\varepsilon(t)\propto t^{-m}, 
\qquad 
m=\frac{3p+2}{p+2}.
\label{eq:epsilon power law}
\end{equation}
For a Saffman spectrum ($p=3$), corresponding to $E(\kappa)\sim \kappa^{2}$ as $\kappa\to0$, one obtains $m=2.2$. For a Loitsyanskii spectrum ($p=5$), associated with $E(\kappa)\sim \kappa^{4}$ at low wavenumber, the decay exponent becomes $m=2.43$ \citep{loitsyanskii1939some,batchelor1956large}.\\

At sufficiently low void fractions, ($\langle \phi \rangle \lesssim \mathcal{O}(10^{-2}$)), the dispersed phase exerts only weak feedback on the carrier-phase turbulence. In this dilute weak-feedback regime, bulk turbulence statistics such as kinetic energy, dissipation rate, and spectral structure remain close to their single-phase counterparts, although localized interfacial forcing and energy exchange may still occur near individual bubbles. This behaviour is consistent with classical experimental and regime analyses of dilute bubbly flows \citep{Lance1991,serizawa1975turbulence,elghobashi1994predicting,elghobashi2006overview}. We quantify this weak-coupling condition by comparing the interfacial forcing exerted by the dispersed phase with the inertial forcing of the carrier turbulence. For a bubble with characteristic relative velocity \(U_s\), the hydrodynamic force has the inertial drag scale
\begin{equation*}
    F_b\sim C_D\rho U_s^2d^2\sim\rho U_s^2d^2,
\end{equation*}
with \(C_D=\mathcal{O}(1)\) for finite- to high-Reynolds-number bubbles \citep{clift1978bubbles,magnaudet2000motion,zeng2021investigation}. In the present high-turbulence regime, a characteristic turbulent bubble--liquid relative velocity,  $U_s\sim u_d\sim(\varepsilon d)^{1/3}$. Since the bubble number density scales as \(n\sim\phi/d^3\), the corresponding volumetric interfacial forcing becomes

\begin{equation*}
        f_b\sim n F_b
    \sim \frac{\phi\rho U_s^2}{d}
    \sim \phi\rho\varepsilon^{2/3}d^{-1/3}.
\end{equation*}

The carrier-phase inertial forcing follows from the convective term in the Navier--Stokes equations and scales with the integral scale as $f_t\sim \rho\frac{u'^2}{\ell}$, consistent with the large-eddy estimate \(\varepsilon\sim C_\varepsilon u'^3/\ell\)~\citep{lumley1972,pope2000turbulent,vassilicos2015dissipation}. Hence \(u'\sim(\varepsilon\ell)^{1/3}\), and $f_t\sim \rho\varepsilon^{2/3}\ell^{-1/3}$. The ratio of dispersed-phase forcing to carrier-phase turbulent inertia is therefore
\begin{equation*}
    \frac{f_b}{f_t}
    \sim
    \langle \phi \rangle\left(\frac{\ell}{d}\right)^{1/3}.
\end{equation*}
When \(\langle \phi \rangle(\ell/d)^{1/3} \sim \mathcal{O}(10^{-1} - 10^{-2})\ll1\), dispersed-phase forcing is weak relative to carrier-phase turbulent inertia, so the bulk cascade is expected to remain close to the corresponding single-phase decay. This interpretation is consistent with dispersed-phase coupling arguments \citep{elghobashi1994predicting,elghobashi2006overview,balachandar2010turbulent}. Conversely, bubbly-flow experiments and simulations show that turbulence modulation becomes significant when void fraction, slip-driven forcing, wake interactions, or bubble back-reaction increase \citep{lance1991turbulence,mazzitelli2003relevance,Rensen2005,RibouxRissoLegendre2010,Risso2018}. Consequently, under the present dilute, high-turbulence conditions, bubble feedback remains primarily localized near individual interfaces, while measurable bulk modulation is expected only at higher void fractions or when collective wake interactions become important. \\

The theoretical scalings are obtained from a reduced population-balance description of bubble coalescence and breakup in decaying turbulence. Let $f(d,t)$ denote the number density of bubbles of diameter $d$, and define the total number density as
\begin{equation}
    n(t)=\int_0^\infty f(d,t)\,\mathrm{d}d .
\end{equation}
Starting from the standard population balance equation with coalescence and breakup for a spatially homogeneous dispersion
\citep{coulaloglou1977description,ramkrishna2000population,lehr2002bubble}, integration over all bubble sizes gives an exact balance for $n(t)$,
\begin{equation}
\frac{\mathrm{d}n}{\mathrm{d}t}
=
-\frac{1}{2}\iint_{(0,\infty)^2}
K(d,d')f(d)f(d')\,\mathrm{d}d\,\mathrm{d}d'
+
\int_0^\infty
a(d')[\tilde{\nu}(d')-1]f(d')\,\mathrm{d}d' ,
\label{eq:n_balance_summary_exact}
\end{equation}
where $K(d,d')$ is the coalescence kernel for bubbles of sizes $d$ and $d'$, $a(d')$ is the breakup frequency of a parent bubble, and $\tilde{\nu}(d')$ is the mean number of daughter bubbles. For a narrow bubble-size distribution, the size-dependent kernel is replaced by an effective monodisperse kernel $\Gamma(t)=K(d(t),d(t))$. The rationale for this monodisperse simplification is that, in the present high-turbulence regime, both collisions and successful coalescence are dominated by similar-sized bubbles, with the coalescence efficiency highest when bubble sizes are comparable, leading to a
kernel that is sharply peaked near unity size ratios~\citep{Tan_Zhong_Qi_Xu_Ni_2025}. For binary breakup, the breakup contribution is represented by an effective frequency $\mathcal{B}(t)$. The number balance then reduces to
\begin{equation}
    \frac{\mathrm{d}n}{\mathrm{d}t}
    =
    -\frac{1}{2}\Gamma n^2
    +
    \mathcal{B} \alpha(t)n ,
    \label{eq:n_balance_summary_reduced}
\end{equation}
where $\alpha(t)$ is the fraction of bubbles whose diameters exceed the local Hinze scale $d_H(t)$ and are therefore susceptible to turbulent breakup. 
The reduction from Eq. \eqref{eq:n_balance_summary_exact} to Eq. \eqref{eq:n_balance_summary_reduced} assumes dilute conditions, binary interactions, approximately homogeneous statistics, a representative bubble diameter $d(t)$, and an approximately constant void fraction. The closure for $\Gamma$ follows from inertial-range turbulence scaling. The relative velocity across a bubble of size $d$ is estimated as
\begin{equation}
    u_{\mathrm{rel}}(d)\sim (\varepsilon d)^{1/3},
\end{equation}
where $\varepsilon$ is the mean turbulent dissipation rate. Multiplying this velocity scale by a geometric collision cross-section proportional to $d^2$ gives
\begin{equation}
    \Gamma
    =
    h\lambda
    \sim
    \varepsilon^{1/3}d^{7/3},
    \label{eq:Gamma_summary}
\end{equation}
where $\lambda$ is the coalescence efficiency \citep{coulaloglou1977description,prince1990bubble}. The kernel has units of volume per unit time and the coalescence efficiency satisfies $\lambda \in$[0, 1].  Consequently, following coalescence, the bubble-deformed interface relaxes quicker than the turbulent time scale, allowing it to reshape before a subsequent collision occurs. This form is consistent with inertial-range collision models based on Kolmogorov scaling \citep{kolmogorov1991local,batchelor1953theory,moninm} and with population-balance closures for turbulent bubbly dispersions \citep{saffman1956collision,prince1990bubble,luo1996breakup}. Under the present strongly decaying turbulence, the coalescence efficiency is assumed to remain at an order of unity, ($\lambda=\mathcal{O}(1)$), with weak temporal variation. As turbulence decays, the ratio of contact time to film-drainage time does not change appreciably over the short interaction interval; hence, $\lambda$ remains approximately constant and can be absorbed into the prefactor of the kernel scaling~\citep{kumar2026bubble}. The breakup frequency is taken to scale with the inverse eddy-turnover time at the bubble scale,
\begin{equation}
    \mathcal{B}
    \sim
    a(d)
    \sim
    \varepsilon^{1/3}d^{-2/3},
    \label{eq:B_summary}
\end{equation}
consistent with turbulent breakup models and simulations of \citet{martinez1999breakup,garrett2000connection,riviere2021subhinze}. The turbulence decay is written as $\varepsilon(t)\sim t^{-m}$, so that the Hinze scale evolves as $d_H(t)\sim \varepsilon^{-2/5}\sim t^{2m/5}$. \\

\begin{table}
\centering
\small
\setlength{\tabcolsep}{6pt}
\renewcommand{\arraystretch}{1.0}
\begin{tabular}{lcccc}
\hline
\textbf{Quantity} & \textbf{Scaling} & \textbf{$m=2.2$} & \textbf{$m=2.43$} & \textbf{Physical description} \\
\hline

$ d(t) $ 
& $ \sim t^{\beta},\;\; \beta=\dfrac{15+m}{25} $
& $0.688$ & $0.697$
& Growth of  bubble size  \\

$ a_b(t) $
& $ \sim t^{2\beta} $
& $1.376$ & $1.394$
& Single-bubble surface area \\

$ A(t)$
& $ \sim t^{-\beta} $
& $-0.688$ & $-0.697$
& Interfacial area per unit volume \\

$ n(t) $
& $ \sim t^{-3\beta} $
& $-2.064$ & $-2.092$
& Bubble number density  \\

$ \Gamma(t) $
& $  \sim t^{-\frac{m}{3}+\frac{7}{3}\beta} $
& $0.872$ & $0.817$
& Coalescence kernel  \\

$ r(t) $
& $ 
\sim t^{-\frac{m}{3}-\frac{2}{3}\beta} $
& $-1.192$ & $-1.275$
& Coalescence hazard rate \\

$ R(t) $
& $ 
\sim t^{-\frac{m}{3}-\frac{11}{3}\beta} $
& $-3.256$ & $-3.366$
& Volumetric coalescence event rate \\

$ \mathcal{B}(t) $
& $ \sim \varepsilon^{1/3} d^{-2/3}
\sim t^{-\frac{m}{3}-\frac{2}{3}\beta} $
& $-1.192$ & $-1.275$
& Breakup frequency  \\

$ d_H(t) $
& $ \sim \varepsilon^{-2/5} \sim t^{\frac{2m}{5}} $
& $0.880$ & $0.972$
& Hinze breakup threshold scale \\

$ d(t)/d_H(t) $
& $ \sim t^{\beta-\frac{2m}{5}} $
& $-0.192$ & $-0.275$
& Breakup suppression \\

$ \tfrac12-\alpha(t) $
& $ \sim \dfrac{d_H}{d}-1 \sim t^{\frac{2m}{5}-\beta} $
& $0.192$ & $0.275$
& Scaling approximation \\

\hline
\end{tabular}
\caption{Temporal scaling of key bubble-population and multiphase quantities during the transition regime where coalescence and breakup coexist. The turbulent dissipation decays as $\varepsilon(t)\sim t^{-m}$, and the characteristic bubble diameter follows $d(t)\sim t^{\beta}$ with $\beta=(15+m)/25$. Exponents (third and fourth columns) are evaluated for the representative decay rates $m=2.2$ and $m=2.43$. Details are available in ~\citet{vivek_prl}.}
\label{tab:scalings_exponents_m_coal_break}
\end{table}

For the mixed coalescence--breakup regime, the bubble population initially spans both sub-Hinze and super-Hinze sizes. During the early decay stage, breakup and coalescence are approximately balanced, so the bubble population is expected to span both sub-Hinze and super-Hinze sizes. Since \(d_H\) marks the crossover between turbulence-dominated fragmentation
and capillary-stabilized bubbles \citep{hinze1955,martinezbazan1999,deane2002}, we take \(d_{32}\approx d_H\) during the equilibrium bubble-size distribution, where coalescence and breakup jointly maintain a statistically stable population. The initial population is therefore assumed to be centred near the Hinze scale, with finite probability mass on both the sub-Hinze and super-Hinze sides~\citep{garrett2000connection,deane2002scale,riviere2021subhinze,ruth2022experimental}. We therefore take $\alpha(0)\simeq 1/2$ as a neutral initial condition. As the dissipation decays, $d_H$ increases and fewer bubbles remain above the breakup threshold. To capture this loss of breakable bubbles, the early-time number-based bubble-size distribution is approximated as Gaussian about the mean diameter $d(t)$ with width $s(t)\sim d(t)$. Expanding the complementary-error-function expression for the fraction above $d_H$ gives the leading-order estimate and using constant void fraction, $n\sim d^{-3}$, so that
\begin{equation}
    \frac{\mathrm{d}d}{\mathrm{d}t}
    \sim
    \varepsilon^{1/3}d^{1/3}
    \left(\frac{d_H}{d}\right),
    \label{eq:dddt_mixed_summary}
\end{equation}
where the factor $d_H/d$ represents the leading-order Hinze-drift correction to the breakable population, as derived in~\citet{vivek_prl}. Substituting $d(t)\sim t^\beta$, $\varepsilon(t)\sim t^{-m}$, and $d_H(t)\sim t^{2m/5}$ into Eq. \eqref{eq:dddt_mixed_summary} and balancing temporal exponents gives

\begin{equation}
    \beta_m=\frac{15+m}{25} .
    \label{eq:beta_mixed_summary}
\end{equation}
With \(\beta_m\), the mixed-regime bubble-size growth and other key scalings can be evaluated as shown in Table~\ref{tab:scalings_exponents_m_coal_break}. This regime is not simply a competition between two independent processes. Breakup initially produces a larger population of smaller bubbles, increasing $n$ and therefore enhancing the coalescence rate
\begin{equation}
    R_{\mathrm{c}}
    \sim
    \Gamma n^2
    \sim
    \varepsilon^{1/3}d^{7/3}(d^{-3})^2
    \sim
    \varepsilon^{1/3}d^{-11/3}.
\end{equation}
At the same time, the growth of $d_H$ progressively removes bubbles from the breakable population. Thus, breakup indirectly accelerates coalescence-driven growth at early times by sustaining a high number density, while turbulence decay suppresses further fragmentation.

\begin{table}
\centering
\small
\setlength{\tabcolsep}{6pt}
\renewcommand{\arraystretch}{1.0}
\begin{tabular}{lcccc}
\hline
\textbf{Quantity} & \textbf{Scaling} & \textbf{$m=2.2$} & \textbf{$m=2.43$} & \textbf{Physical description} \\
\hline
$ d(t) $ 
& $ \sim t^{\frac{3-m}{2}} $
& $0.40$ & $0.28$
& Growth of characteristic bubble size (e.g. $d_{32}$, $d_{99}$) \\

$ a_b(t) $
& $ \sim t^{3-m} $
& $0.80$ & $0.57$
& Growth of single-bubble surface area \\

$ A(t)$
& $ \sim t^{\frac{m-3}{2}} $
& $-0.40$ & $-0.28$
& Decay of interfacial area per unit volume \\

$ \Gamma(t) $
& $ \sim t^{\frac{7-3m}{2}} $
& $0.20$ & $-0.15$
& Inertial-range encounter kernel strength \\

$ n(t) $
& $ \sim t^{-\frac{9-3m}{2}} $
& $-1.20$ & $-0.85$
& Decay of bubble number density \\

$ r(t) $
& $ \sim t^{-1} $
& $-1.00$ & $-1.00$
& Per-bubble coalescence hazard rate \\

$ R(t) $
& $ \sim t^{\frac{3m-11}{2}} $
& $-2.20$ & $-1.85$
& Volumetric coalescence event rate \\

$ d_H(t) $
& $ \sim t^{\frac{2m}{5}} $
& $0.88$ & $0.97$
& Hinze breakup threshold scale \\

$ d(t)/d_H(t) $
& $\sim t^\frac{15-9m}{10} $
& $-0.48$ & $-0.69$
& Drift into sub-Hinze regime (breakup suppression) \\

$ u_{\mathrm{rel}}(t) $
& $ \sim t^{\frac{1-m}{2}} $
& $-0.60$ & $-0.72$
& Turbulent relative velocity at the bubble scale \\

$ Re_b(t)=\dfrac{u_{\mathrm{rel}} d}{\nu} $
& $ \sim t^{2-m} $
& $-0.20$ & $-0.43$
& Bubble Reynolds number \\

$ We_t(t)=\dfrac{\rho u_{\mathrm{rel}}^2 d}{\gamma} $
& $ \sim t^{\frac{15-9m}{6}} $
& $-0.80$ & $-1.15$
& Turbulent Weber number (breakup tendency) \\

$ \ell_b(t)\sim n^{-1/3} $
& $ \sim t^{\frac{3-m}{2}} $
& $0.40$ & $0.28$
& Mean inter-bubble spacing \\

$ k_d(t)\sim u_{\mathrm{rel}}^2 $
& $\sim t^{1-m} $
& $-1.20$ & $-1.43$
& Bubble-scale turbulent kinetic energy \\

\hline
\end{tabular}
\caption{Temporal scaling of key multiphase quantities and bubble-population statistics in dilute bubbly decaying HIT at $\langle \phi \rangle\ll 1$, where the turbulent dissipation follows $\varepsilon(t)\sim t^{-m}$. The table shows the upper and lower limits of the scalings (exponents in third and fourth columns) evaluated at slowest and fastest decay cases, $m=2.2$ and $m=2.43$. Details are available in~\citet{vivek_prl}.}
\label{tab:scalings_exponents_m_coal}
\end{table}

In the pure-coalescence regime, where \(\alpha(t)\) becomes negligible, Eq. \eqref{eq:n_balance_summary_reduced} reduces to
\begin{equation}
    \frac{\mathrm{d}d}{\mathrm{d}t}
    \sim
    \varepsilon^{1/3}d^{1/3}.
    \label{eq:dddt_pure_summary}
\end{equation}
Using $\varepsilon(t)\sim t^{-m}$ and integrating gives the pure-coalescence growth exponent 
\begin{equation}
    \beta_c=\frac{3-m}{2}.
    \label{eq:beta_pure_summary}
\end{equation}

From Eq.~\eqref{eq:beta_pure_summary}, all key bubbly-flow statistics, physical properties, and PBE closures follow directly, as summarized in Table~\ref{tab:scalings_exponents_m_coal}. The bubble diameter growth exponent in the mixed regime is higher compared to pure coalescence regime
\begin{equation*}
\beta_m-\beta_c
= \frac{27}{50}\left(m-\frac{5}{3}\right).
\end{equation*}
Physically, residual breakup, which is proportional to \(n(d>d_H)\), acts as a transient number-density reservoir. It creates additional smaller bubbles and therefore increases the number of possible coalescing pairs, which scale as \(n^2\), while the growing Hinze scale simultaneously removes the breakable population and drives the system toward the pure-coalescence regime. Once breakup becomes negligible, the number density decays, the volumetric collision rate decreases, and the resulting scaling follows pure coalescence regime. The transition between the two regimes is controlled by the drift of the bubble population relative to the Hinze scale. In the mixed regime,
\begin{equation}
    \frac{d(t)}{d_H(t)}
    \sim
    t^{\beta_m-2m/5}
    =
    t^{(15-9m)/25},
    \label{eq:d_over_dH_mixed_summary}
\end{equation}
whereas in the pure coalescence regime,
\begin{equation}
    \frac{d(t)}{d_H(t)}
    \sim
    t^{\beta_c-2m/5}
    =
    t^{(15-9m)/10}.
    \label{eq:d_over_dH_pure_summary}
\end{equation}
For $m>5/3$, both exponents are negative. The Hinze scale therefore grows faster than the characteristic bubble size, the fraction of super-Hinze bubbles decreases, and the system moves progressively into a sub-Hinze, coalescence-dominated state. This establishes the dynamical basis for a transition from an early mixed coalescence--breakup regime to a later pure-coalescence regime.

\section{Numerical and experimental setup} \label{sec:exp and DNS setup}

\subsection{Direct numerical simulations} \label{sec:DNS section}
The scalings derived above rest on a central physical assumption: in dilute decaying turbulence, the Hinze scale grows faster than the characteristic bubble size, driving the population from an early mixed coalescence--breakup state toward a sub-Hinze, coalescence-dominated regime. We test this mechanism using direct numerical simulations (DNS) of dilute bubbly homogeneous isotropic turbulence, where turbulence decay, bubble deformation, collisions, coalescence, and breakup are resolved without the added effects of walls or mean shear. Unlike the reduced theory, the DNS does not prescribe the bubble-population response, but resolves the coupled bubble--turbulence dynamics directly, providing a controlled and physically resolved test of the predicted scalings before comparison with the spatially developing duct-flow experiments. Here, we study incompressible two-phase decaying turbulent flows. The two phases are treated within a diffuse-interface framework, where interfacial dynamics are resolved using the accurate conservative diffuse-interface (ACDI) phase-field formulation, which ensures boundedness of the phase field and robust interface preservation~\citep{jain2022a,jain2020}. The mixture velocity field satisfies the incompressibility constraint,
\begin{equation}
\frac{\partial u_i}{\partial x_i} = 0,
\label{eq:continuity}
\end{equation}
and the conservation of momentum for the variable-density mixture is given by
\begin{equation}
\frac{\partial (\rho u_i)}{\partial t}
+ \frac{\partial (\rho u_i u_j)}{\partial x_j}
= -\frac{\partial p}{\partial x_i}
+ \frac{\partial \tau_{ij}}{\partial x_j}
+ \frac{\partial (u_i f_j)}{\partial x_j}
+ \gamma \tilde{\kappa} \frac{\partial \phi}{\partial x_i}
+ F_i.
\label{eq:momentum}
\end{equation}
Here, $u_i$ and $u_j$ denote the components of the mixture velocity field, $\rho$ and $p$ denote the mixture density and pressure, respectively, $\gamma$ is the surface-tension coefficient, $F_i$ is an external force per unit volume, and $\tau_{ij}$ represents the viscous stress tensor,
\begin{equation}
\tau_{ij}
= \mu\left(
\frac{\partial u_i}{\partial x_j}
+ \frac{\partial u_j}{\partial x_i}
\right),
\label{eq:stress_tensor}
\end{equation}
where $\mu$ is the mixture dynamic viscosity. Consistent momentum transport across the diffuse interface is ensured through the regularization term $u_i f_j$, where $f_j$ is the mass flux associated with the phase-field regularization,
\begin{equation}
f_i = \rho_1 a_{1,i} + \rho_2 a_{2,i}
= (\rho_1 - \rho_2)a_{1,i}.
\label{eq:regularization}
\end{equation}
Here, $\rho_1$ and $\rho_2$ are the phase densities of phases 1 and 2, respectively, $a_{m,i}$ is the interface-regularization flux for phase $m\in \{1,2\}$, and the final equality follows from the two-phase constraint $\phi_1+\phi_2=1$, which gives $a_{2,i}=-a_{1,i}$. Surface tension effects are incorporated via the continuum-surface-force formulation, where $\phi\equiv\phi_1$ denotes the phase indicator used to locate the interface and $\tilde{\kappa}$ is the numerically evaluated interface curvature. The evolution of the interface is described through the transport of the phase volume fraction $\phi_m$ of phase $m$, governed by
\begin{equation}
\frac{\partial \phi_m}{\partial t}
+ \frac{\partial (\phi_m u_j)}{\partial x_j}
= \frac{\partial a_{m,j}}{\partial x_j},
\label{eq:phi_transport}
\end{equation}
where $a_{m,j}$ denotes the interface-regularization, or sharpening, flux that maintains the equilibrium hyperbolic-tangent profile of the diffuse interface and is defined as \begin{equation} 
a_{m,j} = \overline{\Gamma} \left[ \overline{\varepsilon} \frac{\partial \phi_m}{\partial x_j} - \frac{1}{4}\left(1-\tanh^2\left(\frac{\psi_m}{2 \overline{\varepsilon}}\right)\right) n_{m,j} \right], 
\label{eq:sharpening_flux} 
\end{equation}
$\overline{\Gamma}$ is an interface velocity scale taken as $\max(|\mathbf{u}|)$, where $\mathbf{u}$ is the mixture velocity vector, and $\overline{\varepsilon}$ is the prescribed interface thickness, set to $0.51\Delta$, where $\Delta$ is the grid spacing. To improve the accuracy of geometric quantities at the interface, an auxiliary signed-distance-like function $\psi_m$ is introduced \citep{jain2022a},
\begin{equation}
\psi_m = \overline{\varepsilon}
\ln\!\left(\frac{\phi_m + \delta}{1 - \phi_m + \delta}\right),
\label{eq:psi_def}
\end{equation}
where $\delta$ is a small regularization parameter used to avoid singular values near $\phi_m=0$ and $\phi_m=1$. The function $\psi_m$ varies more smoothly across the interface than $\phi_m$. The corresponding unit normal vector is then computed as
\begin{equation}
n_{m,i} =
\frac{\partial \psi_m / \partial x_i}
{\sqrt{\left(\partial \psi_m / \partial x_k\right)
\left(\partial \psi_m / \partial x_k\right)}}.
\label{eq:normal_vector}
\end{equation}
For the continuum-surface-force term, we take $\phi\equiv\phi_1$, $\psi\equiv\psi_1$, and $n_i\equiv n_{1,i}$, from which the interface curvature is evaluated as $\tilde{\kappa}=-\partial n_i/\partial x_i$.

A set of incompressible two-phase decaying homogeneous isotropic turbulence (HIT) simulations is performed in a triply periodic cubic domain. The governing equations are discretized using a finite-volume formulation on a structured Cartesian grid, in which the velocity components are staggered relative to scalar variables. This arrangement suppresses pressure checkerboarding and promotes discrete kinetic-energy conservation.

Spatial discretization employs a low-dissipation, second-order skew-symmetric scheme with implicit kinetic-energy preservation, which has been shown to remain robust even at very high Reynolds numbers. Time integration is carried out using a fourth-order Runge--Kutta method. The incompressibility constraint is enforced through a modified fractional-step approach, wherein the pressure field projects the intermediate velocity onto a divergence-free space at each sub-time step. All simulations are conducted using the in-house, GPU-accelerated \textsc{ExaFlow3D} solver~\citep{jain2022a,jain2025stationary}. \\

The simulations are initialized using a three-dimensional velocity field with a prescribed energy spectrum. The initial velocity field is generated from an energy spectrum whose turbulence statistics are calibrated against downstream measurements from the highly turbulent duct flow~\citep{kumar2026bubble}. This preserves realistic large-scale turbulence characteristics without directly imposing the experimental velocity field. The dispersed phase is introduced into the turbulent flow field at the initial time, $t_0=4.5$, after which the coupled bubbly flow evolves freely as the turbulence decays.  The Taylor-scale Reynolds number of the carrier turbulence is initially at $\mathrm{Re}_\lambda = 206,$ corresponding to a highly turbulent flow with a well-developed inertial range. The computational domain size and grid resolution are selected based on standard HIT scaling arguments. The integral length scale \(\ell\) is estimated using the classical HIT relation
\begin{equation}
\varepsilon = C_\varepsilon \frac{k^{3/2}}{\ell},
\end{equation}
where \(k\) denotes the turbulent kinetic energy, \(\varepsilon\) the dissipation rate, and \(C_\varepsilon\) is an order-unity constant. Using \(k = 1.19 \, m^2/s^2\) and \(\varepsilon = 225 \, m^2/s^3 \) corresponding to $Re_\lambda$ = 206, together with \(C_\varepsilon \approx 0.5\), yields an integral length scale of \(\ell \approx 3.1~\mathrm{mm}\)~\citep{pope2000turbulent}. The domain size is therefore chosen as $L_{\mathrm{box}} = 5\ell \approx 15.6~\mathrm{mm},$ which ensures adequate scale separation between the largest energy-containing eddies and the domain boundaries. The smallest dynamically relevant length scale is characterized by the Kolmogorov length
\begin{equation}
\eta = \left(\frac{\nu^3}{\varepsilon}\right)^{1/4},
\end{equation}
which evaluates to \(\eta \approx 8~\mu\mathrm{m}\) for the initial flow conditions. The simulations are carried out on a uniform Cartesian grid with $1024^3$ grid points. This resolution ensures full resolution of the smallest turbulent scales and interface scales, and meets DNS resolution criteria for two-phase flows \citep{hatashita2025scalings}.

\begin{wrapfigure}{r}{0.50\columnwidth}
  \vspace{-0.5\baselineskip}
  \raggedleft
  \includegraphics[width=\linewidth]{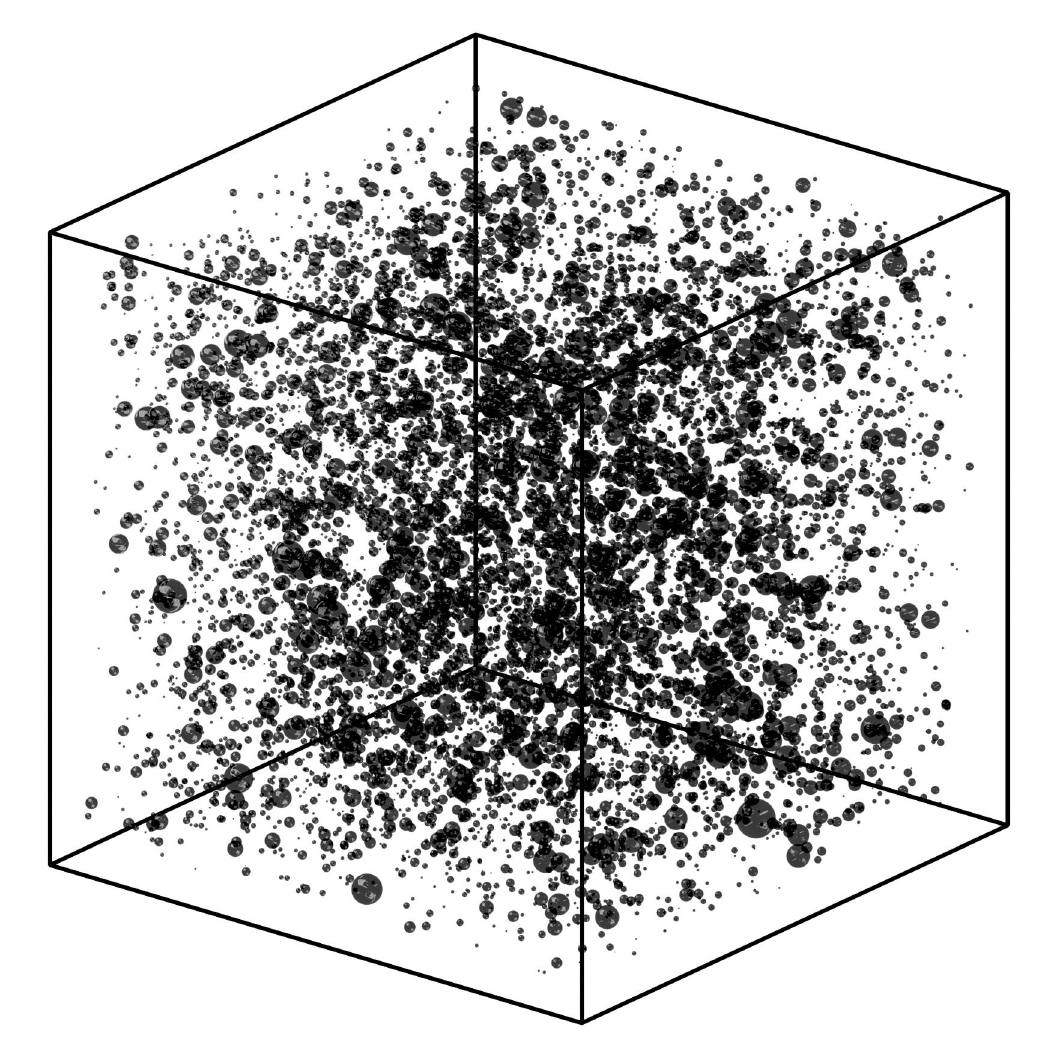}
  \caption{Initial bubble distribution in the computational domain.}
  \label{fig:bubble-distribution}
  \vspace{0.5\baselineskip}
\end{wrapfigure}

The large-scale (integral length scale, $\ell$) Weber number is defined as
\begin{equation}
\mathrm{We}_\ell = \frac{2\rho k^{5/2}}{3\sigma\varepsilon},
\end{equation}
and is set to \(\mathrm{We}_\ell = 63\). The density and viscosity ratios between the two phases are chosen to represent an air--water system. The reference densities, based on a nondimensional domain scaling of \((2\pi)^3\) corresponding to the chosen physical box size \(L_{\mathrm{box}}\), are set to $\rho_{1,0}$ = 1, and $\rho_{2,0}$ = 1.23 $\times 10^{-3}$, while the dynamic viscosities are $\mu_1$ = 4.52 $\times 10^{-4}$, and $\mu_2$ = 8.14 $\times 10^{-6}$. These values yield realistic contrasts in inertial and viscous stresses across the interface while maintaining numerical stability. During the rescaling to the \((2\pi)^3\) nondimensional domain, all key nondimensional parameters, including \(\mathrm{Re}_\lambda\) and \(\mathrm{We}_\ell\), are preserved.  \\

Similar to the turbulent flow field, the initial bubble-size distribution is prescribed based on experimental measurements. The experiments were conducted over sufficiently long sampling durations to ensure statistical convergence of the bubble-size probability density function (PDF), such that further data acquisition did not alter the measured distribution. For each void fraction, the total number of bubbles and their diameters were selected to match the experimentally observed bubble population. At the initial time $t_0 = 0$, the bubbles are introduced as shown in Figure~\ref{fig:bubble-distribution} with a prescribed initial size distribution and are nearly spherical in shape (randomly distributed without touching or overlapping). The bubbles were then initialized in the computational domain by randomly distributing them within the cubic volume, subject to the constraint that no bubbles overlap or are in contact at the initial time.\\

\begin{figure}[t]
    \centering
    \hfill
    \begin{subfigure}[b]{0.32\textwidth}
        \includegraphics[width=\linewidth]{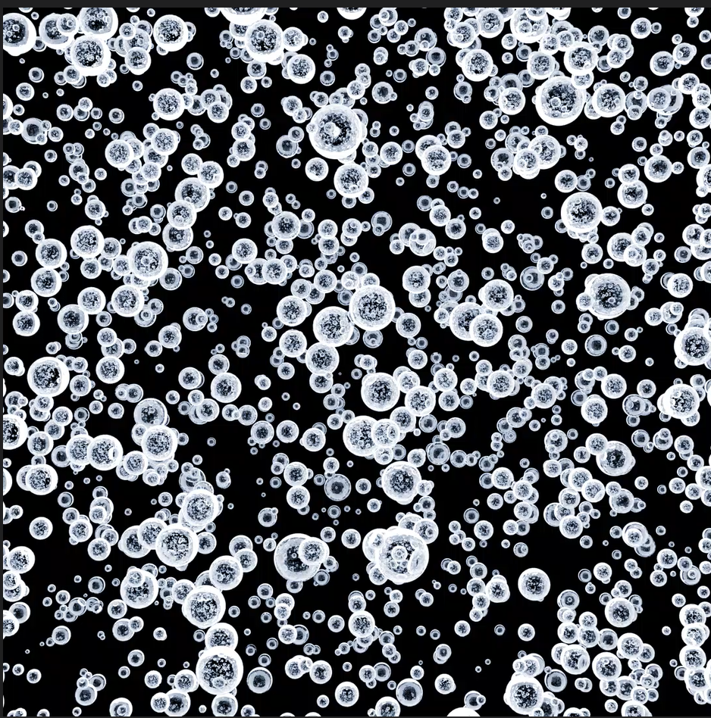}
        \caption{initial distribution ($t_0$)}
        \label{fig:inst-a}
    \end{subfigure}
    \hfill
    \begin{subfigure}[b]{0.32\textwidth}
        \includegraphics[width=\linewidth]{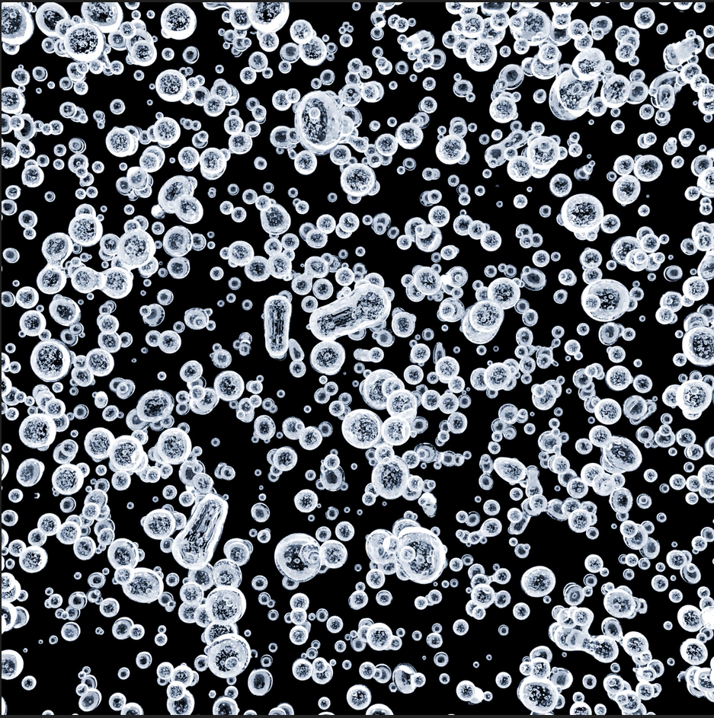}
        \caption{intermediate time ($t_1$)}
        \label{fig:inst-b}
    \end{subfigure}
    \hfill
    \begin{subfigure}[b]{0.32\textwidth}
        \includegraphics[width=\linewidth]{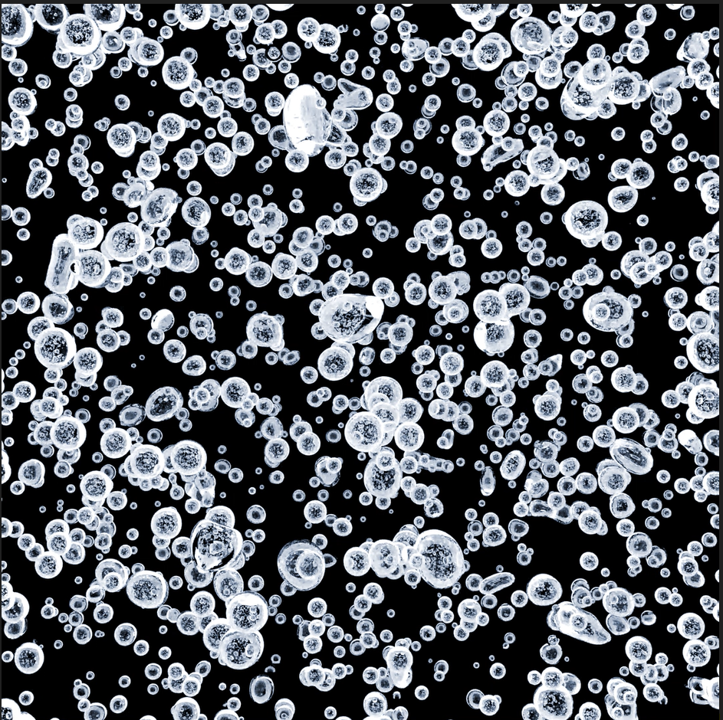}
        \caption{later time ($t_2$)}
        \label{fig:inst-c}
    \end{subfigure}
    \hfill
    \caption{Instantaneous snapshots of the DNS at three stages.}
    \label{fig:combined}
\end{figure}

Figure~\ref{fig:combined} shows instantaneous snapshots of the bubble field in HIT within a $(2\pi)^3$ cubic domain. As time progresses, interactions with turbulent eddies induce bubble deformation and generate relative motion between neighboring bubbles. This relative motion leads to frequent bubble collisions and subsequent coalescence, as observed at time $t_1 > t_0$. With further temporal evolution, continued coalescence increases the characteristic bubble size, as evident from the sequence $t_0 \rightarrow t_1 \rightarrow t_2$, leading to a  corresponding reduction in the total number of bubbles within the domain\footnote{the corresponding time-resolved bubble evolution is provided in supplementary video~\textit{S.mp4}}.

\begin{figure}[htbp]
    \begin{subfigure}[b]{0.48\textwidth}
        \includegraphics[width=\textwidth]{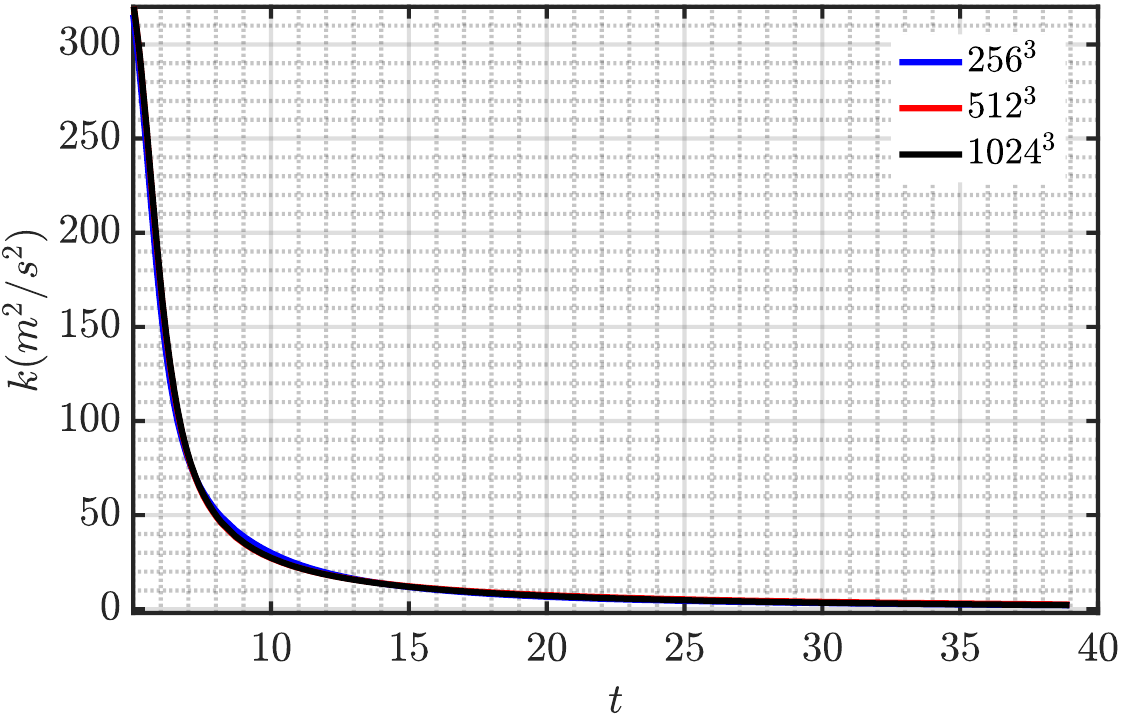}
        \caption{Temporal evolution of $k(t)$ for different grid resolutions.}
        \label{fig:b}
    \end{subfigure}
    \hfill
    \begin{subfigure}[b]{0.48\textwidth}
        \includegraphics[width=\textwidth]{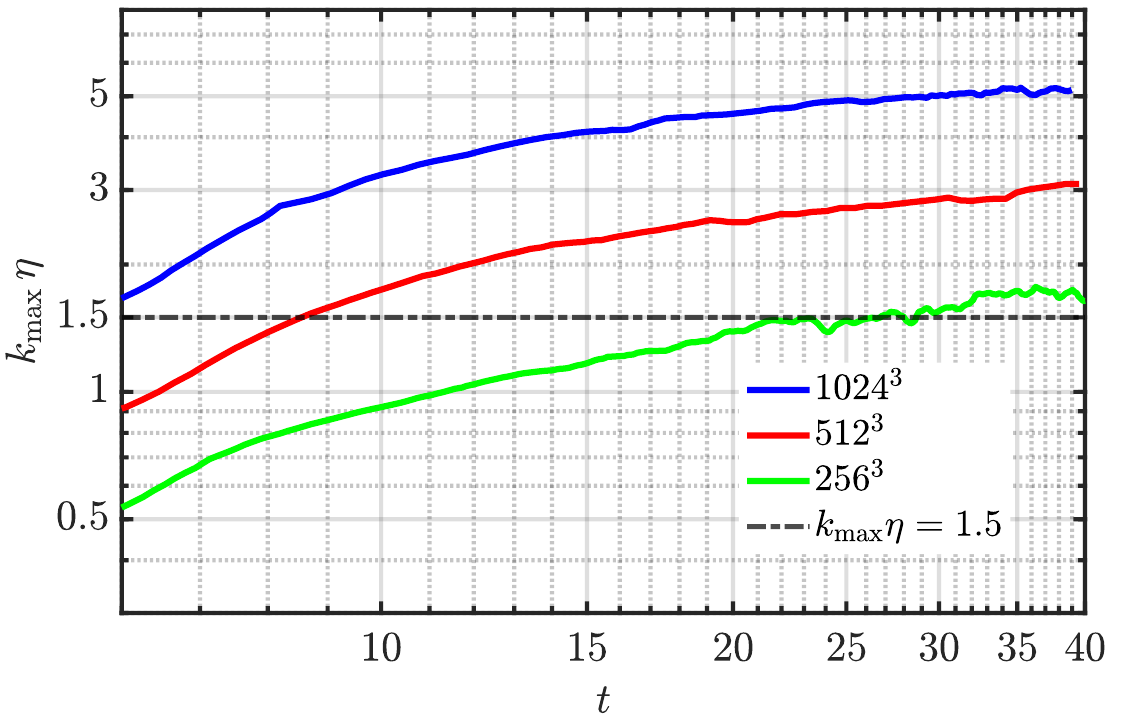}
        \caption{Temporal evolution of $\kappa_\mathrm{max}\eta(t)$  for different grid resolutions.}
        \label{fig:c}
    \end{subfigure}    
    \caption{Grid convergence test.}
    \label{fig:grid convergance}
\end{figure}

Figure~\ref{fig:grid convergance}(a) shows the temporal evolution of the turbulent kinetic energy, $k(t)$, for the $256^3$, $512^3$, and $1024^3$ DNS. The $512^3$ case exhibits excellent agreement with the $1024^3$ over the majority of the simulation, whereas the coarsest grid ($256^3$) displays small deviations, particularly during the early stages. Quantitatively, the maximum deviation in $k(t)$ between the $256^3$ and $1024^3$ simulations reaches approximately $10\%$, indicating insufficient resolution to capture the full range of dynamically relevant turbulent scales. In contrast, the deviation between the $512^3$ and $1024^3$ grids remains small, increasing gradually before reaching approximately $4\%$ and approaching a nearly constant value for $t \gtrsim 35$. Such behaviour is consistent with DNS of homogeneous turbulence, where large-scale and inertial-range statistics converge more rapidly under grid refinement than dissipation-range quantities \citep{pope2000turbulent,MoinMahesh1998}. \\

Figure~\ref{fig:grid convergance}(b) shows the grid resolution quantified by $\kappa_{\max}\eta(t)$, where $\kappa_{\max}=\pi/\Delta$ and the Kolmogorov length scale is $\eta=(\nu^3/\varepsilon)^{1/4}$; here, $\nu$ is the kinematic viscosity. The most stringent requirement occurs during the brief initial transience, when $\varepsilon$ is maximum and $\eta$ is smallest, as typical of decaying homogeneous isotropic turbulence \citep{SagautCambon2008}. For the finest grid, the minimum value is $\kappa_{\max}\eta \approx 1.65$. As turbulence decays, $\varepsilon(t)$ decreases and $\eta(t)$ increases, so $\kappa_{\max}\eta(t)$ rapidly exceeds the accepted DNS threshold $\kappa_{\max}\eta \gtrsim 1.5$ and remains above it throughout the statistical analysis window. For the $256^3$ grid, $\kappa_{\max}\eta$ remains well below unity over much of the simulation, consistent with the underprediction of $k(t)$. The $512^3$ grid initially gives $\kappa_{\max}\eta \approx 1$, but during a period dominated by large-scale motions that remain well resolved even on moderately coarse grids. Thus, despite temporarily marginal resolution of the smallest dissipative scales, the $512^3$ simulation captures the energy-containing eddies and shows only minor deviations in $k(t)$ relative to the $1024^3$ case. This agrees with DNS using low-dissipation central finite-difference schemes, where transient marginal resolution of dissipative scales has negligible influence on large- and inertial-scale dynamics~\citep{LaizetLamballais2009,Moureau2011}. Overall, the convergence of $k(t)$ and the sustained condition $\kappa_{\max}\eta \gtrsim 1.5$ over the analysis window indicate that the $1024^3$ grid is sufficiently resolved. Further refinement would mainly affect the earliest transient dissipation-range dynamics and would not alter the conclusions drawn herein~\citep{pope2000turbulent,MoinMahesh1998,calado2024dynamics}. All bubble-turbulence interaction statistics are therefore evaluated where $\kappa_{\max}\eta \gtrsim 1.5$, ensuring adequate resolution of the dissipative scales governing breakup and coalescence dynamics.

\begin{figure}[htbp]
    \centering
    \includegraphics[width=0.6\textwidth]{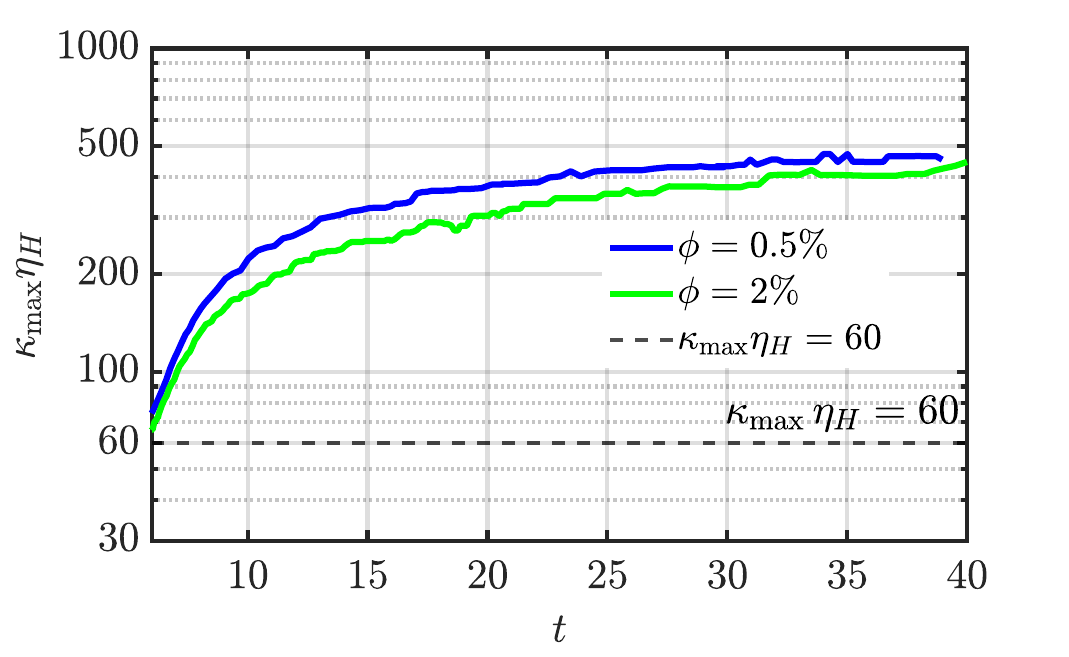}
    \caption{Temporal variation of \(\kappa_{\max}\eta_H\) \((\eta_H\approx d_H)\), showing the grid-convergence ratio for the two-phase flow on the \(1024^3\) grid at \(\phi=0.5\%\) and \(2\%\).}
    \label{fig:grid two phase}
\end{figure}

The adequacy of the interface resolution is further examined using the metric $\kappa_{\max}\eta_H$, where $\kappa_{\max}=\pi/\Delta$ denotes the maximum resolved wavenumber and $\eta_H \approx d_H$ is the Hinze scale associated with turbulence-induced breakup. Figure~\ref{fig:grid two phase} shows the temporal evolution of $\kappa_{\max}\eta_H$ for the $1024^3$ grid at void fractions $\phi=0.5\%$ and $\phi=2\%$. Since the dissipation rate decreases in decaying turbulence, the Hinze scale increases with time, leading to a progressive increase in $\kappa_{\max}\eta_H$ over the course of the simulation. For reference, the baseline value $\kappa_{\max}\eta_H=60$ is indicated in Figure~~\ref{fig:grid two phase}, following the grid-criterion proposed by \citet{Hatashita}, where they observed that key two-phase statistics become effectively resolution-independent beyond this value (not necessarily as a minimum threshold). At the beginning of the analysed interval ($t\approx6$), the present simulations already satisfy this criterion with $\kappa_{\max}\eta_H\approx68$ for $\phi=2\%$ and $\kappa_{\max}\eta_H\approx75$ for $\phi=0.5\%$. As the turbulence decays and the Hinze scale grows, $\kappa_{\max}\eta_H$ increases further, reaching values exceeding $O(10^2)$ and approaching $O(10^3)$ at later times. These results indicate that the interface scales relevant to bubble deformation and breakup remain well resolved throughout the interval used for statistical analysis.

\subsection{Experimental setup and methodologies} \label{sec:experimental section}

\begin{figure}[ht]
\centering
\includegraphics[width=\textwidth]{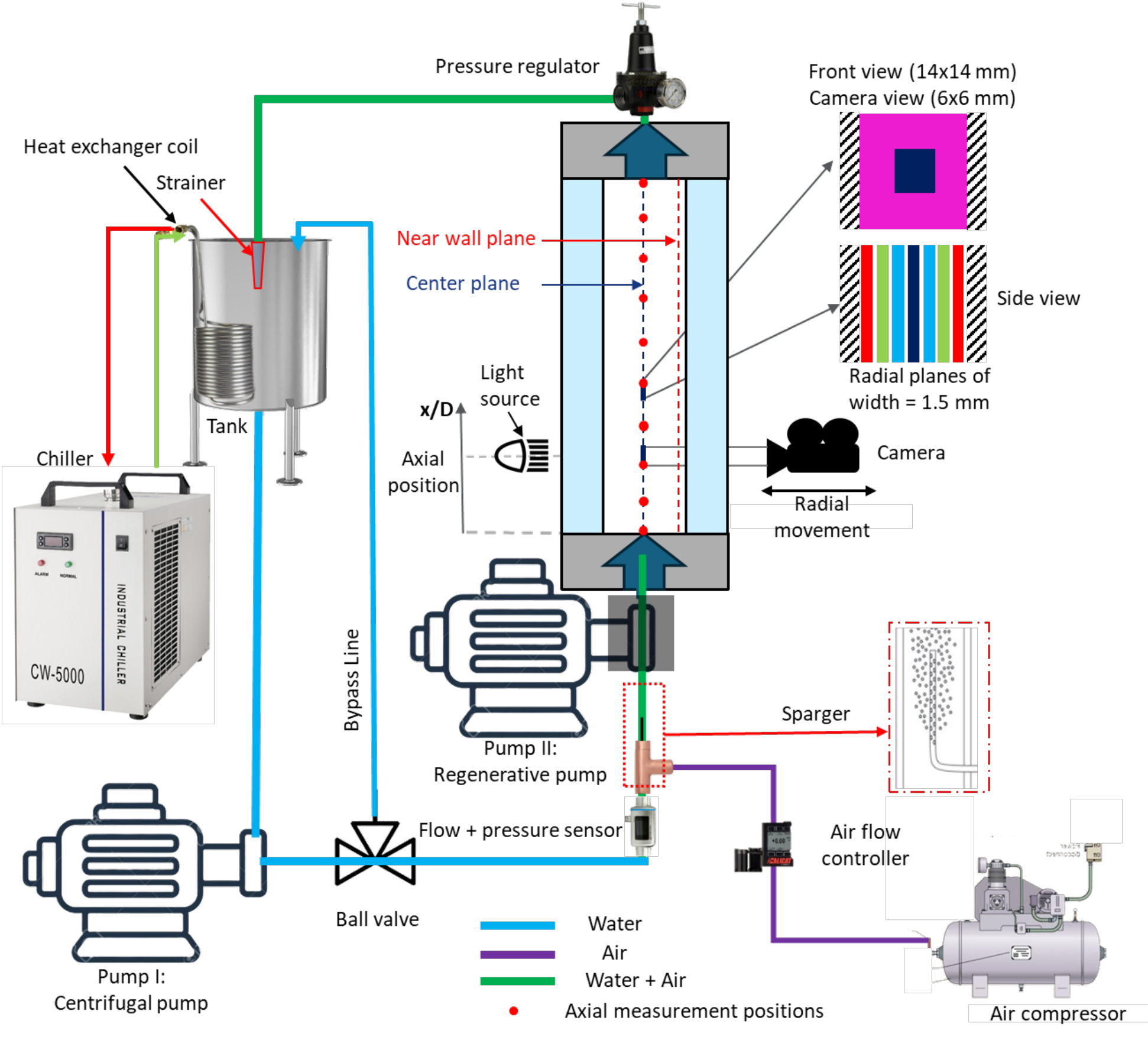}
\caption{Schematic of experimental setup and flow loop. Red markers: ten axial positions. Front view: channel cross-section $14\times14\,\mathrm{mm}$ (magenta); visualisation window $6\times6\,\mathrm{mm}$ (dark blue). Side view: measurements in seven radial planes from near-wall (red) to centre plane (dark blue), each $1.5\,\mathrm{mm}$ wide.}
\label{fig:SetupSchematic}
\end{figure}
The theory and DNS examine bubble-population evolution in an idealized HIT setting, which is essential for isolating the underlying physics of Hinze-scale drift and the transition from mixed coalescence--breakup dynamics to a sub-Hinze, coalescence-dominated regime. Practical bubbly flows, however, are more complex, evolving in spatially developing turbulence with confinement, mean shear, wall effects, and strong near-wall inhomogeneity. It is therefore necessary to test whether the same bubble-population scalings persist in a more realistic configuration, such as a decaying turbulent bubbly flow in a duct. A schematic of our experimental setup is shown in Figure~\ref{fig:SetupSchematic}, highlighting its main components \citep{javadi2026large,kumar2026bubble}. The setup consists of a vertically oriented test section downstream of a regenerative turbine pump (denoted as pump II in Figure~\ref{fig:SetupSchematic}). The test section has internal dimensions of $13.97\pm0.03$\,mm $\times$ $13.97\pm0.03$\,mm and a total height of $60.96\pm0.30$\,cm, corresponding to an aspect ratio of 40. The polycarbonate walls of the test section are $\sim 95\%$ transparent, providing sufficient optical access.

\begin{figure}[htbp]
    \centering
    \begin{subfigure}[t]{0.6\textwidth}
        \centering
        \renewcommand{\arraystretch}{1.0}
        \vspace{0pt}
        \begin{tabular}{|c|c|c|c|c|}
            \hline
             & \textbf{$Q$ ($V$)} &  &  &  Re$_\lambda$ \\
            \textbf{$\mathcal{L}$} & L/min (m/s) & $\phi$ (\%) & \vk{Re ($\times 10^4$)} & ($\mathcal{L}\approx0$) \\
            \hline 
            \multirow{3}{*}{\makecell[c]{\\ 0, 3.6, 8.2,  \\[0.7em]
                12.7, 17.8,  \\[0.7em]                  
                21.8, 26.4,  \\[0.7em]
                30.9, 35.5,  \\[0.7em]
                40.0}}
                &  & 0.5 &  & \\ 
                &  72 (6.1)  & 1   & 9.4 & 583 \\ 
                &    & 2   &  & \\ 
            \cline{2-5}
            & \multirow{3}{*}{87 (7.4)} 
                & 0.5 &  & \\
                &     & 1   & 11.4 & 825 \\
                &     & 2   &  &  \\
            \cline{2-5}
            & \multirow{3}{*}{98 (8.4)} 
                & 0.5 &  & \\
                &     & 1   & 12.9 & 918 \\
                &     & 2   &  & \\ \hline
        \end{tabular}
    \end{subfigure}%
    \hfill
    \begin{subfigure}[t]{0.35\textwidth}
        \centering
        \vspace{3pt}
        \includegraphics[width=0.95\textwidth]{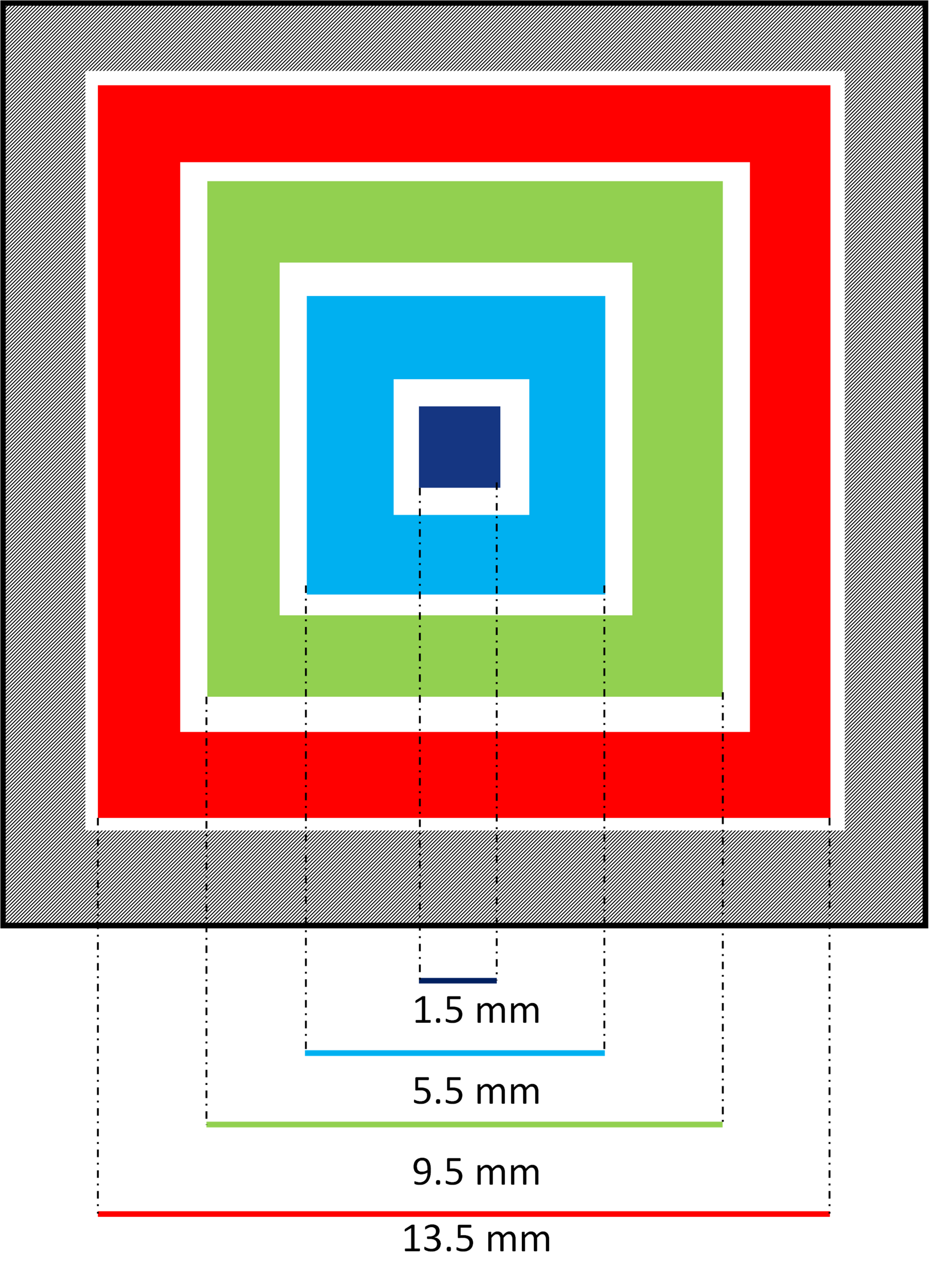}
    \end{subfigure}
    \caption{\textbf{Left:} Experimental conditions, including axial positions ($\mathcal{L}$), flow rates ($Q$), bulk velocity ($V$), void fraction ($\phi$), bulk Reynolds number (Re $= V \text{D} / \nu$), and Taylor Reynolds number \vk{(Re$_\lambda = \mathcal{U} \lambda_T / \nu$)}, where D is the duct hydraulic diameter, $\nu$ is the liquid kinematic viscosity, $\mathcal{U}$ is the root-mean-square (rms) value of the velocity fluctuations, and $\lambda_T$ is the Taylor microscale. \textbf{Right:} Radial measurement locations in a $14 \, \mathrm{mm} \times14 \, \mathrm{mm}$ duct region. Each measurement plane [red (near wall), green, sky blue, dark blue (center)] is $1.5\,\mathrm{mm}$ thick and separated by $0.5\,\mathrm{mm}$ white spacing, and wall to red plane clearance is 0.25\,mm. }
    \label{fig:exp-conditions-and-image}
\end{figure}

Bubble-size measurements were performed at ten streamwise locations, spanning $x/D = \mathcal{L} = 3.6$--40, where $D=13.97$\,mm is the hydraulic diameter of the test section duct, and at four radial positions including the centerline, near-wall region, and intermediate locations. Shadowgraph-based imaging was employed using a high-speed camera fitted with a high-magnification lens and with a halogen backlight. The camera was interfaced with a data acquisition system to record high-resolution images at each measurement location. In parallel, particle shadow velocimetry (PSV)~\citep{estevadeordal2005,khodaparast2013,hessenkemper2018,jassal2025particle} was conducted over a similar visualization window to estimate turbulence statistics, including velocity fields, turbulent kinetic energy $k$, turbulence intensity $\mathcal{I}$, and the dissipation rate $\varepsilon$. More details on the experimental setup and its characterization for use in the investigation of multiphase flows can be found in \citet{kumar2026bubble}.

The experiments were designed to investigate bubble dynamics in a highly turbulent, decaying duct flow over a wide range of operating conditions, as summarized in Figure~\ref{fig:exp-conditions-and-image}(a). Three bulk velocities corresponding to volumetric flow rates of $Q=72$, 87, and 98\,L/min (6.1--8.4\,m/s) were examined, yielding bulk Reynolds numbers of $9.4\times10^4$, $1.14\times10^5$, and $1.29\times10^5$. For each velocity, three inlet void fractions were imposed, defined as $\langle \phi \rangle= Q_{\mathrm{air}}/Q = Q_{\mathrm{air}}/(Q_{\mathrm{air}}+Q_{\mathrm{liquid}}) = 0.5$, 1, and 2\%, where $Q_{\mathrm{air}}$ denotes the air flow rate. The corresponding Taylor-scale Reynolds number near the inlet ranged from 583 to 918. Radial variations were resolved using four discrete interrogation windows, as illustrated in Figure~\ref{fig:exp-conditions-and-image}(b). Complete details of the experimental methods, including shadowgraph bubble imaging, particle shadow velocimetry, pre- and post-processing steps, corrections for limited diagnostic resolution, and associated uncertainties, can be found in~\citet{kumar2026bubble}.

\section{Results and discussion} \label{section:Results and Discussions}
The results are presented in three parts. We first characterize the evolving turbulent flow field in decaying HIT by analysing the temporal decay of \(k(t)\) and \(\varepsilon(t)\), and then relate this behaviour to the spatial decay of turbulence in the pump-generated duct flow. We next examine how the bubble population responds to this evolving turbulence, focusing on the bubble-size distribution, characteristic bubble diameters, interfacial area, and bubble count in both HIT and the decaying duct-flow configuration. Finally, we analyse the coalescence dynamics through the volumetric coalescence rate, per-bubble coalescence hazard, and effective coalescence kernel, and discuss their implications for closure modelling in population-balance equations.

\subsection{Evolution of the turbulent flow field}

\begin{figure}[htbp]
    \begin{subfigure}[b]{0.48\textwidth}
        \includegraphics[width=\textwidth]{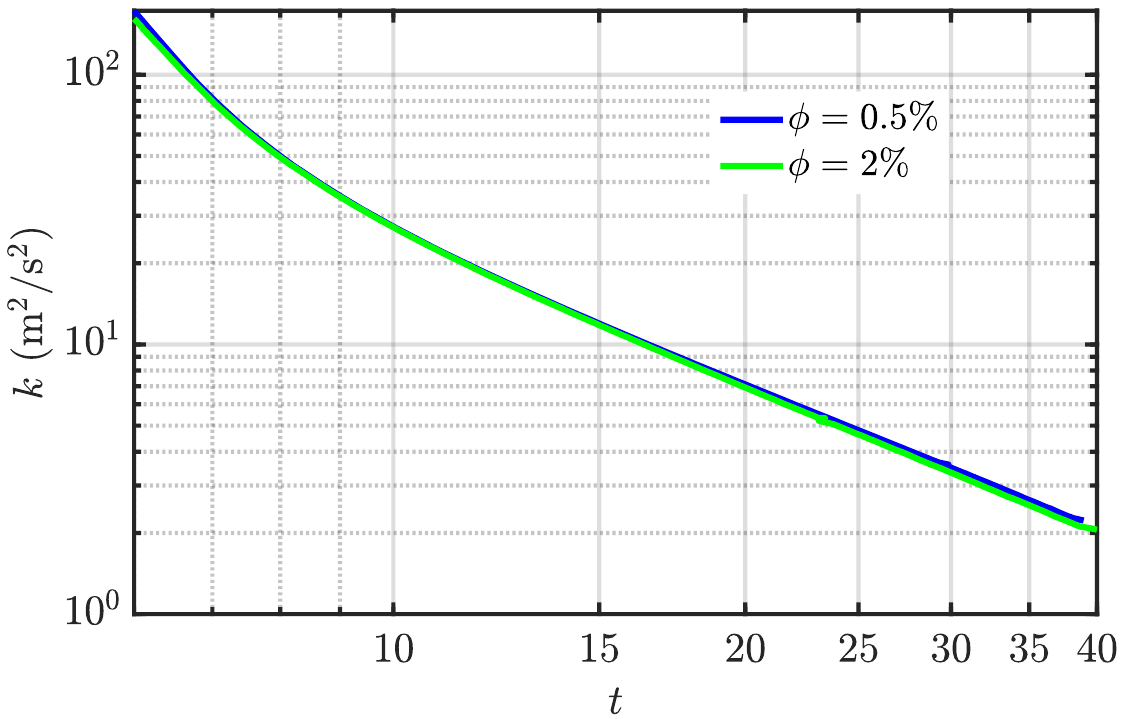}
        \caption{Temporal decay of turbulent kinetic energy, $k(t)$, at $\langle \phi \rangle=0.5\%$ and $2\%$.}
        \label{fig:tke_decay_dns}
    \end{subfigure}
    \hfill    
    \begin{subfigure}[b]{0.48\textwidth}
        \centering
        \includegraphics[width=\textwidth]{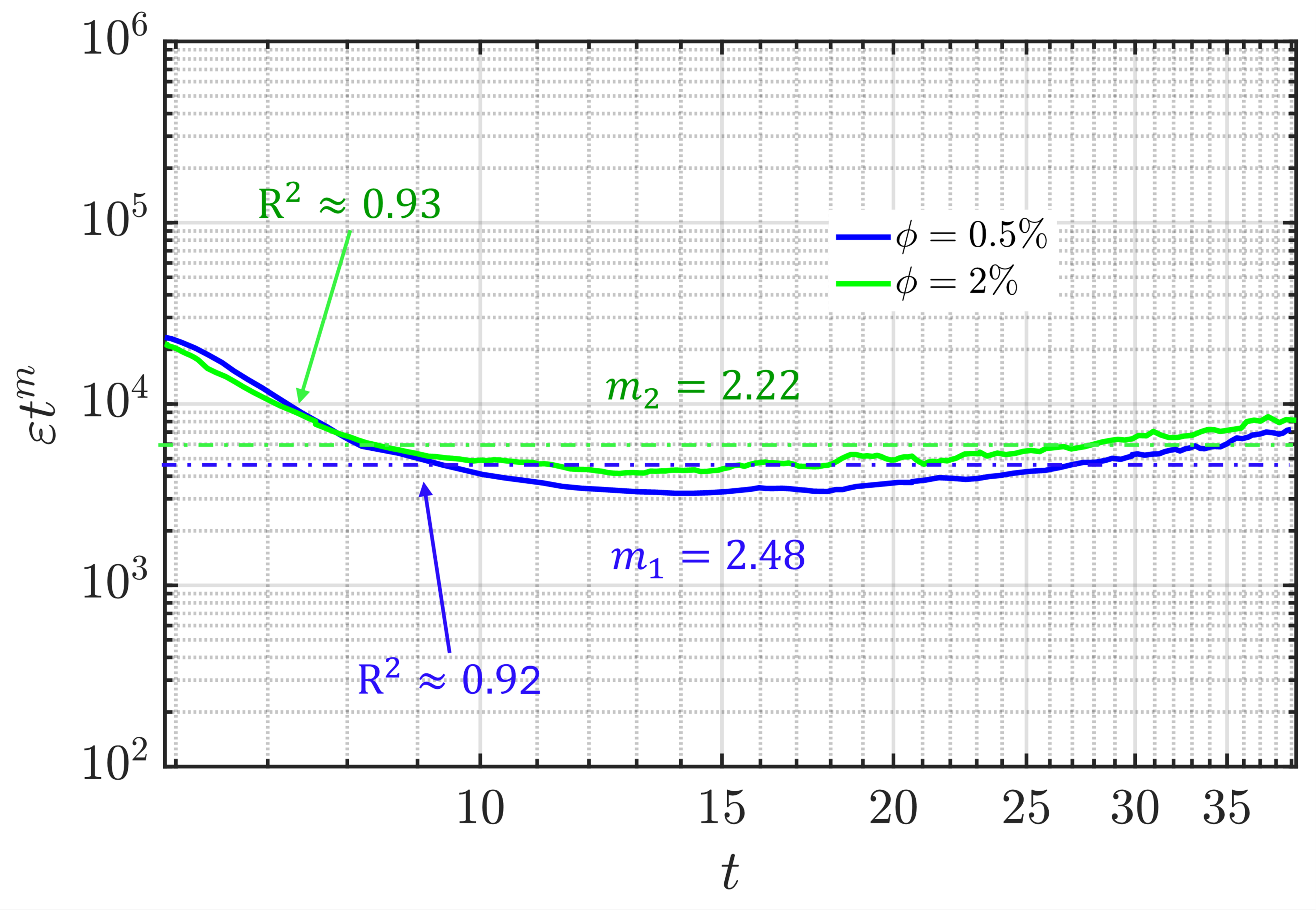}
        \caption{Compensated dissipation rate, $\varepsilon(t)t^m$with $m=2.48$ for $\phi=0.5\%$ and $m=2.22$ for $\phi=2\%$. }
        \label{fig:epsilon_compensated_dns}
    \end{subfigure} 
    \caption{Temporal evolution of turbulent kinetic energy and compensated turbulent dissipation rate, $\varepsilon(t)t^m$, in HIT DNS. The decay exponents are $m=2.48$ for $\langle \phi \rangle=0.5\%$ and $m=2.22$ for $\langle \phi \rangle=2\%$, obtained over the full analysed interval $6\le t\le40$. The corresponding $\mathcal{R}^2$ values are $0.92$ and $0.93$.}
    \label{fig:tke1 and dissipation}
\end{figure}

Figure~\ref{fig:tke1 and dissipation} show the temporal evolution of the turbulent kinetic energy, $k(t)$, and the compensated dissipation rate, $\varepsilon(t)t^m$, for dilute bubbly flows at void fractions $\langle \phi \rangle=0.5\%$ and $2\%$. The HIT DNS data are shown over $6\le t\le40$, where $t\simeq6$ corresponds approximately to the peak of the dissipation rate. The simulation is initialized slightly earlier, but the preceding interval is not used in the scaling analysis because $\varepsilon(t)$ initially grows as the spectrally generated velocity field adjusts dynamically. The interval $t<6$ is therefore excluded because the prescribed initial spectrum has not yet developed a dynamically equilibrated nonlinear cascade, and $\varepsilon(t)$ is increasing rather than freely decaying. Including this interval would mix the initialization transience with the physical decay regime and bias the fitted exponent $m$. We therefore take the dissipation peak as the effective initial condition for the decaying-turbulence analysis.

The exponent $m$ is obtained by fitting $\varepsilon(t)\sim t^{-m}$ over the full analysed range, giving $m=2.48$ for $\langle \phi \rangle=0.5\%$ and $m=2.22$ for $\langle \phi \rangle=2\%$, with corresponding coefficient of determination $\mathcal{R}^2$ values of $0.92$ and $0.93$. The compensated representation shows that $\varepsilon(t)t^m$ remains nearly flat over most of the interval, confirming an effective HIT-like decay, with departures mainly during the early interval $6\lesssim t\lesssim8$ and the latest interval $32\lesssim t\lesssim40$.

The early departure is attributed primarily to the nonlinear adjustment of the spectrally initialized velocity field. Although the velocity field is prescribed to match a target energy spectrum, the nonlinear cascade and small-scale strain field are not initially in dynamic equilibrium. During the initial adjustment, energy is redistributed across wavenumbers, inertial transfer develops, and the dissipation reaches a peak before entering the decay regime. This behaviour is a known feature of freely decaying HIT initialized from prescribed spectra and is not unique to bubbly turbulence \citep{yoffe2018onset,ishihara2009study}. Thus, the rapid decrease of $\varepsilon(t)$ immediately after $t\simeq6$ should be interpreted mainly as the relaxation of the initially non-equilibrated turbulent cascade at $Re_\lambda=O(10^2)$, rather than as a purely interfacial effect.

In the present two-phase simulations, this primary cascade-adjustment mechanism is accompanied by secondary interfacial effects. The bubble population is introduced in nearly spherical form and spans both sub-Hinze and super-Hinze diameters. Super-Hinze bubbles are dynamically out of equilibrium with the surrounding turbulence: turbulent inertial stresses exceed capillary restoring stresses, causing deformation, stretching and, in some cases, fragmentation. During the early post-peak interval $6\lesssim t\lesssim8$, the initially spherical bubbles therefore undergo rapid deformation as they adjust to the surrounding turbulent strain field. Part of the carrier-phase kinetic energy is converted into interfacial energy through surface stretching and deformation, while additional energy is dissipated in the near-interface gradients generated during this relaxation. This interpretation follows the classical Weber-number picture of turbulent breakup \citep{hinze1955fundamentals}, the experimental observation of bubble oscillation and breakup in turbulence \citep{risso1998oscillations}, and recent DNS studies of initially spherical bubbles in HIT \citep{perrard2021bubble,riviere2021subhinze}. Measurements near the Hinze scale further show that deformation and breakup probability are linked to turbulent Weber-number fluctuations \citep{masuk2021simultaneous}.

In single-phase freely decaying HIT, the volume-averaged kinetic-energy budget reduces to ${\rm d}k/{\rm d}t=-\varepsilon$ in the absence of forcing and transport. In the present two-phase system, deformable and breakable interfaces introduce additional pathways for energy exchange between carrier-phase turbulence, interfacial energy and bubble-induced motions \citep{jain:2025}. A schematic volume-averaged budget may be written as
\begin{equation}
    \frac{{\rm d}k}{{\rm d}t}
    =
    -\varepsilon_{\rm eff}
    -
    \frac{1}{\rho_l}\frac{{\rm d}E_\gamma}{{\rm d}t}
    +
    \Pi_{\rm BIT},
    \label{eq:tke_budget_interface_bit}
\end{equation}
where $\rho_l$ is the liquid density, $E_\gamma$ is the interfacial energy per unit volume, and $\Pi_{\rm BIT}$ represents bubble-induced turbulence production. The effective dissipation is written as
\begin{equation}
    \varepsilon_{\rm eff}
    =
    \varepsilon_{\rm HIT}
    +
    \varepsilon_{\rm int},
    \label{eq:effective_dissipation_dns}
\end{equation}
where $\varepsilon_{\rm HIT}$ is the carrier-phase cascade dissipation and $\varepsilon_{\rm int}$ denotes interface-mediated dissipation associated with deformation, breakup, capillary relaxation, coalescence-induced microflows and small-scale gradients near the interface. The term $\Pi_{\rm BIT}$ represents liquid velocity fluctuations generated by bubble wakes, capillary oscillations, post-breakup recoil, relaxation of deformed interfaces and merger-induced motion. Thus, bubbles do not simply remove energy from the carrier phase; they also redistribute part of that energy into local liquid motion, consistent with bubbly-turbulence studies showing that bubbles can generate wake-scale fluctuations while enhancing small-scale dissipation \citep{lance1991turbulence,prakash2016energy}. The interfacial energy is
\begin{equation}
    E_\gamma(t)=\gamma A(t),
    \label{eq:surface_energy_dns}
\end{equation}
where $\gamma$ is the surface tension and $A(t)$ is the interfacial area per unit volume. For an approximately spherical bubble population and contant $\phi$,
\begin{equation}
    A(t) \propto \frac{1}{d_{32}(t)} .
    \label{eq:area_sauter_dns}
\end{equation}
\noindent with $d_{32}$ being the Sauter mean diameter. During the early post-peak interval $6\lesssim t\lesssim8$, the initially spherical bubbles deform rapidly under turbulent strain. Deformation and breakup of the super-Hinze fraction create additional interfacial area and increase $E_\gamma$, so the surface-energy term in Eq. \eqref{eq:tke_budget_interface_bit} acts as a sink of carrier-phase kinetic energy. Physically, part of the turbulent kinetic energy is converted into surface energy as bubbles stretch away from their initial spherical state, while another part is dissipated through the small-scale gradients generated near the interface. Coalescence acts in the opposite direction by reducing interfacial area, but while a finite number of bubbles remains super-Hinze region, deformation and fragmentation provide additional energy-exchange pathways. Although $\Pi_{\rm BIT}$ is positive, the observed rapid early decrease of $k(t)$ indicates that cascade adjustment, surface-energy storage and interface-mediated dissipation together dominate over bubble-induced kinetic-energy return during the initial adjustment.

After this short early interval, $k(t)$ approaches an approximate power-law decay and $\varepsilon(t)t^m$ remains nearly flat over a dominant portion of range of $t$. This occurs because the strongest cascade and interfacial adjustments are transient. As turbulence decays, the Hinze scale increases according to $d_H(t)\sim\varepsilon^{-2/5}$. Since $d_H$ grows faster than the characteristic bubble size for the measured decay exponents, the bubble population drifts progressively toward smaller sizes relative to $d_H$. Consequently, the fraction of super-Hinze bubbles decreases, breakup is progressively suppressed and the system enters a coalescence-dominated regime.

The two void-fraction cases show nearly overlapping turbulent kinetic energy throughout the simulation. A small but persistent separation is nevertheless visible, with the $\langle \phi \rangle=0.5\%$ case remaining slightly above the $\langle \phi \rangle=2\%$ case. This behaviour is consistent with the dilute nature of the suspension. At these void fractions, the carrier turbulence remains the dominant kinetic-energy reservoir, while the dispersed phase introduces a secondary correction through interface deformation, breakup, coalescence and bubble-induced small-scale motion. The leading interfacial-energy scale per unit liquid mass is
\begin{equation}
    e_\gamma \equiv \frac{E_\gamma}{\rho_l}
    \simeq
    \frac{6\gamma\phi}{\rho_l d_{32}} .
    \label{eq:surface_energy_mass_dns}
\end{equation}
Thus, at fixed $d_{32}$, the kinetic energy temporarily stored in interfacial deformation increases approximately with $\langle \phi \rangle$. The higher-void-fraction case is therefore expected to transfer slightly more carrier-phase kinetic energy into interfacial deformation, breakup and near-interface dissipation. At higher void fraction, the larger number of bubbles also increases the number of local strain disturbances, interfacial boundary layers, breakup events, coalescence events and relaxation events, enhancing the conversion of large-scale kinetic energy into bubble-scale motion and near-interface dissipation. The resulting difference in $k(t)$ remains small because $\phi=O(10^{-2})$, consistent with dilute-coupling arguments for dispersed turbulent flows \citep{elghobashi1994predicting,elghobashi2006overview,lance1991turbulence}.

The dissipation rate is more sensitive than $k(t)$ to the dispersed phase because it depends on small-scale velocity gradients. During the early mixed breakup--coalescence stage, deformation and fragmentation of super-Hinze bubbles generate local gradients around the interface. Breakup creates new interface and daughter bubbles, while post-breakup relaxation, capillary oscillations and coalescence-induced merger flows introduce additional small-scale motions. These effects are secondary to the nonlinear spectral adjustment at early times, but they contribute to the early excess dissipation relative to the compensated HIT-like plateau.

A simple estimate of the bubble-induced turbulence contribution follows from the inertial-range velocity difference at the bubble scale, $u_d \sim(\varepsilon d)^{1/3}$. The corresponding bubble-induced kinetic-energy production rate per unit mass may be written as
\begin{equation}
    \Pi_{\rm BIT} \sim C_{\rm BIT}\langle \phi \rangle\frac{u_d^3}{d}
    \sim C_{\rm BIT}\phi\varepsilon ,
    \label{eq:bit_scaling_dns}
\end{equation}
where $C_{\rm BIT}$ is an efficiency coefficient depending on deformability, slip, breakup, shape oscillations, local flow topology, and the bubble-size distribution. Similarly, the additional interface-mediated dissipation may be represented as $\varepsilon_{\rm int} \sim C_{\rm int}\phi\varepsilon$, where $C_{\rm int}$ depends on interfacial strain, deformation, breakup, coalescence-induced relaxation, and the bubble-size distribution. Eq.~\eqref{eq:bit_scaling_dns} shows that increasing $\phi$ increases both bubble-induced liquid fluctuations and dissipation associated with interfacial gradients, provided that the changes in the size-distribution-dependent efficiencies remain secondary. Bubble-induced motions feed energy
preferentially into bubble-scale and wake-scale motions, which are dissipated 
efficiently and therefore affect $\varepsilon(t)$ more strongly than the integral quantity $k(t)$ ~\citep{lance1991turbulence,shawkat2007spectra,prakash2016energy}.

This distinction explains why the higher-void-fraction case can maintain a larger instantaneous dissipation level while exhibiting a smaller fitted decay exponent. At $\langle \phi \rangle=0.5\%$, interfacial feedback is weaker and the dissipation follows the freely decaying carrier-phase cascade more closely, giving a faster decay close to the Batchelor--Loitsyanskii limit. At $\langle \phi \rangle=2\%$, the larger initial bubble population produces stronger bubble-scale motion, coalescence-induced relaxation and near-interface dissipation. As coalescence proceeds over the analysed interval, the bubble number decreases and the total interfacial area is reduced, releasing surface energy through capillary relaxation and merger-induced microflows. Only a fraction of this released surface energy appears as carrier-phase motion; the remainder is dissipated viscously near the interface. This persistent interfacial-energy pathway sustains bubble-scale gradients and near-interface dissipation, so the measured dissipation can remain larger while decaying more slowly, giving a smaller exponent $m$. Thus, the void fraction affects both the magnitude of dissipation and its decay rate, and these two effects need not vary in the same way.

The fitted exponents lie close to the classical dissipation-decay range expected for freely decaying HIT. Saffman-type turbulence gives $k\sim t^{-6/5}$ and hence $\varepsilon=-{\rm d}k/{\rm d}t\sim t^{-11/5}$, whereas Batchelor--Loitsyanskii-type turbulence gives $k\sim t^{-10/7}$ and $\varepsilon\sim t^{-17/7}$, corresponding to $m\simeq2.2\text{--}2.43$ \citep{saffman1967large,loitsyanskii1939some,batchelor1956large}. The agreement indicates that, after the initial cascade and interfacial adjustments, the carrier-phase cascade remains broadly HIT-like, while the bubbles provide a weak, time-dependent interfacial correction.

The compensated curves need not remain perfectly flat over the full interval. At late times, $32\lesssim t\lesssim40$, the local decay exponent decreases and approaches values closer to $m\simeq2$, consistent with finite-time curvature as $\mathrm{Re}_\lambda$ decreases and the integral scale grows relative to the computational box. For Saffman-type turbulence, $\ell(t)\sim t^{2/5}$, whereas for Batchelor--Loitsyanskii-type turbulence, $\ell(t)\sim t^{2/7}$ \citep{saffman1967large,batchelor1956large}. Since the computational domain was chosen with $L_{\rm box}\simeq7\ell$ (obtained from DNS) at the beginning of the analysed decay, these scalings imply that $\ell/L_{\rm box}$ increases from approximately $0.14$ to $\sim 0.25\text{--}0.29$ by the end of the analyzed times. Thus, the flow is not fully box-limited, but the largest eddies occupy an increasingly significant fraction of the periodic domain, making the assumptions of asymptotically high $Re_\lambda$ and $\ell\ll L_{\rm box}$ only marginally satisfied. This distinction is important because complete saturation of the integral scale does not by itself explain a slower dissipation decay; in the fully saturated limit, finite-domain arguments predict $k\sim t^{-2}$, i.e. faster energy decay than the Saffman and Batchelor--Loitsyanskii limits \citep{skrbek2000decay,thornber2016impact}. The present behaviour is therefore better interpreted as a finite-time transition in which finite-domain and finite-Reynolds-number effects produce curvature in the decay. \citet{meldi2017turbulence} showed that saturation and finite-Reynolds-number effects strongly affect the large-scale features of freely decaying HIT before complete saturation is reached, while \citet{thornber2016impact} showed that the integral length scale is more sensitive to domain size than the kinetic-energy decay rate, with noticeable discrepancies once the integral length becomes a non-negligible fraction of the box size. The estimated late-time value $\ell/L_{\rm box} \sim 0.25\text{--}0.29$ is therefore in the range where finite-domain effects should become visible. Consistently, \citet{anas2020finite} found that Saffman decay remains close to the ideal exponent only while $Re_\lambda\gtrsim10$ and $\ell\lesssim0.25L_{\rm box}$, and also reported intervals where the local energy-decay exponent falls below the Saffman value.

In the present bubbly flow, this finite-time and finite-domain curvature is further reinforced by interfacial energy exchange. As $d_H(t)\sim\varepsilon^{-2/5}$ grows, the bubble population drifts into the sub-Hinze regime, breakup is suppressed and coalescence becomes dominant. Coalescence reduces interfacial area and releases surface energy. Only a fraction of this released energy appears as carrier-phase motion through capillary relaxation or merger-induced microflows; the remainder is dissipated viscously near the interface. Because the background turbulent dissipation has already decayed substantially, even weak interfacial-energy exchange becomes more visible in the total dissipation budget. This provides a secondary mechanism for the observed late-time flattening of $\varepsilon(t)$, in addition to finite-$Re_\lambda$ and finite-domain effects. Thus, the compensated representation separates three stages: a short post-peak adjustment dominated by nonlinear spectral energy transfer, a dominant HIT-like interval in which $\varepsilon(t)t^m$ is nearly flat, and a weak late-time flattening caused by finite-time/finite-domain effects with secondary coalescence-mediated interfacial energy release.

\begin{figure}[htbp]
    \begin{subfigure}[b]{0.36\textwidth}
        \includegraphics[width=\textwidth]{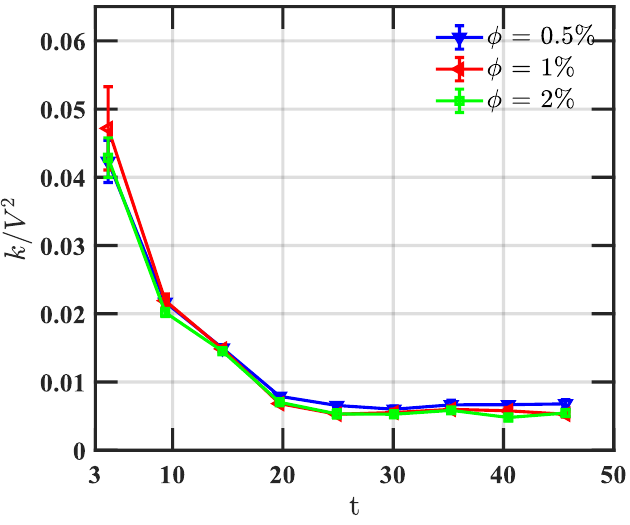}
        \caption{$V$=6.1\,m/s}
        \label{fig:a}
    \end{subfigure}
    \begin{subfigure}[b]{0.312\textwidth}
        \includegraphics[width=\textwidth]{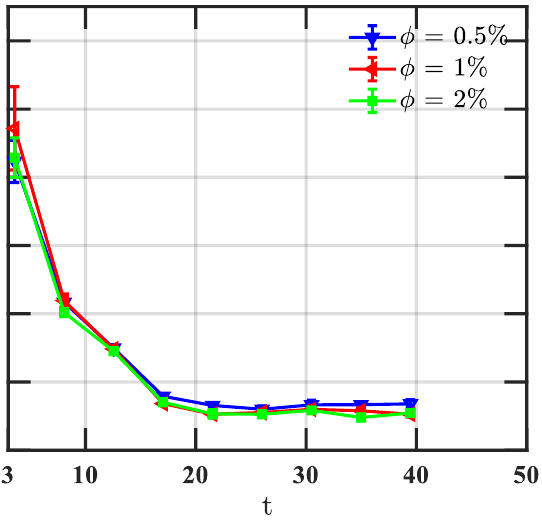}
        \caption{$V$=7.4\,m/s}
        \label{fig:b}
    \end{subfigure}
    \begin{subfigure}[b]{0.312\textwidth}
        \includegraphics[width=\textwidth]{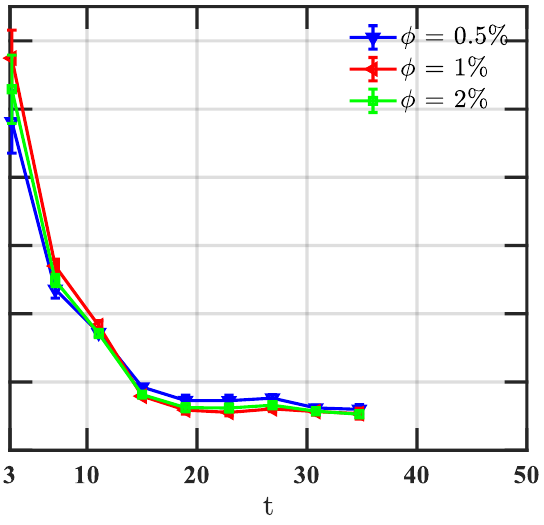}
        \caption{$V$=8.4\,m/s}
        \label{fig:c}
    \end{subfigure}    
    \caption{Axial variation of $k$ along the duct centerline for three bulk velocities $V(\text{flowrates } Q)$ and three void fractions $\phi$.}
    \label{fig:tke1}
\end{figure}

\begin{figure}[htbp]
    \centering
    \begin{subfigure}[b]{0.378\textwidth}
        \centering
        \includegraphics[width=\textwidth]{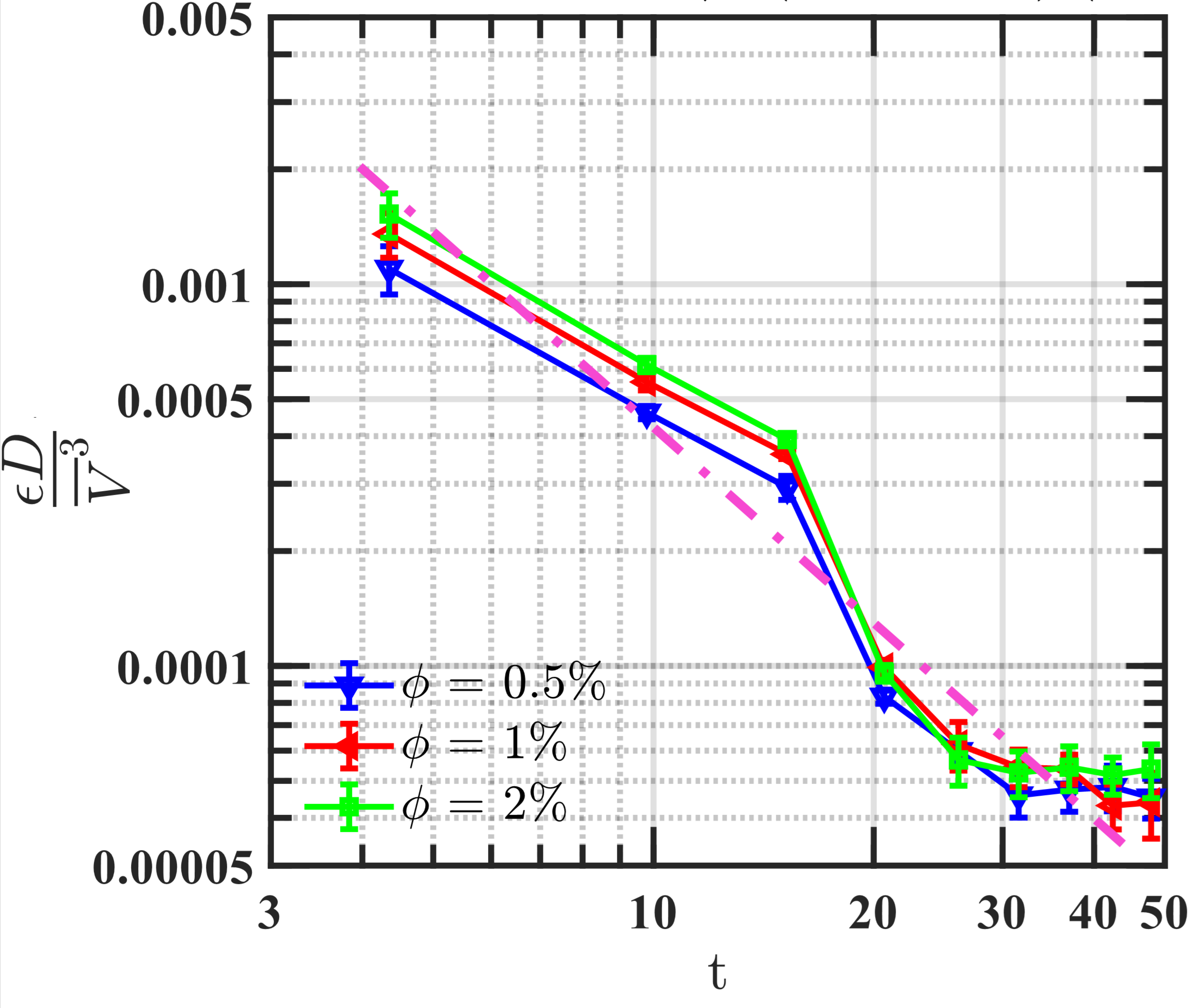}
        \caption{$V$=6.1\,m/s \vk{(slope: 1.85$\pm$0.1, $\mathcal{R}^2$: 0.94 - 0.97)}}
        \label{fig:a}
    \end{subfigure}
    \begin{subfigure}[b]{0.30\textwidth}
        \centering
        \includegraphics[width=\textwidth]{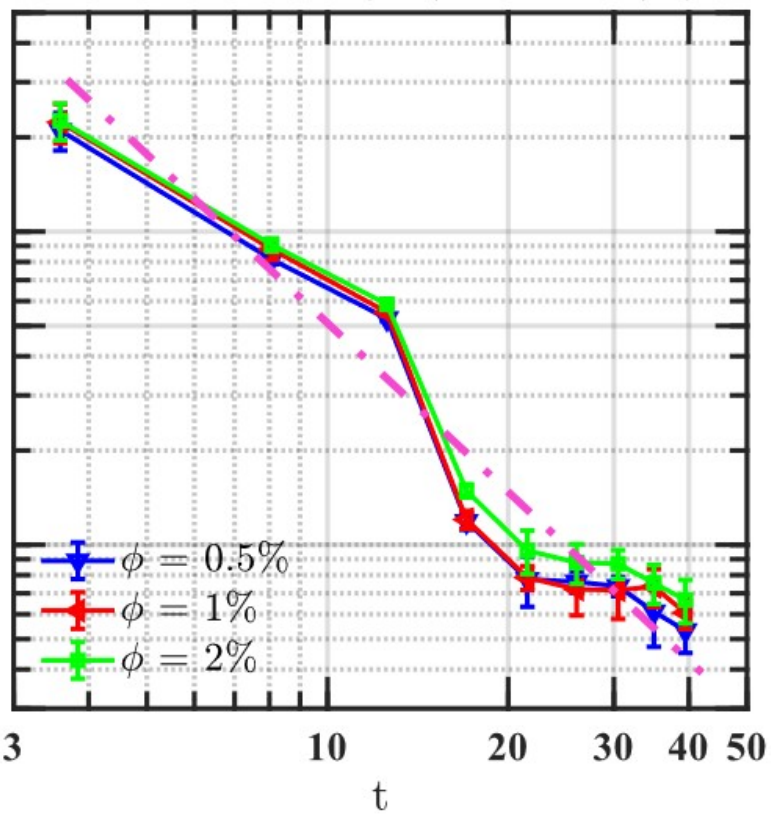}
        \caption{$V$=7.4\,m/s \vk{(slope: 1.95$\pm$0.1,$\mathcal{R}^2$: 0.93 - 0.97)}}
        \label{fig:b}
    \end{subfigure}
    \begin{subfigure}[b]{0.30\textwidth}
        \centering
        \includegraphics[width=\textwidth]{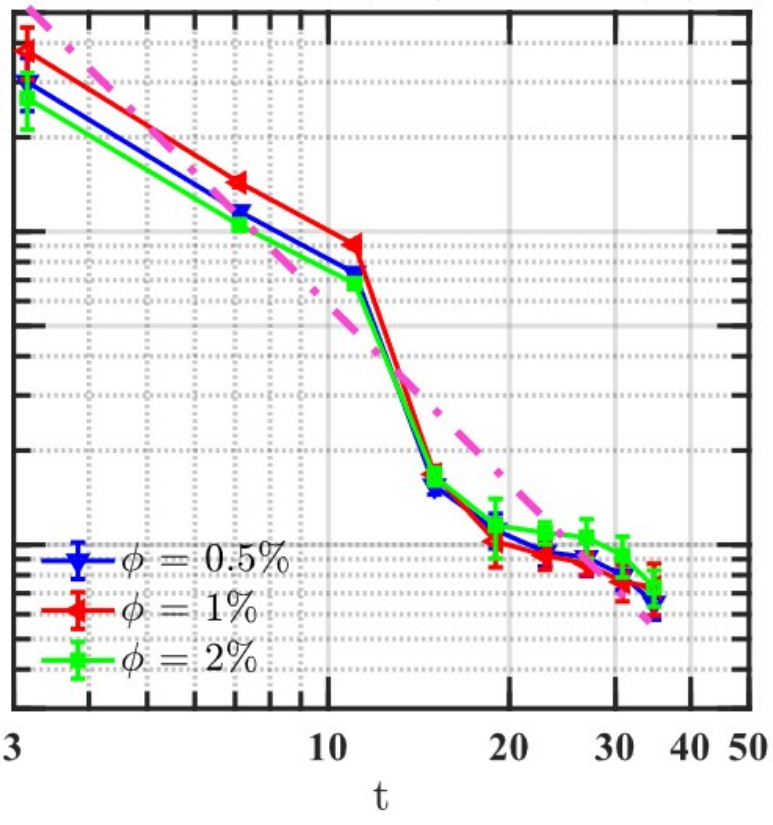}
        \caption{$V$=8.4\,m/s \vk{(slope: 2.1$\pm$0.1, $\mathcal{R}^2$: 0.93 - 0.95)}}
        \label{fig:c}
    \end{subfigure}    
    \caption{Axial variation of $\varepsilon$ along the duct centerline, for three bulk velocities $V(\text{flowrates } Q)$ and three void fractions $\phi$. Note the use of log-log scale. \vk{The range of the coefficient of determination ($\mathcal{R}^2$) is provided.}}
    \label{fig:diss1}
\end{figure}

Furthermore, the high-Reynolds-number decaying duct-flow measurements of \citet{kumar2026bubble} provide a practical, spatially developing counterpart to the present temporal DNS and show how the DNS mechanism extends beyond ideal HIT. In that configuration, turbulence is generated by a regenerative pump and then decays downstream in a square duct at $\mathrm{Re} \sim O(10^5)$, $\mathrm{Re}_\lambda\sim O(10^3)$ and $\langle \phi \rangle\sim O(1\%)$, with turbulence intensity exceeding $30\%$ near the duct inlet. Near the inlet, the pump-generated flow is close to homogeneous in the duct core: the mean streamwise velocity is nearly flat across most of the cross-section, while the r.m.s. fluctuations remain relatively uniform in the core before increasing toward the wall. This core behaviour provides the basis for comparing the experimentally measured centreline spatial decay with the HIT-like temporal decay observed in the DNS. \citet{kumar2026bubble} reported preliminary experimental results in which the turbulence statistics were plotted against the dimensionless axial coordinate $\mathcal{L}=x/D$. For comparison with the present temporal DNS, the experimental abscissa is recast as a convective time, $t=\frac{\mathcal{L}}{\mathcal{V}}, \mathcal{V}=\frac{V}{\overline{V}}, \overline{V}=\frac{V_1+V_2+V_3}{3}$, while the ordinates are retained as $k/V^2$ and $\varepsilon D/\overline{V}^3$, consistent with~\citet{kumar2026bubble}; as seen in Figures \ref{fig:tke1} and \ref{fig:diss1}, respectively. This transformation provides a common temporal basis for comparing the spatially developing duct flow with freely decaying HIT. Since the mapping is linear for each bulk velocity, it does not change the fitted decay exponents; it only expresses the downstream evolution in terms of the convective time experienced by the bubble population.

The comparison reveals both a common driving mechanism and important configuration-dependent differences. In the DNS, the bubbles are introduced directly into HIT as an initially spherical population spanning both sub-Hinze and super-Hinze diameters. The early response is therefore an initial-value adjustment: super-Hinze bubbles deform and break, interfacial area is created, and energy is rapidly exchanged among carrier-phase turbulence, interfacial deformation, and bubble-induced motions. This produces the sharp early decrease and curvature observed in $k(t)$ and $\varepsilon(t)$. In the duct measurements, by contrast, the bubbles enter the optical measurement region only after passing through the regenerative pump and upstream development section. They are therefore already conditioned by intense pump-driven turbulence, fragmentation, deformation and partial relaxation before the first measurement station. Consequently, the early decay of both $k/V^2$ and $\varepsilon D/\overline{V}^3$ is smoother than in the DNS and does not exhibit the same abrupt initial transience. The experimental decay is nevertheless substantial: the centreline $k$ and $\varepsilon$ decrease by more than an order of magnitude from the inlet to the farthest measurement station, followed by a gentler power-law-like tail as the core flow approaches a quasi-equilibrium state~\citep{kumar2026bubble}. Increasing the bulk velocity raises both $k$ and $\varepsilon$ and extends the downstream development length, while the effect of void fraction over $\langle \phi \rangle=0.5\%$--$2\%$ remains comparatively weak in the measured turbulence statistics. Despite these differences, the downstream evolution follows the same physical direction as the DNS: turbulence decays, $d_H\sim\varepsilon^{-2/5}$ increases, the fraction of super-Hinze bubbles decreases, breakup is progressively suppressed and the population evolves toward a coalescence-dominated state~\citep{hinze1955fundamentals,kumar2026bubble}. The measured duct-flow dissipation exponent, $m\simeq1.85$--$2.10$, is slightly smaller than the ideal HIT-like DNS values because confinement, mean shear, near-wall production and the growing wall boundary layer continue to sustain turbulence~\citep{pope2000turbulent}. Thus, the DNS isolates the intrinsic bubble--turbulence mechanism in a homogeneous flow, whereas the duct measurements show that the same Hinze-scale drift and transition toward coalescence-dominated dynamics persist in a realistic pump-driven, spatially developing turbulent flow; the main effect of the experimental complexity is to modify the effective dissipation exponent rather than the underlying interaction mechanism.

\begin{figure}[htbp]
    \begin{subfigure}[b]{0.49\textwidth}
        \includegraphics[width=\textwidth]{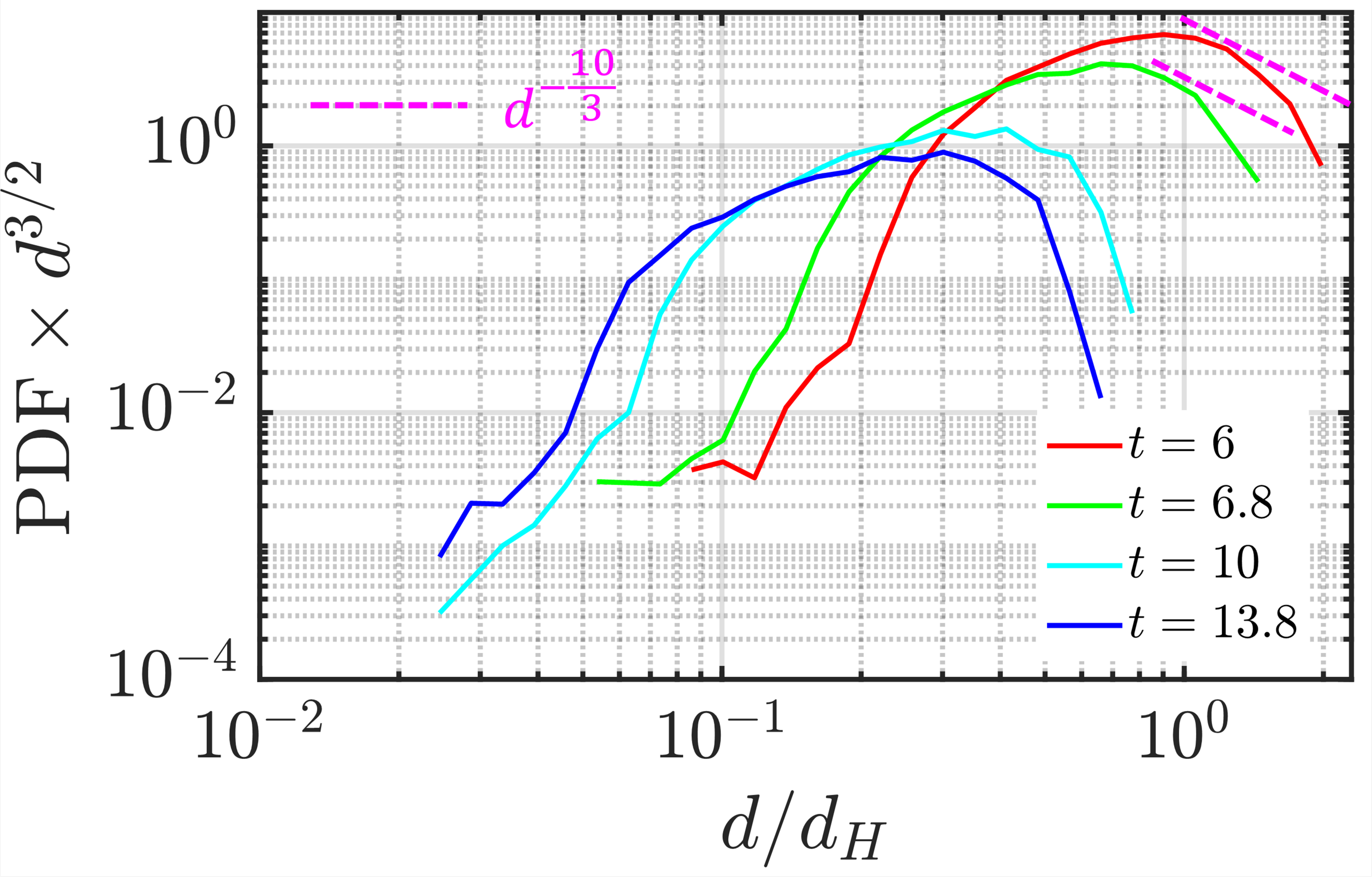}
        \caption{$\langle \phi \rangle=$0.5\%}
        \label{fig:a}
    \end{subfigure}
    \hfill
    \begin{subfigure}[b]{0.49\textwidth}
        \includegraphics[width=\textwidth]{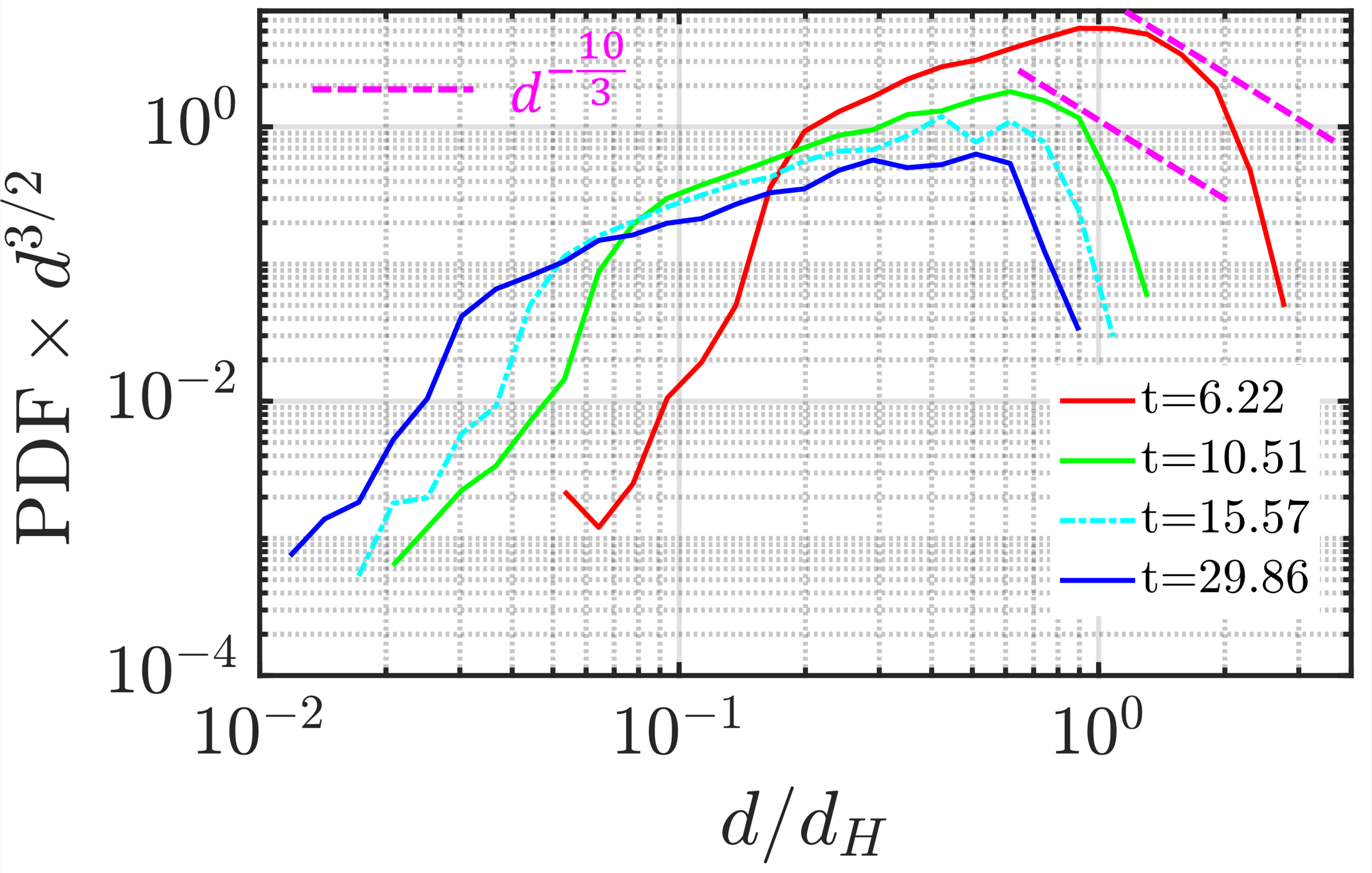}
        \caption{$\langle \phi \rangle=$2\%}
        \label{fig:b}
    \end{subfigure}  
    \caption{Temporal evolution of the compensated bubble-diameter probability distribution, $\text{PDF}\times d^{3/2}$, for (a) $\phi=0.5\%$ and (b) $\phi=2\%$. A horizontal region corresponds to PDF$\sim d^{-3/2}$ coalescence scaling.}
    \label{fig:pdf}
\end{figure}

\subsection{Bubble-size distribution, interfacial area, and bubble count}

Figure~\ref{fig:pdf} shows the temporal evolution of the compensated bubble-diameter probability density, PDF$\times d^{3/2}$ from our DNS, for $\langle \phi \rangle=0.5\%$ and $2\%$. This representation is useful because a horizontal region corresponds PDF$\,\sim d^{-3/2}$, whereas departures from a plateau indicate deviations from this scaling. At the early time \(t\simeq6\), both void fractions exhibit a two-range structure, with bubbles distributed on both sides of the Hinze scale. The compensated plateau occurs close to $d/d_H\simeq1$ and extends into the sub-Hinze range, indicating a PDF$\, \sim d^{-3/2}$ branch near the transition scale. This branch should not be interpreted as the inertial-breakup cascade. Rather, it corresponds to the sub-Hinze/capillary branch of the distribution, associated with capillary-mediated pinch-off, relaxation and sub-Hinze bubble production in turbulent fragmentation studies, and with coalescence-dominated redistribution once sustained breakup is suppressed in the present decaying system \citep{deane2002scale,garrett2000connection,crialesi2023interaction,farsoiya2023role}. The early PDF therefore reflects a mixed regime: sub-Hinze bubbles populate the $d^{-3/2}$ branch, while super-Hinze bubbles remain susceptible to deformation and breakup.

The distinction between these two branches follows directly from the Hinze criterion \citep{hinze1955fundamentals,masuk2021simultaneous,riviere2021subhinze}. For $d<d_H$, the turbulent Weber number is subcritical, $\mathrm{We}(d)<We_c$, and turbulent inertial stresses are not strong enough to drive sustained fragmentation. Bubble interactions in this range therefore redistribute mass primarily through coalescence and capillary relaxation rather than through an inertial breakup cascade. In contrast, for $d>d_H$, the Weber number is supercritical, $\mathrm{We}(d)>We_c$, and turbulent eddies can stretch, deform and fragment bubbles. The super-Hinze part of the PDF therefore steepens relative to the compensated $d^{-3/2}$ plateau. The super-Hinze scaling, which is in this upper branch approaches the inertial-fragmentation scaling $\sim d^{-10/3}$, consistent with turbulent intertial bubble breakup observed in turbulent dispersions \citep{garrett2000connection,soligo2019breakage,riviere2021subhinze}. At the extreme right end, the distribution can become even steeper because very few bubbles occupy the largest-size bins; this terminal steepening reflects limited sampling of rare large bubbles rather than a separate physical scaling \citep{farsoiya2023role}.

As time increases, $\varepsilon$ decays and the Hinze scale grows as $d_H\sim\varepsilon^{-2/5}$, as derived from the classical turbulent breakup criterion \citep{hinze1955fundamentals}. Since the characteristic bubble sizes grow more slowly than $d_H$ for the measured decay exponents, the distribution shifts progressively toward smaller $d/d_H$. This drift can be quantified by the probability mass carried by super-Hinze bubbles,

\begin{equation}
    \mathcal{P}_{d>d_H}(t)
    =
    \int_{d_H(t)}^{\infty} P(d,t)\,{\rm d}d .
    \label{eq:super_hinze_probability}
\end{equation}

At early times, $\mathcal{P}_{d>d_H}$ is finite, so the PDF contains both a sub-Hinze $d^{-3/2}$ branch and a breakup-active super-Hinze tail. As $d_H$ increases, $\mathcal{P}_{d>d_H}$ rapidly decreases, providing a direct population-level measure of the loss of the breakup pathway. In the present DNS, the super-Hinze probability mass becomes negligible at approximately $t\simeq7$ for $\langle \phi \rangle=0.5\%$ and $t\simeq8$ for $\langle \phi \rangle=2\%$. The slightly later transition at higher void fraction reflects the larger number of initially large bubbles and the stronger interaction rate in the more crowded dispersed phase. Beyond this time, no sustained super-Hinze tail remains, breakup events become negligible, and the distribution evolves predominantly through coalescence.

The post-transition PDFs clarify the regime change. With increasing time, the $d^{-3/2}$ scaling is observed progressively farther below the Hinze scale, over ranges with $d/d_H<1$ and eventually $d/d_H\ll1$. At the same time, the $d^{-10/3}$ super-Hinze branch nearly disappears because the population has moved below the breakup threshold. The remaining large-diameter end of the PDF becomes very steep, corresponding to apparent terminal slopes of order $\sim d^{-4}$ to $\sim d^{-5}$. These steep terminal tails should not be interpreted as renewed inertial fragmentation; they arise because only a small number of bubbles populate the largest bins once the system becomes sub-Hinze. Similar finite-number effects in the extreme tail have been reported in bubble-size distributions and can produce slopes steeper than the physically dominant scaling~\citep{farsoiya2023role}. The robust late-time feature is therefore the persistence of the $d^{-3/2}$ branch over the main populated range, together with the collapse of the super-Hinze breakup tail.

A further feature of Figure~\ref{fig:pdf} is the systematic reduction of the compensated peak height with time. This reduction should not be interpreted simply as a decrease of the peak probability in the uncompensated PDF, because the plotted quantity is PDF$\times d^{3/2}$ and the abscissa is normalized by the time-dependent Hinze scale. As $\varepsilon$ decays, $d_H$ increases and the same physical bubble population is progressively displaced toward smaller $d/d_H$. The compensated peak therefore decreases and shifts leftward as probability mass is redistributed away from the super-Hinze tail and into the sub-Hinze coalescence range. At the same time, the near-horizontal portion of the compensated PDF becomes more apparent over the dynamically populated sub-Hinze interval, although the rendered data do not by themselves establish a strictly monotonic increase of plateau width. Rather, the dominant trend is the collapse of the breakup-active tail and the progressive organization of the main populated range around the $d^{-3/2}$ scaling.

The apparent translation of the compensated PDFs can be understood from the relative motion of two length scales: the characteristic bubble size and the Hinze scale. In the coalescence-dominated regime, the bubble population grows as $d_{32}\sim t^\beta$, while the decay of turbulence gives $d_H\sim \varepsilon^{-2/5}\sim t^{2m/5}$ from Hinze scaling \citep{hinze1955fundamentals}. Therefore,
\begin{equation}
    \frac{d_{32}}{d_H}
    \sim
    t^{\beta-2m/5}.
    \label{eq:d32_dH_drift_pdf}
\end{equation}
For the late-time decay exponents, $m\simeq2.22$--$2.48$, the pure-coalescence scaling $\beta=(3-m)/2$ gives $\beta\simeq0.26$--$0.39$, whereas $2m/5\simeq0.98$--$1.02$. Thus, $\beta-2m/5<0$, so $d_{32}/d_H$ decreases rapidly even though the physical bubble size grows by coalescence. This relative drift explains the leftward motion of the PDF in $d/d_H$, the reduction and displacement of the compensated peak, and the collapse of the super-Hinze tail.

This drift is even stronger during the earliest adjustment period. As discussed, $\varepsilon(t)$ decays very rapidly during the initial adjustment period, giving a local effective exponent close to $m_{\rm eff}\simeq3$ over this short interval. In that case, $d_H\sim t^{2m_{\rm eff}/5}\sim t^{6/5}$, while coalescence-driven growth remains much slower. Hence the breakup threshold sweeps rapidly through the distribution before coalescence can fully re-equilibrate the population. The result is a rapid loss of super-Hinze probability mass, followed by slower coalescence-controlled broadening within the sub-Hinze range. The compensated spectra therefore do not represent an exact self-similar collapse; rather, they translate and reorganize coherently in $d/d_H$ because rapid Hinze-scale drift precedes slower coalescence-controlled redistribution.

\begin{figure}[htbp]
    \begin{subfigure}[b]{0.48\textwidth}
        \includegraphics[width=\textwidth]{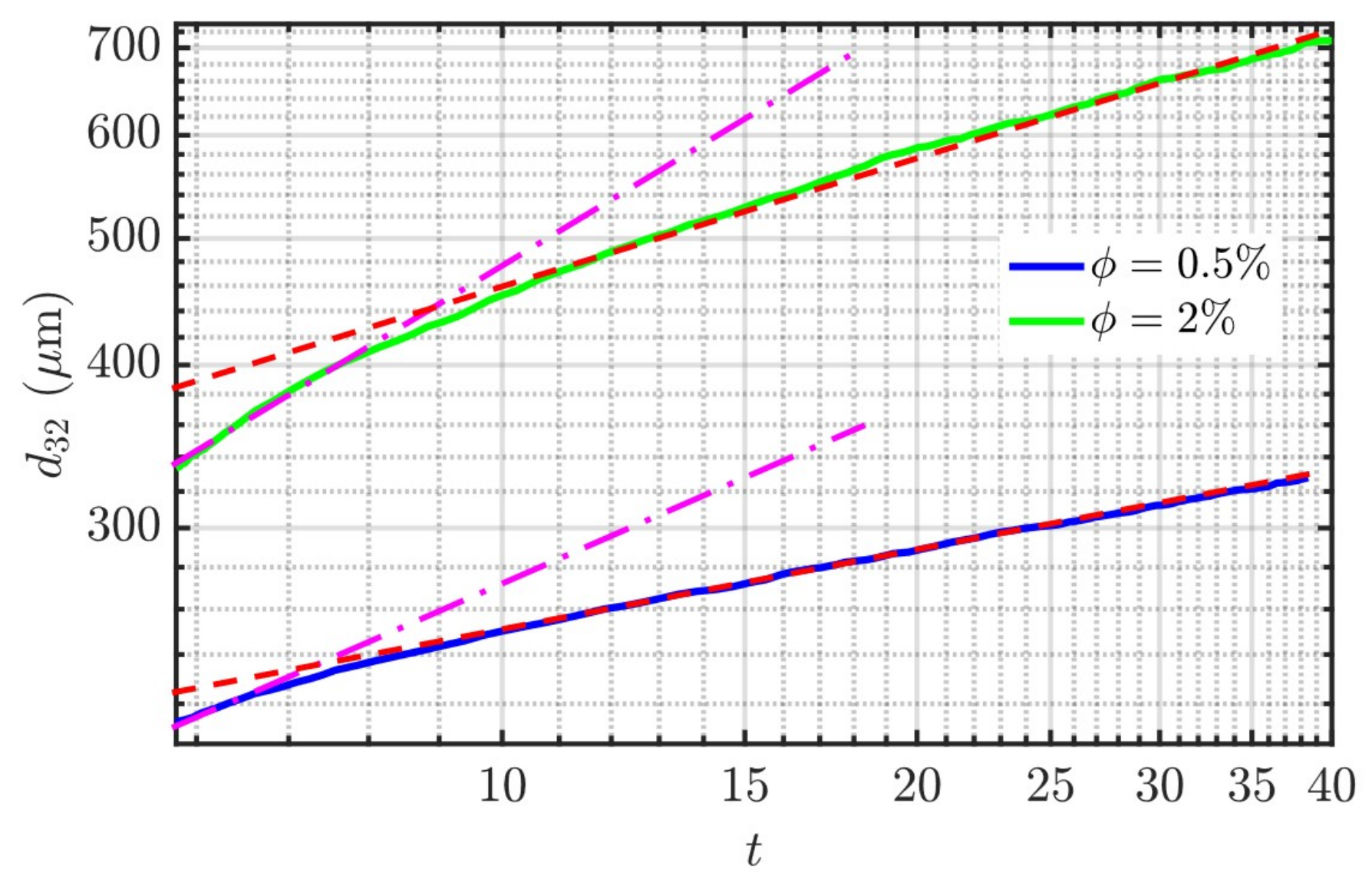}
       \caption{For the late-time regime, the fitted slopes are $0.24$ for $\langle \phi \rangle=0.5\%$ at $t>7$ and $0.35$ for $\langle \phi \rangle=2\%$ at $t>8$, with corresponding $\mathcal{R}^2$ values of $0.99$ and $0.99$, respectively. For the early-time regime, the fitted slopes are $0.55$ for $\langle \phi \rangle=0.5\%$ at $t<7$ and $0.66$ for $\langle \phi \rangle=2\%$ at $t<8$, both having same $\mathcal{R}^2$ values of $0.99$.}
        \label{fig:b}
    \end{subfigure}
    \begin{subfigure}[b]{0.48\textwidth}
        \includegraphics[width=\textwidth]{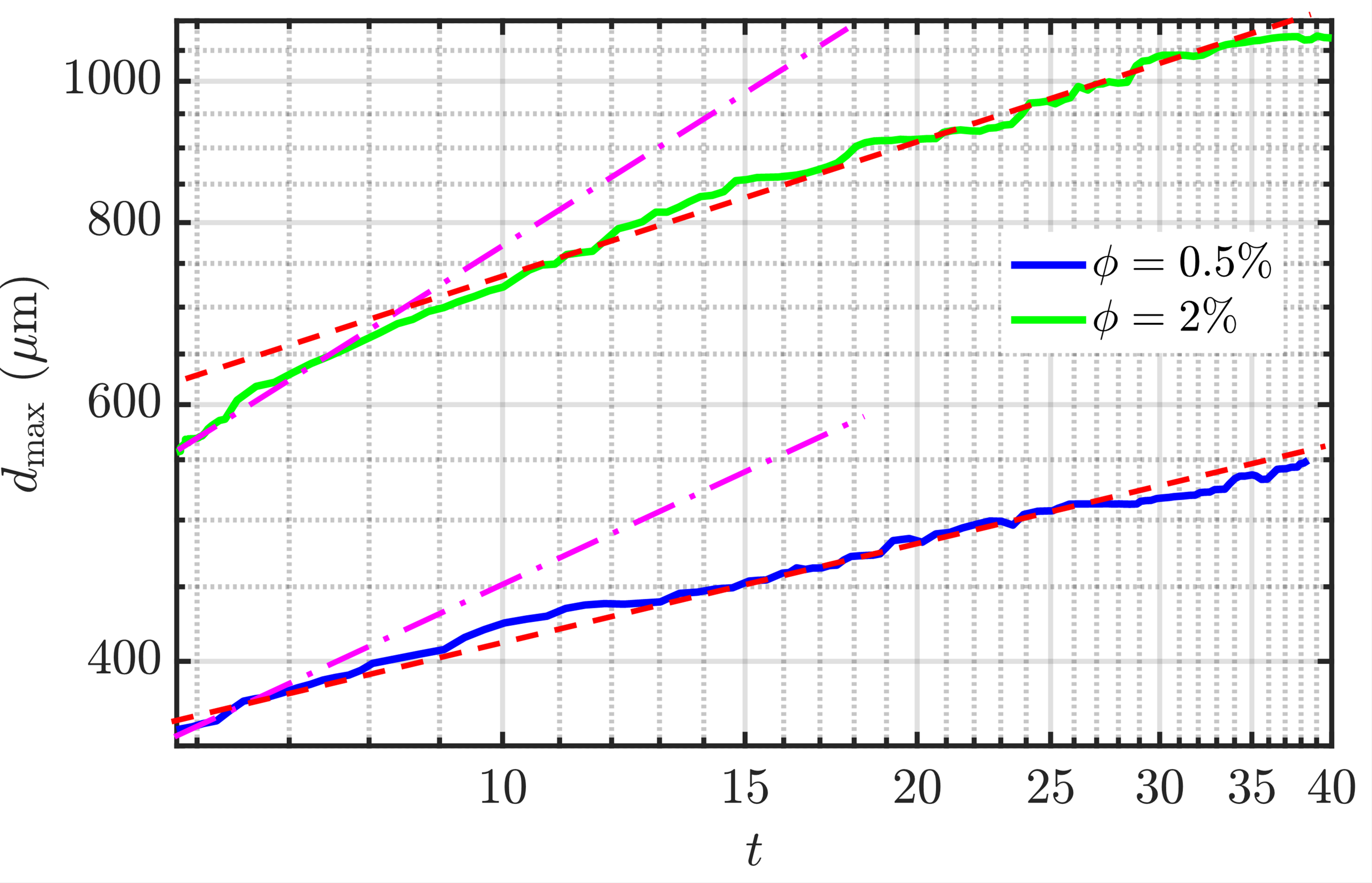}
        \caption{In the late-time regime, the fitted slopes are $0.23$ for $\langle \phi \rangle=0.5\%$ at $t>7$ and $0.36$ for $\langle \phi \rangle=2\%$ at $t>8$, with corresponding $\mathcal{R}^2$ values of $0.98$ and $0.97$, respectively. In the early-time regime, the fitted slopes are $0.51$ for $\langle \phi \rangle=0.5\%$ at $t<7$ and $0.68$ for $\langle \phi \rangle=2\%$ at $t<8$, with corresponding $\mathcal{R}^2$ values of $0.98$ and $0.99$, respectively.}
        \label{fig:c}
    \end{subfigure}    
    \caption{Temporal evolution of $d_{32}$, and $d_\mathrm{max}(\approx d_{99.5} \approx d_{99.8}$; average of the largest 1\% of bubbles).}
    \label{fig:diameter}
\end{figure}

The DNS evolution is consistent with the spatial PDF evolution reported in the high-Reynolds-number decaying duct-flow measurements of \citet{kumar2026bubble}. In that study, the bubble-diameter PDF near the duct inlet exhibited a dual power-law structure, with a breakup-associated $d^{-10/3}$ tail and a coalescence-associated or capillary/sub-Hinze $d^{-3/2}$ range, before evolving within a few hydraulic diameters toward a single $d^{-3/2}$ scaling characteristic of a predominantly sub-Hinze, coalescence-dominated population. The same study showed that the cumulative distribution plotted against $d/d_{32}$ broadened with axial position and void fraction, while the normalized extreme-to-mean diameter ratio increased from the breakup-dominated value and approached a quasi-self-similar coalescence-dominated state \citep{kumar2026bubble}. This comparison also clarifies the distinction between temporal HIT and spatially developing duct flow: in the DNS, initially spherical bubbles are introduced directly into HIT, so the early dual-range PDF reflects the immediate adjustment of a prescribed population spanning both sides of the Hinze scale, whereas in the duct experiment the population entering the measurement section has already passed through the pump and is partially conditioned by intense upstream fragmentation, deformation and coalescence. Nevertheless, when interpreted through the evolving ratio $d/d_H$, both systems exhibit the same sequence: an initial mixed breakup--coalescence regime, rapid reduction of $\mathcal{P}_{d>d_H}$, loss of the super-Hinze tail, and convergence toward a $d^{-3/2}$ coalescence-dominated distribution. This agreement indicates that the PDF evolution is governed primarily by the relative drift between the bubble population and the growing Hinze scale, rather than by the specific flow configuration. \\ 
Figures~\ref{fig:grid two phase} and~\ref{fig:diameter} show the temporal evolution of the characteristic bubble diameters in the DNS, including the Hinze scale $d_H$($=\eta_H$), the Sauter mean diameter $d_{32}$ and the upper-tail diameter $d_{99.8}$. The Hinze scale is not a measured bubble size, but a virtual inertial breakup threshold evaluated from the instantaneous turbulent dissipation using Eq.~\eqref{eq:hinze scale}. Since $d_H\sim\varepsilon^{-2/5}$, the decay of $\varepsilon(t)$ causes $d_H$ to increase with time. After a short initial transience, similar to $\varepsilon(t)$, $d_H$ follows a power-law growth for both void fractions, with power law slopes of 0.97 and 0.89 at $\langle \phi \rangle$ of 0.5\% and 2\%, respectively. This Hinze-scale drift is close to HIT as shown in Tables~\ref{tab:scalings_exponents_m_coal_break} and~\ref{tab:scalings_exponents_m_coal}, which ranges from 0.88 to 0.97. The growth of $d_H$ is central to the mechanism governing bubble interactions because it determines whether breakup dominates, coalescence dominates, or both processes coexist in a mixed breakup--coalescence regime. \\
The bubble diameters are represented in form of $d_{32}$ and $d_{99.8}$ which are obtained directly from the resolved phase field by identifying connected gas-phase regions in the computational domain at each output time \citep{nathan2025accurate}. Individual bubbles are segmented from the volume-fraction field using a connected-component algorithm, and the bubble volume $V_b$ is computed by integrating the gas fraction over all cells belonging to each connected structure. An equivalent spherical diameter is then defined as $d=\left(6V_b/{\pi}\right)^{1/3}$, which preserves the true bubble volume and is independent of instantaneous deformation.

The initial distribution was deliberately chosen (from experiment) to straddle this moving breakup threshold value of $d_H$: $d_{32}$ was close to $d_H$, while the upper tail which is maximum bubble diameter extended beyond $d_H$. The subsequent evolution confirms that the relevant control is not the absolute growth of the bubble population, but its motion relative to the Hinze scale. Although both $d_{32}$ and $d_{99.8}$ increase with time, $d_H$ grows faster. It rapidly becomes larger than $d_{32}$ and subsequently exceeds the largest resolved bubbles at approximately $t\simeq7.1$ for $\langle \phi \rangle=0.5\%$ and $t\simeq8.3$ (exept few transient random bubbles) for $\langle \phi \rangle=2\%$. The measured breakup activity shuts off over the same time interval, providing direct evidence that the disappearance of breakup is controlled by the loss of the super-Hinze population. The later crossing at $\langle \phi \rangle=2\%$ is consistent with its initially broader upper tail, and with a larger initial ratio of $d_{99.8}/d_H$ than in the $\langle \phi \rangle=0.5\%$ case. Thus, the higher-void-fraction case retains a finite super-Hinze population for a longer time before the growing Hinze scale sweeps through the distribution.

The evolution of $d_{32}$ provides a direct test of the predicted transition from an early mixed coalescence--breakup regime to a later pure-coalescence regime. The early regime should not be interpreted as breakup dominated. The PDF evolution shows that the population is already drifting toward the sub-Hinze range as $d_H$ increases, so coalescence is the dominant net process. However, a finite super-Hinze population remains present and undergoes deformation and occasional breakup. This residual breakup has an important indirect effect. In the reduced number balance derived in Eq.~\ref{eq:beta_mixed_summary}, the coalescence sink scales as $\Gamma n^2$, whereas the breakup source scales as $B \alpha(t)n$, where $\alpha(t)$ is the fraction of bubbles above the Hinze scale. Thus, even a comparatively weak breakup contribution can increase the number density of smaller bubbles and amplify subsequent coalescence, because the coalescence rate depends quadratically on $n$. The mixed regime is therefore coalescence dominated but breakup assisted: breakup supplies additional collision partners, while the increasing Hinze scale simultaneously removes bubbles from the breakable population. In the early mixed regime, the fitted exponents for \(d_{32}\) are approximately 0.55 for \(\langle \phi \rangle=0.5\%\) and 0.66 for \(\langle \phi \rangle=2\%\). The theoretical scaling \(d(t)\sim t^{\beta_m}\), with \(\beta_m=(15+m)/25\), gives \(\beta_m\simeq0.69\)--\(0.70\) using the DNS dissipation exponents. The lower slope values stem from the steep PDF distribution near the super-Hinze tail, which accelerates super-Hinze bubble removal.

The agreement is strongest for $\langle \phi \rangle=2\%$, where the larger initial number density increases the collision frequency and strengthens the breakup-assisted coalescence pathway. A second contribution may arise from deformation of initially spherical super-Hinze bubbles. Turbulent stresses rapidly stretch these bubbles before breakup or merger, increasing their instantaneous interfacial area and effective collision cross-section relative to an idealized spherical-bubble population. Together, the enhanced number density and deformation-enhanced encounter area explain why the mixed regime can exhibit a larger apparent growth exponent than the later pure-coalescence regime. 

Once the super-Hinze probability mass becomes negligible, breakup becomes negligible and the population enters the pure-coalescence regime. The supply of newly generated small bubbles then vanishes, the number density decays, and the volumetric coalescence rate decreases. In this limit, the Smoluchowski-type balance developed in Eq.~\eqref{eq:beta_mixed_summary} reduces to $d(t)\sim t^{\beta_c}$, where $\beta_c=\frac{3-m}{2}$. For the measured DNS decay exponents, this gives $\beta_c$ in the same range. The DNS yields $d_{32}\sim t^{0.24}$ for $\langle \phi \rangle=0.5\%$ and $d_{32}\sim t^{0.35}$ for $\langle \phi \rangle=2\%$, with $\mathcal{R}^2\simeq0.99$. These values confirm the predicted slowing of bubble growth after breakup is suppressed. Small deviations from the mixed-regime prediction are expected because the reduced theory absorbs coalescence efficiency into a prefactor, whereas the resolved DNS contains finite deformation, capillary relaxation and shape-dependent collision outcomes. Nevertheless, the spherical-bubble approximation used in the reduced theory remains
appropriate for the leading-order scaling because the interfacial relaxation time is shorter
than the large-eddy turnover time. The integral-scale eddy turnover time, \(\tau_{\ell}=\ell/\mathcal{U}\), is \(\mathcal{O}(10^{-2}\,\mathrm{s})\), whereas the bubble interfacial relaxation times, namely the capillary--inertial time scale
\(\tau_{\gamma}=(\rho d^{3}/\gamma)^{1/2}\) and the viscous--capillary time scale \(\tau_{\mu}=\mu d/\gamma\), are \(\mathcal{O}(10^{-3}\,\mathrm{s})\) \citep{kumar2026viscosity}. Thus, bubbles can relax toward near-spherical shapes over times shorter than the large-eddy evolution time, supporting the use of spherical-bubble
geometry in the scaling estimates. A small number of strongly deformed bubbles may have lower coalescence efficiency than ideal spherical bubbles because film drainage and contact geometry are altered. Such effects can reduce the observed growth exponent relative to the ideal breakup-assisted prediction, especially for $\langle \phi \rangle=0.5\%$, where the mixed regime is shorter and the initially super-Hinze population is smaller.

The transition between these regimes is controlled by the drift of the bubble population relative to the Hinze scale. The theory predicts, $\frac{d(t)}{d_H(t)} \sim t^{\beta-2m/5}$,
with a negative exponent for $m>5/3$ in both the mixed and pure-coalescence regimes. This holds for both DNS cases. Therefore, even though $d_{32}$ and $d_{99.8}$ increase in physical size, $d_H$ grows faster, the ratio $d/d_H$ decreases, and the population is driven progressively into the sub-Hinze regime. This is the central mechanism linking the PDF evolution in Figure~\ref{fig:pdf} and the diameter evolution in Figure~\ref{fig:diameter}: the super-Hinze tail disappears after the early transience, while both the mean and upper-tail diameters continue to grow by coalescence below the moving breakup threshold.

This interpretation is consistent with classical and modern descriptions of turbulent breakup and coalescence. The Hinze criterion identifies the scale above which turbulent inertial stresses overcome capillary restoring forces \citep{hinze1955fundamentals,hesketh1987bubble,salibindla2020lift}. Breakup models based on eddy--bubble interactions, stochastic energy transfer and critical Weber-number arguments predict fragmentation when turbulent fluctuations are sufficiently energetic relative to surface tension \citep{coulaloglou1977description,lehr2002bubble,martinez1999breakup}. Recent simulations and experiments further show that breakup and deformation are not determined by mean dissipation alone, but depend on local Weber-number fluctuations, finite deformation time and the instantaneous interaction between turbulent eddies and the interface \citep{vela2022memoryless,masuk2021simultaneous,calado2024dynamics,riviere2021subhinze}. In a decaying flow, these processes become explicitly time dependent because $\varepsilon(t)$, $d_H(t)$ and the active eddy population evolve simultaneously. The present DNS therefore extends the classical Hinze picture from a quasi-stationary threshold to a moving threshold that reorganizes the entire bubble population.

The growth of $d_{32}$ is governed by coalescence between bubbles in the dynamically populated part of the distribution. In population-balance closures, the coalescence kernel is usually written as
\begin{equation}
    \Gamma(d_1,d_2)=h(d_1,d_2)\lambda(d_1,d_2),
\end{equation}
where $h$ is the collision frequency and $\lambda$ is the coalescence efficiency \citep{coulaloglou1977description,prince1990bubble,chesters1991modelling,lehr2002bubble,liao2010review}. The present scaling uses the inertial-range estimate $u_{\rm rel}\sim(\varepsilon d)^{1/3}$, giving an effective encounter kernel $\Gamma\sim\varepsilon^{1/3}d^{7/3}$ for comparable-sized bubbles. This assumption is supported by recent three-dimensional measurements in homogeneous turbulence, which show that more than $98\%$ of bubble collisions occur for size ratios $\xi=d_1/d_2<2$ \citep{Tan_Zhong_Qi_Xu_Ni_2025}. Comparable-sized collisions are favoured because bubbles of similar diameter have similar inertial response and are preferentially sampled by similar strain-dominated flow structures. Film-drainage models also predict larger coalescence efficiency for comparable radii because symmetric interfacial films drain more rapidly than highly asymmetric films~\citep{chesters1991modelling,coulaloglou1977description,liao2010review}. Thus, mean-mean collisions dominate the growth of $d_{32}$.

Figure~\ref{fig:diameter}(b) shows that the upper-tail metric $d_{99.8}$ follows the same qualitative behaviour as $d_{32}$. It grows rapidly during the early mixed regime and then follows a slower power law once breakup is suppressed. The fitted exponents for $d_{99.8}$ are close to those of $d_{32}$, with similarly high $\mathcal{R}^2$ values. This agreement indicates that the upper tail and the mean diameter are governed by the same coalescence-controlled mechanism rather than by renewed fragmentation. In physical terms, the largest bubbles are formed through successive mergers involving already large structures. Because both encounter frequency and coalescence efficiency are strongest for comparable-sized pairs, large--large coalescence controls the growth of $d_{99.8}$ in the same way that mean--mean coalescence controls $d_{32}$. The close correspondence between $d_{32}$ and $d_{99.8}$ also shows that the upper tail evolves coherently with the bulk distribution, rather than developing as an independent population of rare fragments.

The same mechanism is observed in the high-Reynolds-number decaying duct-flow measurements. Figure~11 of ~\citet{kumar2026bubble} shows the axial evolution of $d_{32}$, $d_{99.8}$ and $d_H$ measured near the duct centre for different bulk velocities and void fractions. The Hinze scale is evaluated locally from the measured dissipation using Eq.~\eqref{eq:hinze scale}. As the axial coordinate $\mathcal{L}$ increases, $\varepsilon$ decreases by approximately $90\%$, so $d_H\propto\varepsilon^{-2/5}$ increases downstream. If $\varepsilon\sim\mathcal{L}^{-m}$ with $m\simeq1.85$--$2.10$, then $d_H\sim\mathcal{L}^{2m/5}$, corresponding to exponents of approximately $0.74$--$0.84$. Thus, even a large decrease in dissipation produces a comparatively moderate but systematic increase in the Hinze threshold.

Near the duct inlet, $d_{99.8}$ remains close to $d_H$, indicating that the bubble-size distribution spans both sub-Hinze and super-Hinze sizes. Bubbles larger than $d_H$ are therefore susceptible to turbulent breakup in this near-inlet region. Farther downstream, $d_H$ increases faster than $d_{99.8}$, and bubbles larger than the Hinze scale become rare beyond $\mathcal{L}>8.2$. This marks the transition to a coalescence-dominated regime in which breakup is largely suppressed. The Sauter mean diameter $d_{32}$ also increases monotonically with $\mathcal{L}$ and follows a power law away from the inlet, but remains smaller than both $d_{99.8}$ and $d_H$ throughout the measurement domain. This hierarchy, $d_{32}<d_{99.8}<d_H$ downstream, is the experimental counterpart of the DNS observation that the bubble population is progressively driven into the sub-Hinze regime. The relevant DNS and experimental length scales further support the use of
inertial-range collision and Hinze-type arguments. In the duct-flow measurements
of \citet{kumar2026bubble}, \(\eta\approx 2\times10^{-5}\,\mathrm{m}\),
\(\lambda_T\approx0.7\)--\(1.5\,\mathrm{mm}\), and
\(\ell\sim10\)--\(15\,\mathrm{mm}\), while
\(d_{32}\approx0.2\)--\(1.2\,\mathrm{mm}\) and
\(d_{99.8}\approx0.5\)--\(1.5\,\mathrm{mm}\). The DNS is performed at
smaller absolute length scales, but is constructed to preserve the relevant
inertial, capillary, and bubble-scale ordering. Thus, in both the idealized DNS
and the duct-flow experiment, the characteristic hierarchy is approximately
\begin{equation*}
    \eta \ll d \sim \lambda_T \ll \ell .
    \label{eq:lengthscale_hierarchy}
\end{equation*}
The bubbles therefore interact primarily with inertial- and Taylor-scale eddies
rather than with Kolmogorov-scale viscous motions, consistent with the use of
\(u_{\rm rel}(d)\sim(\varepsilon d)^{1/3}\) and the local Hinze scale as the
breakup threshold.

The experimentally fitted growth exponents for $d_{32}$ and $d_{99.8}$ range from approximately $0.44$ to $0.53$, with high coefficients of determination as reported in ~\citet{kumar2026bubble}. These values are close to the pure-coalescence prediction as $\beta_{\rm th}=\frac{3-m}{2}$,
which gives $\beta_{\rm th}\simeq0.45$--$0.58$ for $m\simeq2.10$--$1.85$. The comparison is significant because the experimental flow is spatially developing, confined and wall-bounded, whereas the DNS isolates the mechanism in temporal HIT. Despite these differences, both systems validate the same theoretical mechanism: turbulence decay increases $d_H$, the super-Hinze population vanishes, and both $d_{32}$ and $d_{99.8}$ grow more slowly through coalescence within the sub-Hinze range.

This result has direct implications for closure modelling. Classical breakup and coalescence kernels are typically formulated for statistically stationary or locally quasi-stationary turbulence, using a representative dissipation rate and empirical constants calibrated in specific configurations such as stirred tanks, bubble columns or pipe flows \citep{coulaloglou1977description,prince1990bubble,luo1996breakup,lehr2002bubble,liao2010review}. In strongly decaying turbulence, however, the breakup threshold itself evolves because $d_H(t)\sim\varepsilon(t)^{-2/5}$. The present results show that the dominant control parameter is not only the instantaneous value of $d/d_H$, but also the direction and rate of its drift. A bubble population that is initially near the Hinze scale can rapidly become sub-Hinze even while its physical diameter increases. Standard closures that prescribe breakup and coalescence rates solely from the local instantaneous $\varepsilon$ may therefore miss the history-dependent transition from mixed breakup--coalescence to pure coalescence. Incorporating Hinze-scale drift, or equivalently the evolution of the super-Hinze probability mass, provides a route to more predictive population-balance closures for high-Reynolds-number, non-stationary bubbly flows.

The same closure issue arises in several practical systems. In multiphase pumps, injectors, flotation cells, bubble columns, aeration devices and gas--liquid reactors, turbulence is often generated intensely over a short region and then decays downstream. Under such conditions, the bubble population does not evolve under a fixed breakup threshold; instead, the threshold moves as the turbulence decays. The present DNS and experiments show that this moving-threshold effect can suppress breakup and drive coalescence-dominated growth even in flows that are initially energetic enough to fragment bubbles. Thus, the observed agreement between theory, DNS and experiment is not only a validation of the scaling framework, but also a physically interpretable closure principle: in evolving turbulence, bubble-size evolution is governed by the relative drift between the population and the Hinze scale, rather than by a static balance between breakup and coalescence.

Finally, the theory identifies a limiting condition for maintaining self-similar growth relative to the Hinze scale. In a pure-coalescence regime, $d\sim\mathcal{L}^{(3-m)/2}$, whereas $d_H\sim\mathcal{L}^{2m/5}$. Equating these exponents gives $m=5/3$. At this value, the characteristic bubble size and the Hinze scale grow in parallel. For $m>5/3$, as in both the DNS and the experiments, $d_H$ grows faster than the bubble population. The ratio $d/d_H$ therefore decreases, breakup is progressively suppressed, and the system evolves toward a sub-Hinze, coalescence-dominated state.\\

The key DNS observation is the crossing of the bubble population by the growing Hinze scale. The initial distribution was chosen from the experiments and deliberately straddled the Hinze scale: initially, $d_{32}$ was close to $d_H$, while the upper tail and the maximum diameter exceeded $d_H$. During the subsequent decay, both $d_{32}$ and $d_{99.8}$ increase in physical size, but $d_H$ grows faster. The Hinze scale first becomes larger than $d_{32}$ and then exceeds the largest resolved bubbles at approximately $t\simeq7.1$ for $\langle \phi \rangle=0.5\%$ and $t\simeq8.3$ for $\langle \phi \rangle=2\%$. Breakup shuts off over the same interval. This correspondence directly links breakup suppression to the loss of the super-Hinze population, rather than to a decrease of bubble size in absolute units. The later crossing for $\langle \phi \rangle=2\%$ follows from its broader initial upper tail and larger initial $d_{99.8}/d_H$ compared with the $\langle \phi \rangle=0.5\%$ case. Thus the transition is caused by the breakup threshold moving faster than the bubble population, while the bubbles themselves continue to grow.

The early interval is therefore not breakup dominated. It is coalescence dominated, but breakup assisted. In the reduced number balance of Eq.~\eqref{eq:beta_mixed_summary}, coalescence scales as $\Gamma n^2$, whereas breakup scales as $Bf(t)n$, where $\alpha(t)$ is the fraction of bubbles above $d_H$. Because the coalescence contribution is quadratic in number density, even weak residual breakup can increase the population of smaller bubbles and amplify later coalescence. Breakup can therefore accelerate net bubble growth by feeding collision partners, even when coalescence controls the net evolution of the distribution. A further DNS-resolved contribution may come from deformation of initially spherical super-Hinze bubbles in HIT. Turbulent stresses stretch these bubbles before breakup or merger, increasing their interfacial area and effective collision cross-section; this is a plausible enhancement of encounter frequency, although its effect is not isolated as a separate closure here.

The mixed-regime scaling derived in Eq.~\eqref{eq:beta_mixed_summary} predicts
\begin{equation}
    d(t)\sim t^{\beta_m}, \qquad \beta_m=(15+m)/25 .
\end{equation}
For the DNS decay exponents, this gives $\beta_m\simeq0.69$--$0.70$. The early fitted $d_{32}$ exponents are approximately $0.55$ for $\langle \phi \rangle=0.5\%$ and $0.66$ for $\langle \phi \rangle=2\%$, with high $\mathcal{R}^2$. The agreement is strongest for $\langle \phi \rangle=2\%$, where the larger initial number density and the broader super-Hinze tail strengthen the breakup-assisted coalescence pathway. The lower exponent for $\langle \phi \rangle=0.5\%$ is consistent with its shorter mixed regime and smaller initially super-Hinze population, so the breakup-assisted supply of additional collision partners is exhausted earlier.

Once the super-Hinze probability mass collapses, breakup becomes negligible and the system enters a pure-coalescence regime. The supply of newly generated small bubbles vanishes, $n$ decays, and the volumetric coalescence rate decreases. The Smoluchowski-type balance from the theory predicts
\begin{equation}
    d(t)\sim t^{\beta_c}, \qquad \beta_c=(3-m)/2 .
\end{equation}
The DNS decya exponent gives $d_{32}\sim t^{0.26}$ for $\langle \phi \rangle=0.5\%$ and $d_{32}\sim t^{0.39}$ for $\langle \phi \rangle=2\%$. This confirms the predicted slowing of bubble growth after breakup becomes negligible. The important physical point is that removing breakup does not make growth faster. Once breakup stops feeding number density, coalescence becomes progressively slower because the number of available collision partners decreases. Small deviations from the ideal theory are expected because the DNS retains finite deformation, capillary relaxation, shape-dependent collision outcomes and possible reductions in coalescence efficiency for strongly deformed bubbles, for which film drainage and contact geometry differ from ideal spherical collisions.

This transition can be stated generally as a drift criterion. For a characteristic bubble size $d(t)\sim t^\beta$,
\begin{equation}
    d(t)/d_H(t)\sim t^{\beta-2m/5}.
\end{equation}
For $m>5/3$, the exponent is negative in both the mixed and pure-coalescence regimes, a condition satisfied in the DNS. Thus, even while $d_{32}$ and $d_{99.8}$ grow, $d_H$ grows faster and the population moves into the sub-Hinze regime. This mechanism connects directly to the PDF evolution: the super-Hinze tail disappears, while the mean and upper-tail diameters continue to grow by coalescence below the moving breakup threshold. The transition is therefore a dynamic one, controlled by the relative drift of two evolving scales, not a static Hinze balance.

This interpretation is consistent with, but distinct from, classical stationary descriptions of turbulent breakup. Classical Weber/Hinze-based arguments and subsequent breakup models based on eddy--bubble interactions, stochastic energy transfer and critical Weber-number criteria predict fragmentation when turbulent fluctuations are sufficiently energetic relative to surface tension \citep{Kolmogorov1949,hinze1955fundamentals,hesketh1987bubble,salibindla2020lift,coulaloglou1977description,luo1996breakup,lehr2002bubble,martinez1999breakup}. Recent simulations and experiments show that breakup and deformation also depend on local Weber-number fluctuations, finite deformation time and the instantaneous interaction between eddies and the interface \citep{vela2022memoryless,masuk2021simultaneous,calado2024dynamics,riviere2021subhinze}. In the present decaying turbulence, however, \(\varepsilon(t)\), \(d_H(t)\), and the active eddy population evolve simultaneously. Consequently, the breakup-active portion of the size distribution contracts in time, shifting the dynamics from fragmentation-assisted evolution toward coalescence-dominated growth.

The coalescence pathway is governed by both collision frequency and coalescence efficiency ($\Gamma =h \lambda$). The present scaling uses $u_{\rm rel}\sim(\varepsilon d)^{1/3}$ and hence $\Gamma\sim\varepsilon^{1/3}d^{7/3}$ for comparable-sized bubbles. This assumption is supported by recent three-dimensional measurements in homogeneous turbulence, which show that more than $98\%$ of bubble collisions occur for size ratios $\xi=d_1/d_2<2$ \citep{Tan_Zhong_Qi_Xu_Ni_2025}. Comparable-sized collisions are favoured because bubbles of similar diameter have similar inertial response and sample similar strain-dominated flow structures. Film-drainage models also predict higher efficiency for comparable radii because the intervening liquid film drains more readily than in highly asymmetric encounters. Thus mean--mean collisions control the growth of $d_{32}$, while large--large collisions control the growth of $d_{99.8}$.

The upper-tail diameter $d_{99.8}$ follows the same qualitative evolution as $d_{32}$. It grows rapidly in the early mixed regime and slows after breakup is suppressed. The fitted exponents are close to those of $d_{32}$, with high $\mathcal{R}^2$, showing that the upper tail is governed by the same coalescence-controlled mechanism rather than by renewed fragmentation. The largest bubbles form through successive mergers involving already large structures, so the upper tail evolves coherently with the bulk distribution. Once the super-Hinze population vanishes, the upper tail does not remain breakup controlled; it becomes part of the same coalescence cascade as the mean population.

The corresponding characteristic-diameter measurements of \citet{kumar2026bubble} provide a quantitative check of the diameter scaling. Figure~11 of \citet{kumar2026bubble} shows the axial evolution of \(d_{32}\), \(d_{99.8}\), and the locally evaluated Hinze scale \(d_H\) near the duct centre. Since the measured dissipation follows \(\varepsilon\sim\mathcal{L}^{-m}\) with \(m\simeq1.85\)--\(2.10\), the Hinze scale grows as \(d_H\sim\mathcal{L}^{2m/5}\), giving exponents \(0.74\)--\(0.84\). The measured growth exponents of \(d_{32}\) and \(d_{99.8}\) are approximately \(0.44\)--\(0.53\), with high coefficients of determination, close to the pure-coalescence prediction \(\beta_{\rm th}=(3-m)/2=0.45\)--\(0.58\). Downstream of the near-inlet adjustment region, where bubbles larger than \(d_H\) become rare beyond \(\mathcal{L}>8.2\), the hierarchy \(d_{32}<d_{99.8}<d_H\) is established. Thus, the duct-flow diameter data provide the spatial counterpart of the DNS result: bubble sizes increase by coalescence while the population becomes progressively sub-Hinze.

This comparison indicates that Hinze-scale drift is not a DNS artefact of ideal HIT. The same organizing mechanism survives in a wall-bounded, spatially developing, pump-driven flow, although the effective exponent changes because of confinement, mean shear, wall effects and inhomogeneity. At fixed bulk velocity $V$, increasing $\langle \phi \rangle$ increases the bubble number density and therefore the collision frequency. Higher $\langle \phi \rangle$ also increases the probability of bubble--bubble interactions and can broaden the upper tail, making $d_{99.8}/d_H$ larger near the inlet and maintaining a finite super-Hinze population over a longer downstream distance. Once the flow enters the coalescence-dominated region, however, the fitted growth exponents of $d_{32}$ and $d_{99.8}$ remain only weakly dependent on $\langle \phi \rangle$. This suggests that void fraction changes the amplitude, collision frequency and upper-tail probability, while the downstream exponent is controlled primarily by the decay rate of $\varepsilon$ through Hinze-scale drift.

At fixed void fraction, changing $V$ modifies the initial turbulence level and therefore the initial Hinze scale. Increasing the bulk velocity increases the initial turbulence intensity and dissipation, and the larger initial $\varepsilon$ reduces $d_H$ because $d_H\sim\varepsilon^{-2/5}$. More of the inlet distribution can therefore lie near or above the Hinze scale, making near-inlet breakup more likely. Stronger turbulence can generate more small bubbles near the inlet and increase the initial bubble count. Downstream, however, turbulence decay again increases $d_H$ and suppresses breakup. Thus higher $V$ mainly changes the early adjustment region, the initial population amplitude and the distance over which the distribution becomes sub-Hinze. Once the flow is in the sub-Hinze coalescence-dominated regime, the fitted growth exponents remain close to the pure-coalescence prediction. This supports the view that the dominant asymptotic control is not the absolute turbulence intensity alone, but the decay of $\varepsilon$ and the resulting drift of $d/d_H$.

These scaling predictions and their validation are critical for PBE closures and can be applied directly to model source terms. Classical breakup and coalescence kernels assume statistically stationary or locally quasi-stationary turbulence and evaluate rates from a representative dissipation rate \citep{coulaloglou1977description,prince1990bubble,luo1996breakup,lehr2002bubble,liao2010review}. In strongly decaying turbulence, $d_H(t)\sim\varepsilon(t)^{-2/5}$ evolves, so the control parameter is not only instantaneous $d/d_H$, but also the direction and rate of drift of $d/d_H$. A population initially near the Hinze scale can become sub-Hinze even while physical diameters increase. The closure problem is therefore not merely inaccurate constants. Stationary kernels treat breakup and coalescence as rates evaluated at a fixed local state, whereas in decaying turbulence the state itself is moving through regime space. Incorporating Hinze-scale drift, or equivalently the super-Hinze probability mass $\mathcal{P}_{d>d_H}$, provides a route toward closures that retain this history dependence.

The preceding diameter analysis also defines a critical decay exponent. In the pure-coalescence limit, comparing ($d\sim\mathcal{L}^{(3-m)/2}$) with ($d_H\sim\mathcal{L}^{2m/5}$) gives ($m_c=5/3$). For the present DNS and experimental cases, ($m>m_c$), so the breakable tail is expected to be depleted and the subsequent evolution should be reflected not only in the diameter statistics, but also in population-level measures.  \\

We therefore test the same mechanism using two quantities that are independent of any single diameter fit: the interfacial area ($A(t)$) and the bubble count ($n(t)$). The quantity $A(t)$ measures the gas--liquid interface available for transfer processes, while ($n(t)$) counts disconnected gas bodies. Coalescence transfers gas into fewer larger bubbles and reduces both quantities at approximately fixed void fraction, whereas breakup creates new interface and new components. Their evolution therefore tests whether the observed diameter growth reflects a genuine coalescence-dominated redistribution of gas volume.

These interfacial-area and bubble-count diagnostics provide direct constraints for IATE and PBE closures. Interfacial area controls gas--liquid mass transfer, heat transfer, reaction rates and process efficiency \citep{kulkarni2007mass,kantarci2005bubble,garcia2009bioreactor,hibiki1999experimental,ishii2010thermo,Chen2021}, and it is a primary variable in interfacial-area transport equations \citep{hibiki1999experimental,wu1998one,ishii2010thermo,Chen2021}. Population-balance models similarly evolve number density or the bubble-size distribution using breakup and coalescence kernels \citep{ramkrishna2000population,coulaloglou1977description,prince1990bubble,chesters1991modelling,luo1996breakup,lehr2002bubble,liao2010review}. These source terms are usually evaluated from instantaneous local variables such as $\varepsilon$, $\phi$, $d$, local Weber number and empirical interaction frequencies. Classical turbulent breakup is organized by inertial--capillary balance \citep{Kolmogorov1949,hinze1955fundamentals,hesketh1987bubble,martinez1999breakup,salibindla2020lift}, with additional sensitivity to local Weber-number fluctuations, finite deformation time and eddy--interface interactions \citep{vela2022memoryless,masuk2021simultaneous,calado2024dynamics,riviere2021subhinze}. In decaying turbulence, however, the local breakup frequency is insufficient; the closure must also consider the distribution that remains above the moving Hinze scale.

The relevant breakable population ($\mathcal{P}_{d>d_H}$) is quantified by the super-Hinze probability mass as discussed in Eq.~\eqref{eq:super_hinze_probability}. This quantity controls whether breakup can still create new interface and new bubbles. Once $\mathcal{P}_{d>d_H}\rightarrow0$, breakup-source terms should shut down, even though turbulent fluctuations remain, and the population should reduce to coalescence-controlled decay. The novelty here is to show this drift not only through $d_{32}$, $d_{99.8}$ and the PDFs, but also through the resolved interfacial geometry, the experimental interfacial-area decay and the population topology.

In the DNS, $A(t)$ and $n(t)$ are extracted independently of $d_{32}$. The interfacial area is estimated directly from the diffuse-interface field using the magnitude of the gas-phase indicator gradient. For a connected
bubble \(b\),
\begin{equation}
    A_b(t)=\int_{\Omega_b} |\nabla \chi_g|\,{\rm d}V,
    \qquad
    A(t)=\sum_b A_b(t),
\end{equation}
where \(\chi_g\) is the gas-phase indicator field, with \(\chi_g=1\) in
the gas phase and \(\chi_g=0\) in the liquid phase, and \(\Omega_b\) is the diffuse interfacial region associated with bubble \(b\). In this formulation, \(|\nabla \chi_g|\) acts as a regularized interfacial delta function, so its volume integral gives the bubble surface area. This definition naturally accounts for bubble deformation and topology changes. This grid-consistent measure retains deformation, wrinkling and transient topology changes, and therefore avoids spherical or ellipsoidal assumptions. The bubble count is obtained from connected gas regions, so $n(t)$ is a direct topological statistic: mergers remove connected components, whereas breakup creates them.

\begin{figure}[htbp]
    \begin{subfigure}[b]{0.49\textwidth}
        \includegraphics[width=\textwidth]{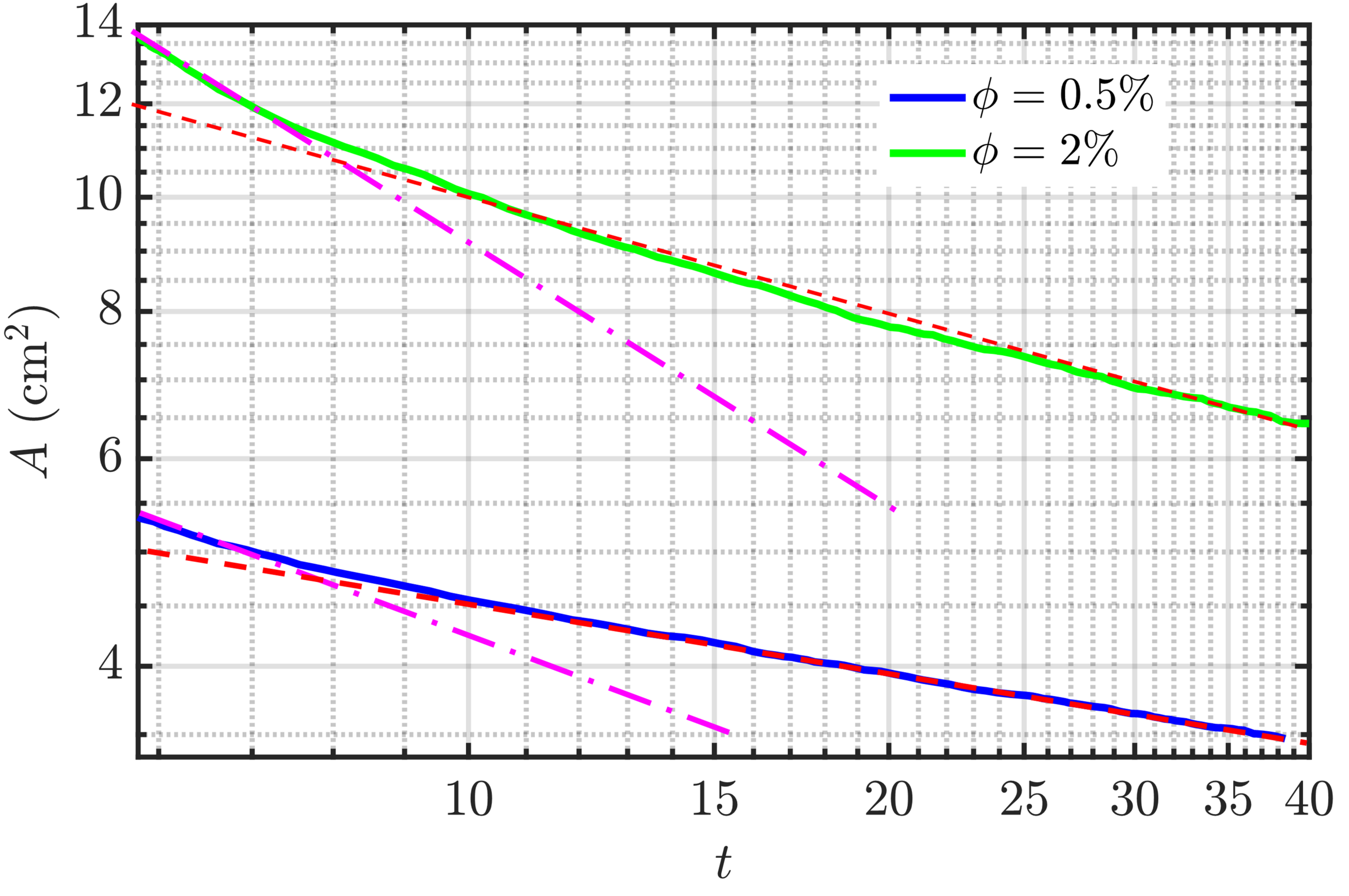}
        \caption{$A(t)$, with slopes $-0.24$ (0.5\%) for $t>7$ and $-0.37$ (2\%) for $t>8$ , and corresponding $\mathcal{R}^2$ values of $0.99$ and $0.98$. }
        \label{fig:a}
    \end{subfigure}
    \hfill
    \begin{subfigure}[b]{0.49\textwidth}
        \includegraphics[width=\textwidth]{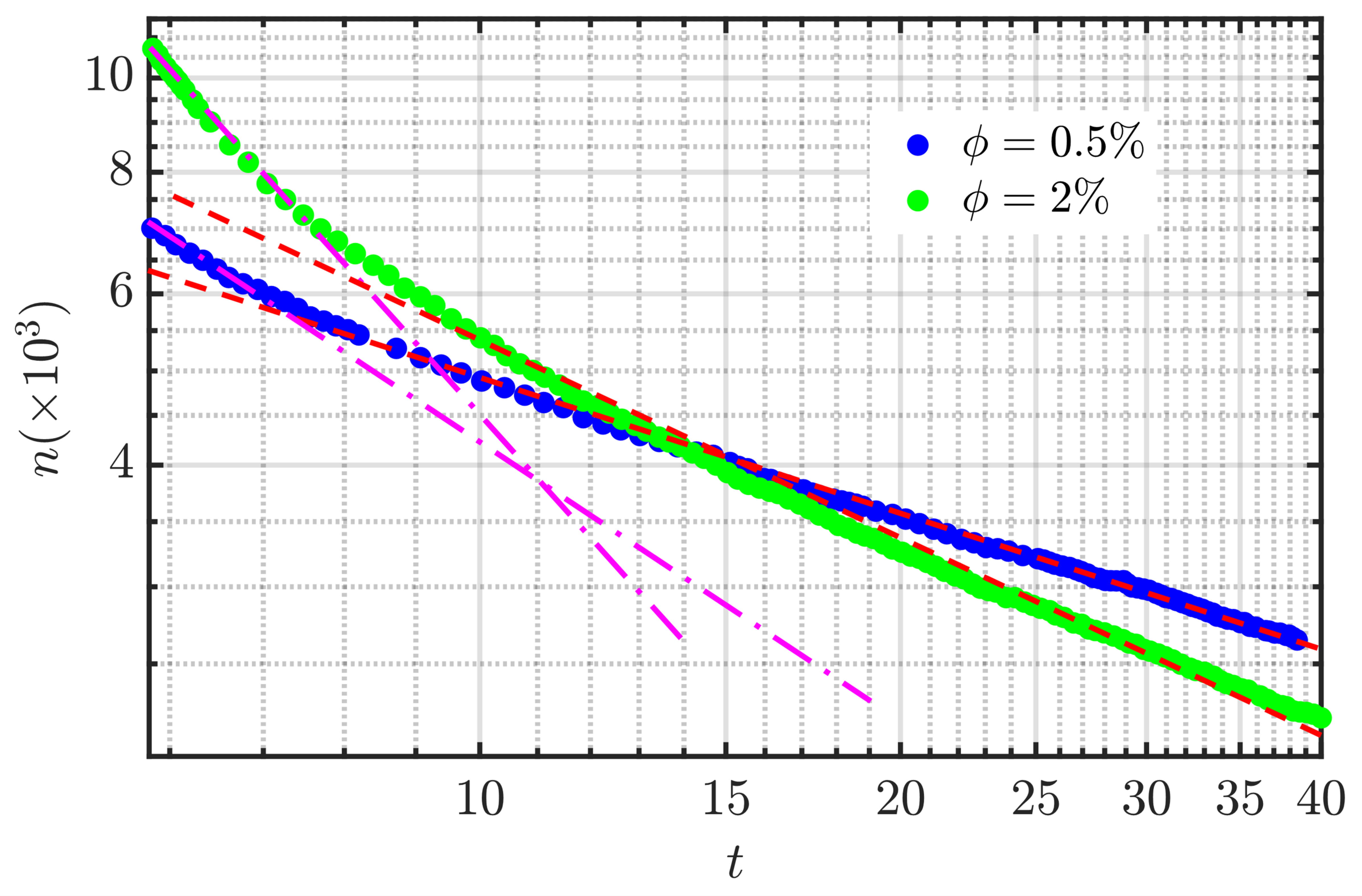}
        \caption{$n(t)$, with slopes $-0.56$ (0.5\%) for $t>7$ and $-0.79$ (2\%) for $t>8$ , and corresponding $\mathcal{R}^2$ values of $0.98$ and $0.95$. }
        \label{fig:b}
    \end{subfigure}  
    \caption{Temporal evolution of interfacial area $A(t)$ and total bubble count $n(t)$ in HIT DNS.}
    \label{fig:A and n}
\end{figure}

Figure~\ref{fig:A and n}(a) shows that $A(t)$ decreases monotonically for both $\langle \phi \rangle=0.5\%$ and $2\%$. The initial decrease is rapid because coalescence already dominates the net population budget, even though a finite super-Hinze population can still deform and occasionally break. Once breakup becomes negligible, $A(t)$ follows a power law. The fitted late-time slopes are approximately $-0.24$ for $\langle \phi \rangle=0.5\%$ for $t>7$ and $-0.37$ for $\langle \phi \rangle=2\%$ for $t>8$, with $\mathcal{R}^2\simeq0.99$ and $0.98$, respectively. In the pure-coalescence regime, the theory in Eq.~\eqref{eq:beta_mixed_summary} gives $A\sim d^{-1}$ at approximately fixed void fraction. Since $d\sim t^{(3-m)/2}$,
\begin{equation}
    A(t)\sim t^{\theta_A}, \qquad \theta_A=(m-3)/2 .
\end{equation}
For the measured DNS decay exponents this gives values close to $-0.26$ and $-0.39$, consistent with the simulations. The decay of $A(t)$ shows that merger removes gas--liquid interface faster than breakup creates it.

The remaining deviations from the ideal area scaling are consistent with finite deformation, capillary relaxation, polydispersity and event-identification uncertainty during rapid topology changes. Early super-Hinze bubbles may be stretched before they merge or fragment, whereas later sub-Hinze bubbles relax while coalescence removes interface. The stronger decay at $\langle \phi \rangle=2\%$ is consistent with larger collision frequency, although the two DNS cases are not sufficient to infer a separate exponent law in $\langle \phi \rangle$.

The bubble count in Figure~\ref{fig:A and n}(b) gives the corresponding topological measure. For both void fractions, $n(t)$ decreases monotonically and follows a power law after the mixed early interval, approximately for $t>7$ at $\langle \phi \rangle=0.5\%$ and $t>8$ at $\langle \phi \rangle=2\%$. The fitted slopes are about $-0.56$ and $-0.79$, with $\mathcal{R}^2\simeq0.98$ and $0.95$. At fixed gas volume, pure coalescence gives $n\sim d^{-3}$ and hence
\begin{equation}
    n(t)\sim t^{\theta_n}, \qquad \theta_n=(3m-9)/2 .
\end{equation}
The measured slopes are somewhat lower magnitude than the ideal predictions, approximately $-0.78$ and $-1.1$ for the corresponding DNS decay exponents. This is expected because $n(t)$ is especially sensitive to small bubbles and to segmentation during rapid topology changes. Together, the DNS results show that the HIT-based theory predicts more than one diameter metric: it captures the coupled redistribution of gas volume across size, area and connected components as the super-Hinze tail disappears.

The duct-flow measurements provide the spatially developing counterpart to the temporal DNS. The experimental analysis below is new: \citet{kumar2026bubble} reported the duct configuration, turbulence decay, bubble size distribution (BSD) evolution, characteristic diameters, Hinze scale, PDF scaling and void-fraction redistribution, whereas the present subsection extracts $A(t)$ and $n(t)$ from the same measurements and treats them as the interfacial-area and number-density variables required by IATE/PBM closures. These quantities are reconstructed from measured bubble-size PDFs and local void fraction in a centreline measurement volume, whereas the DNS uses $|\nabla\alpha|$ and connected components directly. This distinction explains why agreement with the ideal theory is close but not exact.

\begin{figure}[htbp]
\centering

\begin{minipage}[t]{0.48\textwidth}
    \vspace{0pt}
    \centering
    \includegraphics[width=\textwidth]{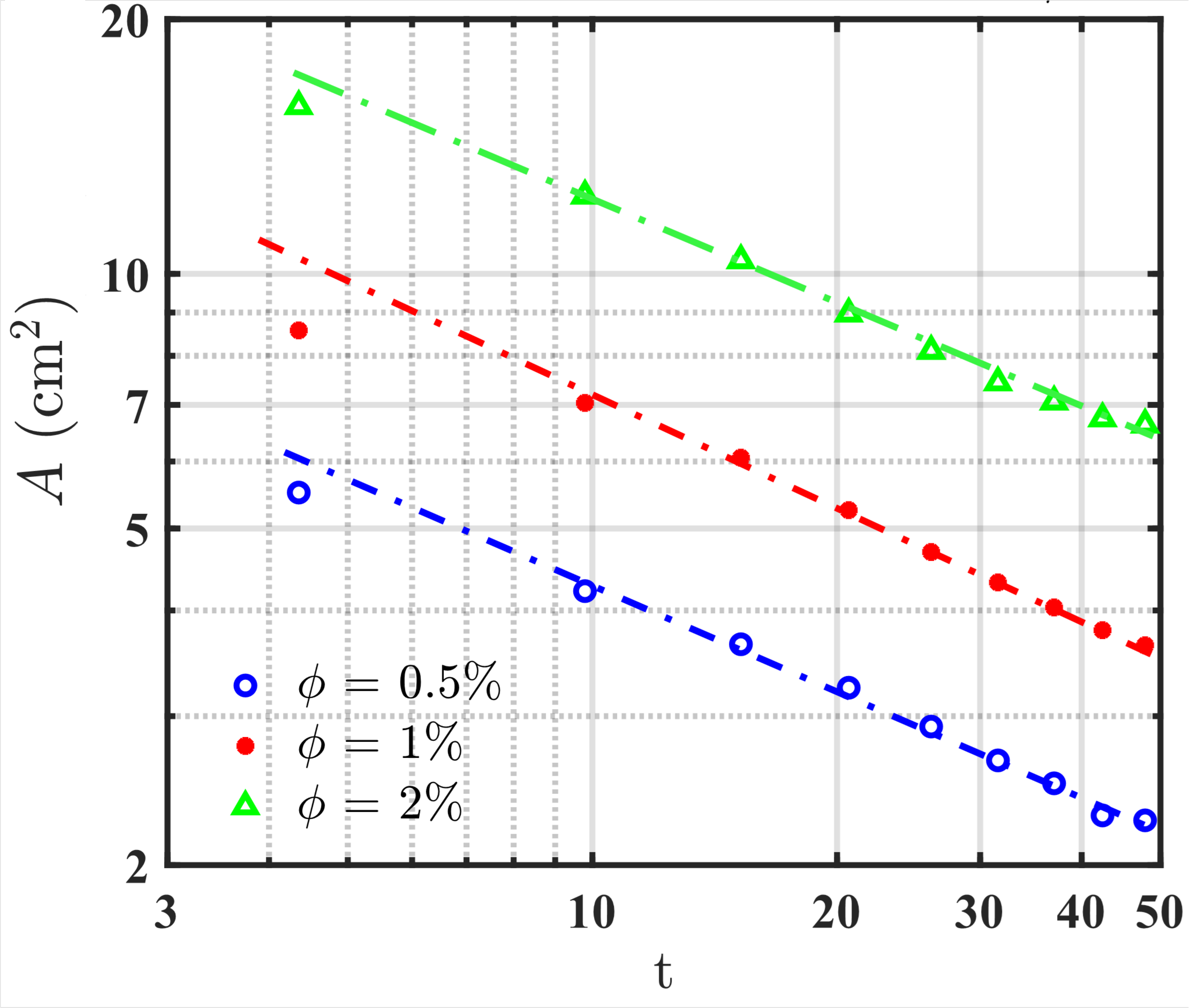}
    \subcaption{$V = 6.1\,\mathrm{m\,s^{-1}}$}
    \label{fig:area-v61}
\end{minipage}
\hfill
\begin{minipage}[t]{0.48\textwidth}
    \vspace{0pt}
    \centering
    \includegraphics[width=\textwidth]{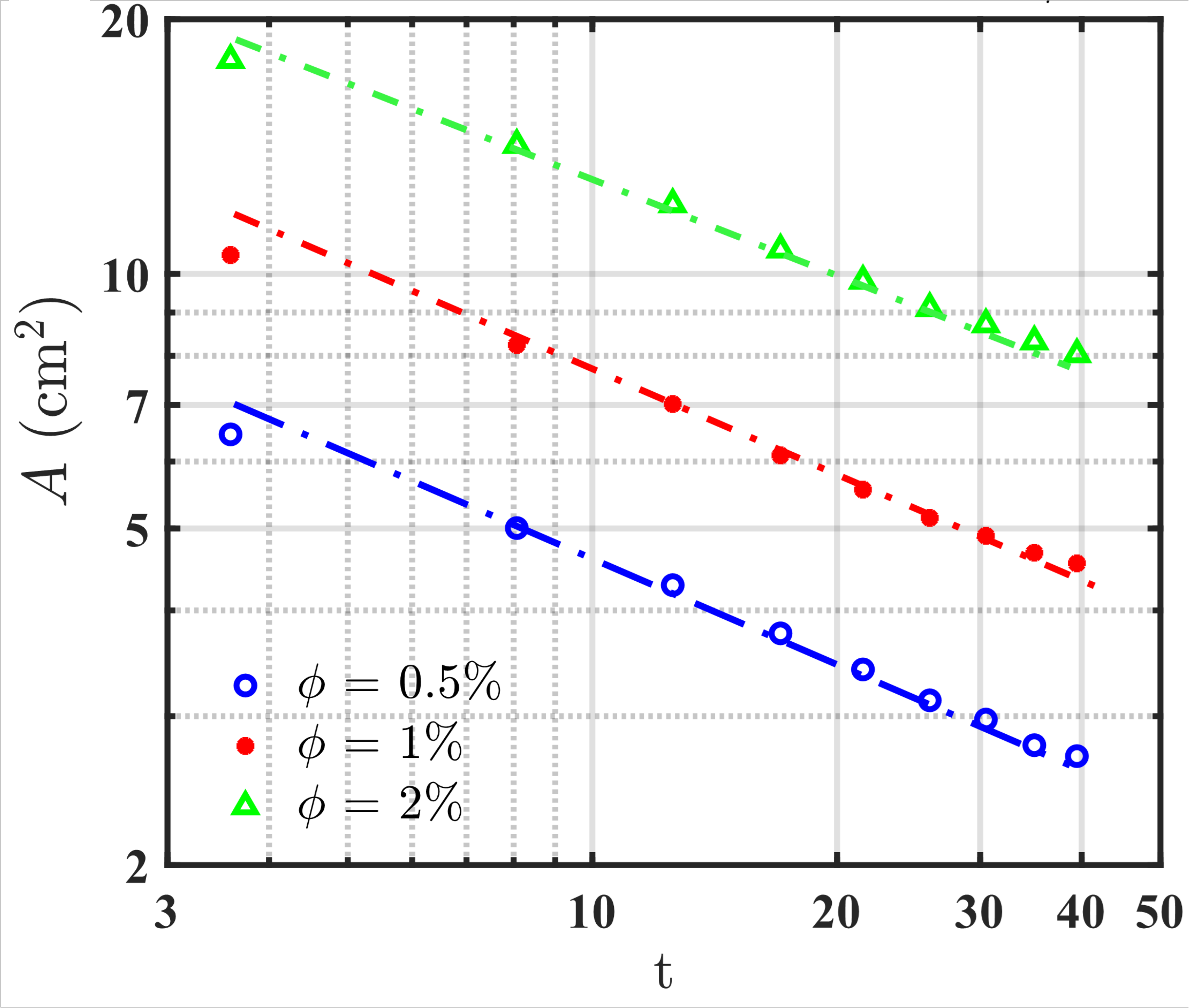}
    \subcaption{$V = 7.4\,\mathrm{m\,s^{-1}}$}
    \label{fig:area-v74}
\end{minipage}

\vspace{1em}

\begin{minipage}[t]{0.48\textwidth}
    \vspace{0pt}
    \centering
    \includegraphics[width=\textwidth]{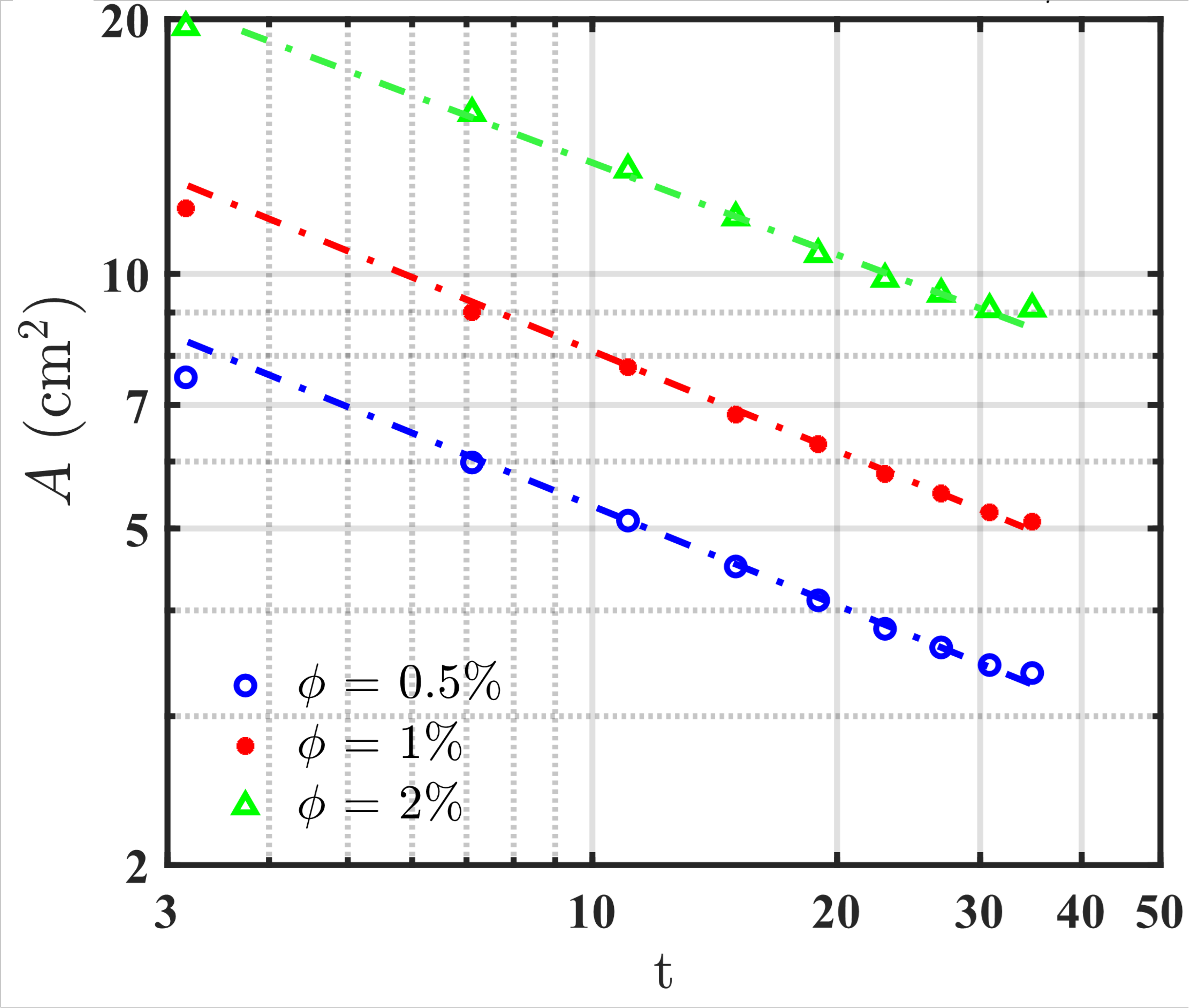}
    \subcaption{$V = 8.4\,\mathrm{m\,s^{-1}}$}
    \label{fig:area-v84}
\end{minipage}
\hfill
\begin{minipage}[t]{0.48\textwidth}
    \vspace{0pt}
    \centering
    \subcaption{Fitting parameters for $A(t)$, excluding the first data point.}
    \label{fig:area-fit-table}
    \vspace{0.5em}

    \renewcommand{\arraystretch}{1.25}
    \begin{tabular}{c c c c}
        \hline
        & $V_1$ & $V_2$ & $V_3$ \\ 
        $\phi$ (\%) & $\theta_A$ ($\mathcal{R}^2$) & $\theta_A$ ($\mathcal{R}^2$) & $\theta_A$ ($\mathcal{R}^2$) \\
        \hline
        $\phi_1$ & -0.44 (0.97) & -0.42 (0.96) & -0.40 (0.98) \\
        $\phi_2$ & -0.43 (0.98) & -0.41 (0.98) & -0.41 (0.98) \\
        $\phi_3$ & -0.43 (0.98) & -0.40 (0.99) & -0.39 (0.98) \\
        \hline
    \end{tabular}
\end{minipage}

\caption{Temporal evolution of the interfacial area $A(t)$ for three bulk velocities: (a) $V=6.1\,\mathrm{m\,s^{-1}}$, 
(b) $V=7.4\,\mathrm{m\,s^{-1}}$, and (c) $V=8.4\,\mathrm{m\,s^{-1}}$, each shown for void fractions $\phi=0.5\%$, $1\%$, and $2\%$. In all cases, $A(t)$ exhibits a power-law decay beyond the initial transient region. Panel (d) reports the fitted power-law exponents $\theta_A$ and coefficients of determination $\mathcal{R}^2$, obtained by excluding the first data point. The interfacial area (A) is calculated within a control volume, $V_c$ $(\approx 6~\mathrm{mm})^3$, near the duct centre and containing the bubbles detected in that region. }
\label{fig:interfacial area_exp}

\end{figure}

Figure~\ref{fig:interfacial area_exp} shows the newly analysed experimental interfacial area, computed by summing individual bubble-area contributions inferred from the measured PDF. For all three bulk velocities, $V=6.1$, $7.4$ and $8.4\,\mathrm{m\,s^{-1}}$, and all void fractions, $\langle \phi \rangle=0.5\%$, $1\%$ and $2\%$, $A(t)$ decreases downstream, or equivalently with convective time. The first measurement station is excluded from the fits because near-inlet breakup persists there. Beyond this adjustment region, $A(t)$ follows a power law with $\mathcal{R}^2>0.96$ and fitted exponents between approximately $-0.39$ and $-0.44$.

The experimental area exponents are close to, although slightly smaller magnitude than, the pure-coalescence prediction. For $m\simeq1.85$--$2.10$, the scaling $\theta_A=(m-3)/2$ gives exponents from about $-0.58$ to $-0.45$. The weaker measured decay is consistent with expected effects of spatial development, confinement, mean shear, wall effects, finite optical sampling and polydispersity. Small and intermediate bubbles contribute strongly to area per unit gas volume, so a monodisperse theory is expected to overpredict the magnitude of the decay. The agreement in sign, magnitude and fit quality nevertheless indicates that the duct flow loses active gas--liquid interface downstream even while $d_{32}$ and $d_{99.8}$ increase by coalescence.

\begin{figure}[htbp]
\centering

\begin{minipage}[t]{0.48\textwidth}
\vspace{0pt}
\centering
\includegraphics[width=\textwidth]{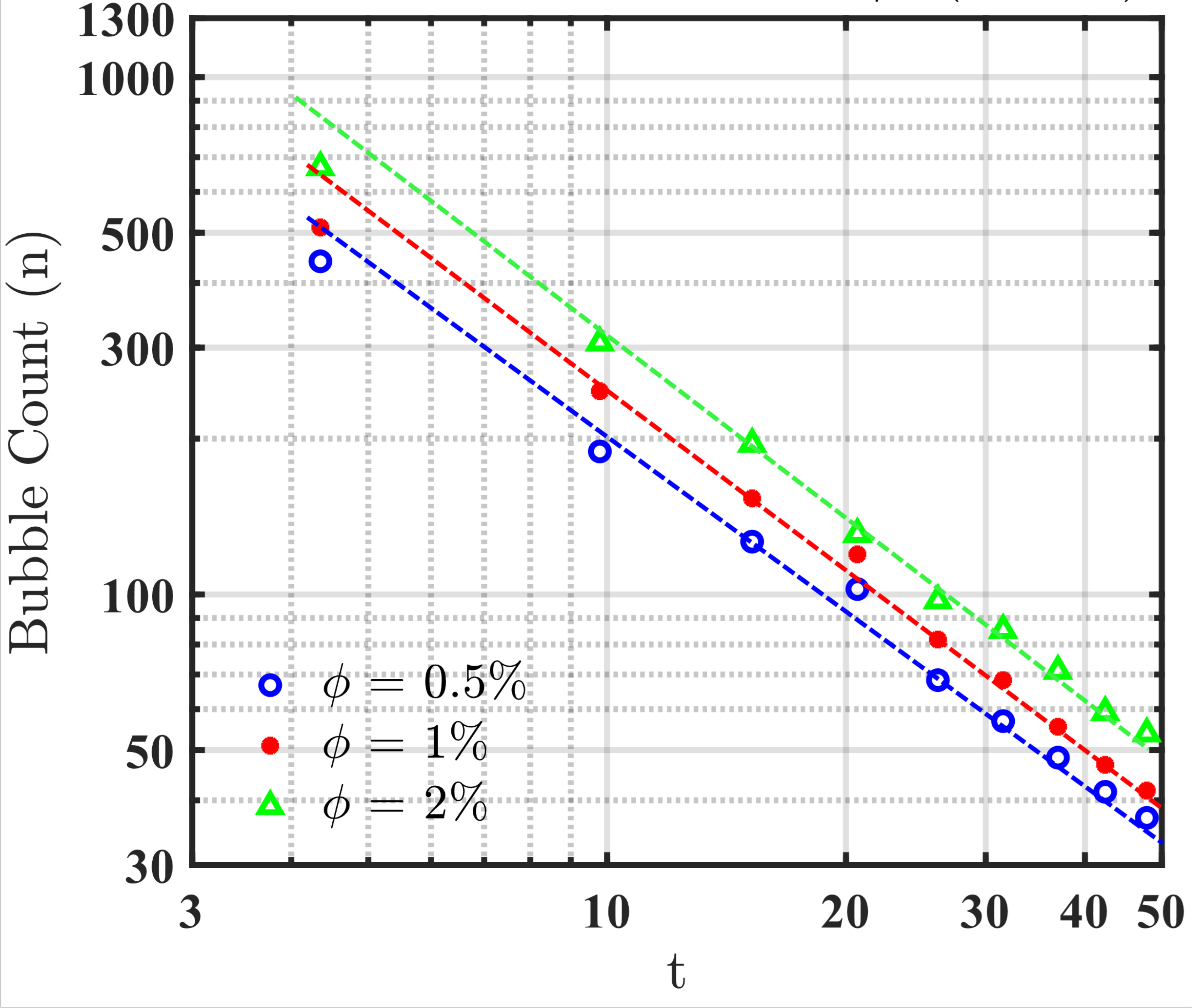}
\subcaption{$V = 6.1,\mathrm{m,s^{-1}}$}
\label{fig:bubble-count-v61}
\end{minipage}
\hfill
\begin{minipage}[t]{0.48\textwidth}
\vspace{0pt}
\centering
\includegraphics[width=\textwidth]{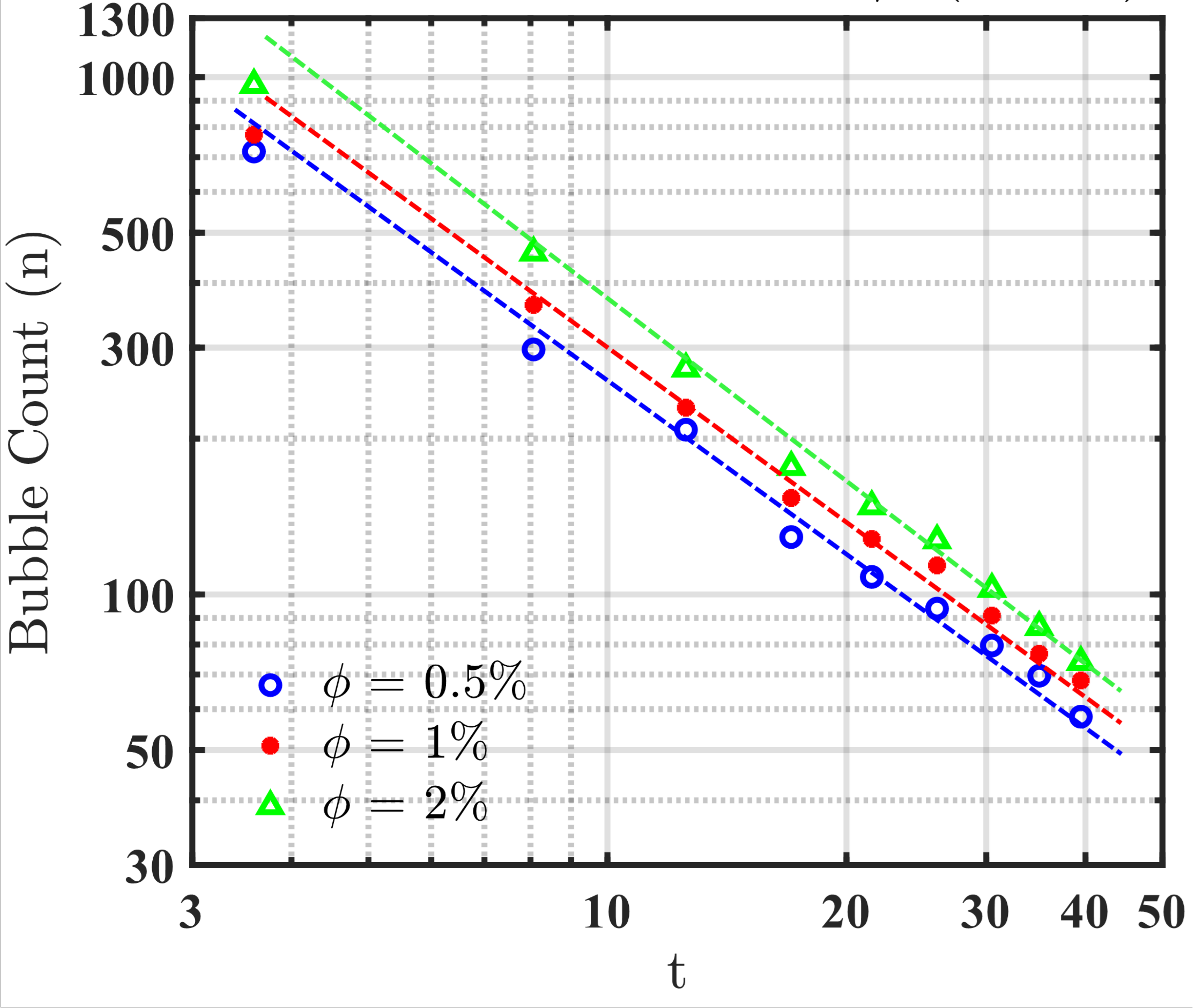}
\subcaption{$V = 7.4,\mathrm{m\,s^{-1}}$}
\label{fig:bubble-count-v74}
\end{minipage}

\vspace{1em}

\begin{minipage}[t]{0.48\textwidth}
\vspace{0pt}
\centering
\includegraphics[width=\textwidth]{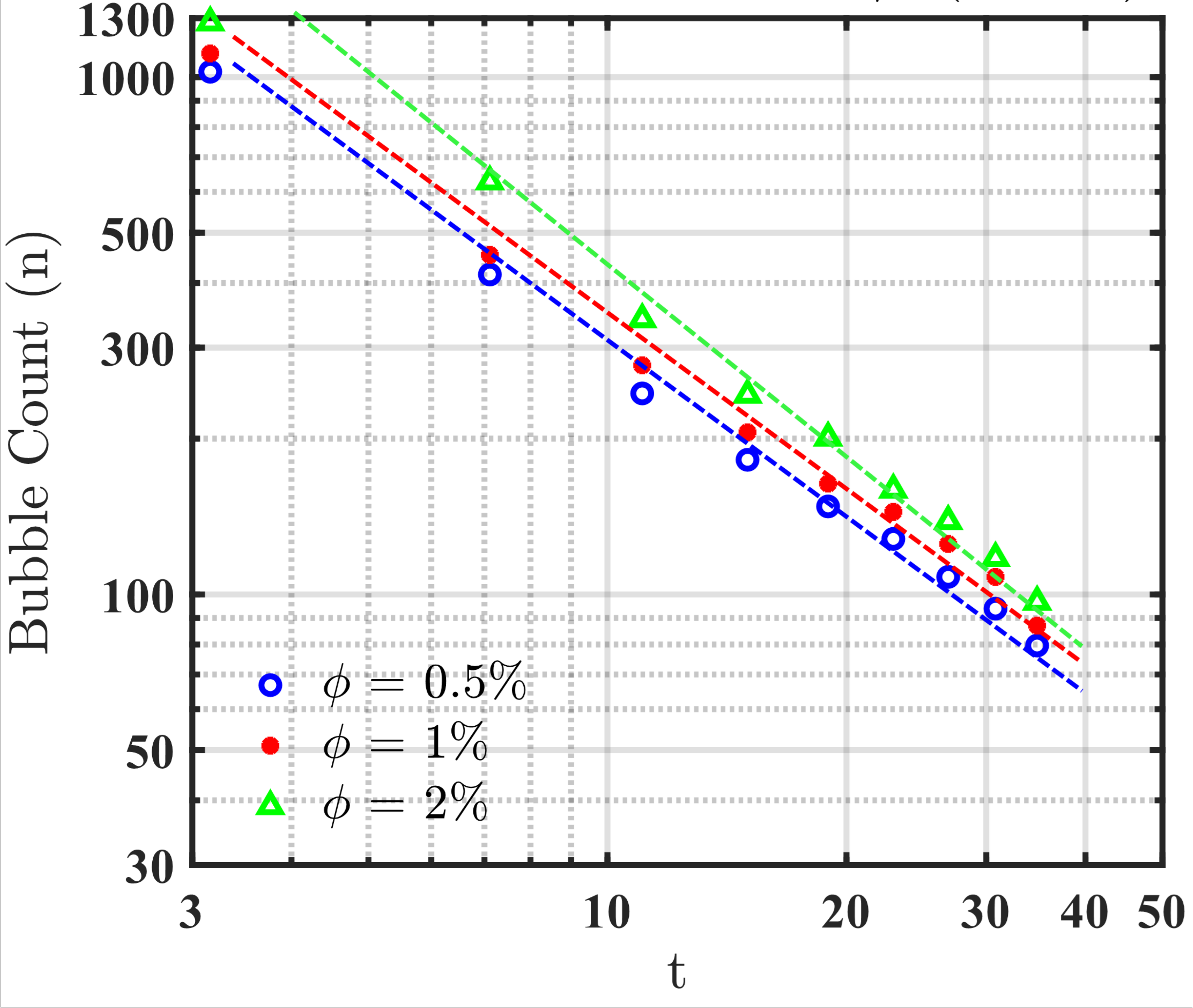}
\subcaption{$V = 8.4\,\mathrm{m\,s^{-1}}$}
\label{fig:bubble-count-v84}
\end{minipage}
\hfill
\begin{minipage}[t]{0.48\textwidth}
\vspace{0pt}
\centering

\subcaption{Fitting parameters for $n(t)$, excluding the first data point.}
\label{fig:bubble-count-fit-table}
\vspace{0.5em}

\renewcommand{\arraystretch}{1.25}
\begin{tabular}{c c c c}
    \hline
    & $V_1$ & $V_2$ & $V_3$ \\ 
    $\phi$ (\%) & $\theta_n$ ($\mathcal{R}^2$) & $\theta_n$ ($\mathcal{R}^2$) & $\theta_n$ ($\mathcal{R}^2$) \\
    \hline
    $\phi_1$ & -1.25 (0.98) & -1.24 (0.96) & -1.25 (0.99) \\
    $\phi_2$ & -1.27 (0.97) & -1.22 (0.98) & -1.23 (0.97) \\
    $\phi_3$ & -1.27 (0.97) & -1.26 (0.96) & -1.30 (0.97) \\
    \hline
\end{tabular}

\end{minipage}

\caption{Temporal evolution of the total bubble count $n(t)$ near the center of the domain.
Panels (a--c) show $n(t)$ for three bulk velocities,
$V=6.1$, $7.4$, and $8.4,\mathrm{m\,s^{-1}}$, respectively, each shown for void fractions
$\langle \phi \rangle=0.5\%$, $1\%$, and $2\%$. Panel (d) summarizes the fitted power-law exponents
$\theta_n$ and coefficients of determination $\mathcal{R}^2$, obtained from log--log fits after excluding the first data point.}
\label{fig:exp bubble count}
\end{figure}

The newly analysed experimental bubble count provides the corresponding topological test. Figure~\ref{fig:exp bubble count} shows $n(t)$ in a cubic control volume of side length $6\,\mathrm{mm}$ located near the duct centre. At each station, a statistically converged BSD is obtained from the image ensemble and the bubble count is reconstructed using the measured PDF and known $\langle \phi \rangle$. The count decreases monotonically for every velocity and void fraction. The first point is again excluded because near-inlet breakup persists. Downstream of this adjustment region, $n(t)$ follows power laws with exponents between $-1.22$ and $-1.3$ and $\mathcal{R}^2>0.95$.

The measured count exponents are slightly smaller magnitude than the ideal pure-coalescence prediction. For $m\simeq1.85$--$2.10$, the scaling $\theta_n=(3m-9)/2$ gives values from approximately $-1.73$ to $-1.35$. Given finite imaging resolution, finite control volume, polydispersity and sensitivity to the smallest bubbles, the observed range is consistent with coalescence-controlled decay. This newly analysed count decay is the experimental topological counterpart of the interfacial-area decay.

The operating-parameter trends mainly affect amplitudes and the near-inlet adjustment. At fixed \(V\), increasing \(\phi\) increases the absolute values of \(A(t)\) and \(n(t)\) through increased gas holdup and number density, but the fitted exponents remain only weakly dependent on $\langle \phi \rangle$ once the flow becomes predominantly sub-Hinze. At fixed $\langle \phi \rangle$, increasing $V$ raises initial $\varepsilon$, lowers the initial Hinze scale, promotes near-inlet breakup and shifts both curves upward. Downstream, however, the exponents are organized primarily by the decay of $\varepsilon$ and the associated Hinze-scale drift rather than by void fraction or absolute inlet turbulence level alone.

The combined DNS and newly analysed experimental $A(t)$ and $n(t)$ results provide closure-level constraints. Existing IATE and PBM closures contain breakup and coalescence source terms for interfacial area and number density, but many such closures evaluate those terms at a fixed local state. In decaying turbulence, the drift of the distribution relative to $d_H(t)$ is not a small correction. It determines whether breakup can still create new bubbles and interfacial area. A non-stationary closure should therefore weight breakup sources by the breakable population, for example
\begin{equation}
    S_A^{br},\,S_n^{br}\propto \mathcal{P}_{>H}(t),
\end{equation}
where $S_A^{br}$ and $S_n^{br}$ denote the breakup contributions to interfacial area and bubble count, respectively. Equivalently, the source terms may depend on a drift variable such as $d_c(t)/d_H(t)$, where \(d_c\) is a representative bubble size, for example \(d_{32}\). Once \(\mathcal{P}_{>H}\) becomes negligible, breakup source terms should vanish or become negligible, and the model should reduce to the coalescence-controlled limits
\begin{equation}
    A(t)\sim t^{(m-3)/2}, \qquad
    n(t)\sim t^{(3m-9)/2}.
\end{equation}
These relations provide asymptotic constraints for non-stationary IATE/PBM source terms in decaying turbulence. The issue is not merely calibration of empirical constants. Stationary closures treat source terms at a fixed local state, whereas the present flow moves through regime space as $\varepsilon(t)$ decays and $d_H(t)$ grows. The new $A(t)$ and $n(t)$ analyses show that Hinze-scale drift reorganizes not only characteristic size, but also available interface and population topology. This is why the present results matter for closure modelling: they provide DNS- and experiment-supported scaling constraints for source terms in non-stationary turbulent bubbly flows. \\

\subsection{Coalescence rates and implications for population-balance closures}

The preceding discussions established the geometric and population-level consequences of Hinze-scale drift: the bubble population moves into the sub-Hinze range, $d_{32}$ and $d_{99.8}$ increase, while interfacial area and bubble count decrease. We now test the dynamic source terms themselves. In the coalescence-dominated limit of the reduced number balance derived in Eq.~\eqref{eq:n_balance_summary_reduced}, the theory predicts distinct temporal laws for $R(t)$, $r(t)$, $\Gamma(t)$. Here $R(t)$ is the total rate at which coalescence removes bubbles from the population, $r(t)=R(t)/n(t)$ is the per-bubble coalescence hazard, or coalescence clock, and $\Gamma(t)=R(t)/n^2(t)$ is the effective pairwise coalescence kernel, with any constant numerical factor absorbed into its definition. This decomposition separates total population loss, per-bubble removal probability and pairwise interaction strength. 

These rates are source-term diagnostics for population-balance closures and, through their effect on bubble number, interfacial area and the bubble-size distribution, these rates constrain interfacial-area-transport closures. Population-balance models evolve the bubble-size distribution or number density through breakup and coalescence kernels~\citep{ramkrishna2000population,coulaloglou1977description,prince1990bubble,chesters1991modelling,luo1996breakup,lehr2002bubble,liao2010review}. Interfacial-area-transport equations treat interfacial area concentration as a transported structural variable with coalescence and breakup source terms~\citep{wu1998one,hibiki1999experimental,ishii2010thermo,Chen2021}. In most applications these closures are evaluated from an instantaneous local state, typically through $\varepsilon$, $\phi$, $d$, a Weber-number criterion and model-dependent breakup or coalescence efficiencies. Classical Hinze arguments and modern studies of local Weber-number fluctuations, finite deformation time and eddy-interface interactions clarify the local physics of breakup~\citep{Kolmogorov1949,hinze1955fundamentals,hesketh1987bubble,martinez1999breakup,salibindla2020lift,vela2022memoryless,masuk2021simultaneous,riviere2021subhinze,calado2024dynamics}. In decaying turbulence, however, the closure also needs the fraction of the population that remains breakable. A compact way to express this missing state variable is the super-Hinze probability mass as shown in Eq.~\eqref{eq:super_hinze_probability}. Once $\mathcal{P}_{d>d_H}$ becomes negligible, breakup can no longer create new bubbles or interfacial area at an appreciable rate, and the measured $R$, $r$ and $\Gamma$ should recover the pure-coalescence scalings in Tables~\ref{tab:scalings_exponents_m_coal_break} and~\ref{tab:scalings_exponents_m_coal}.

In our DNS, these quantities are measured directly from event-resolved bubble lineages. Bubbles are identified as connected gas-phase regions and characterized by their volume-equivalent diameter, volume and centroid. Successive snapshots are linked using mass conservation and bounded centroid displacement, following established volume-conserving lineage strategies~\citep{chan2021identifying}. Breakup is recorded when one parent connects to multiple children with conserved total volume, whereas coalescence is recorded when multiple parents form a single child satisfying the same volume constraint. Because the classification is based on connectivity and volume rather than sphericity, the method remains insensitive to bubble deformation and provides direct event-resolved estimates of $R(t)$, $r(t)$ and $\Gamma(t)$.

\begin{figure}[htbp]
    \begin{subfigure}[b]{0.312\textwidth}
        \includegraphics[width=\textwidth]{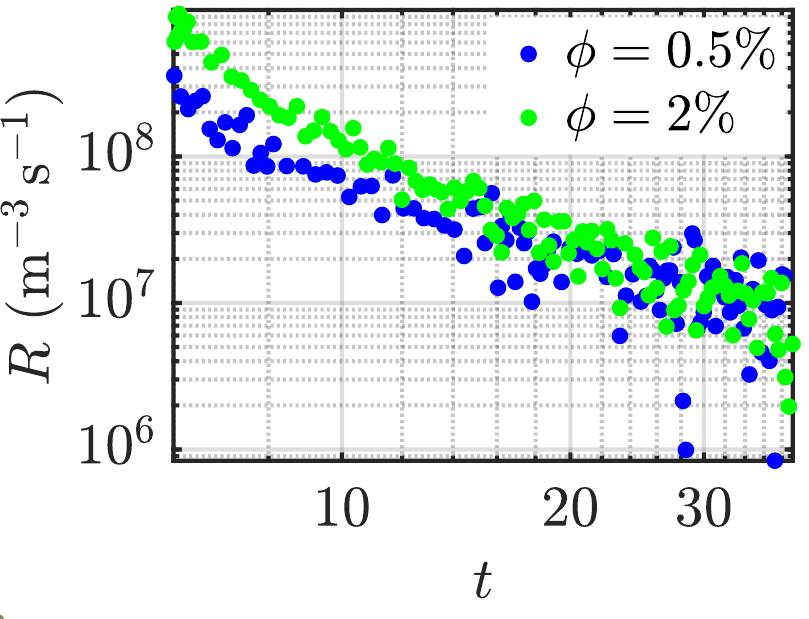}
        \caption{$R$, with slopes $-1.59$ (0.5\%) for $t>7$ and $-1.9$ (2\%) for $t>8$, and corresponding $\mathcal{R}^2$ values of $0.89$ and $0.93$.}
        \label{fig:a}
    \end{subfigure}
    \begin{subfigure}[b]{0.312\textwidth}
        \includegraphics[width=\textwidth]{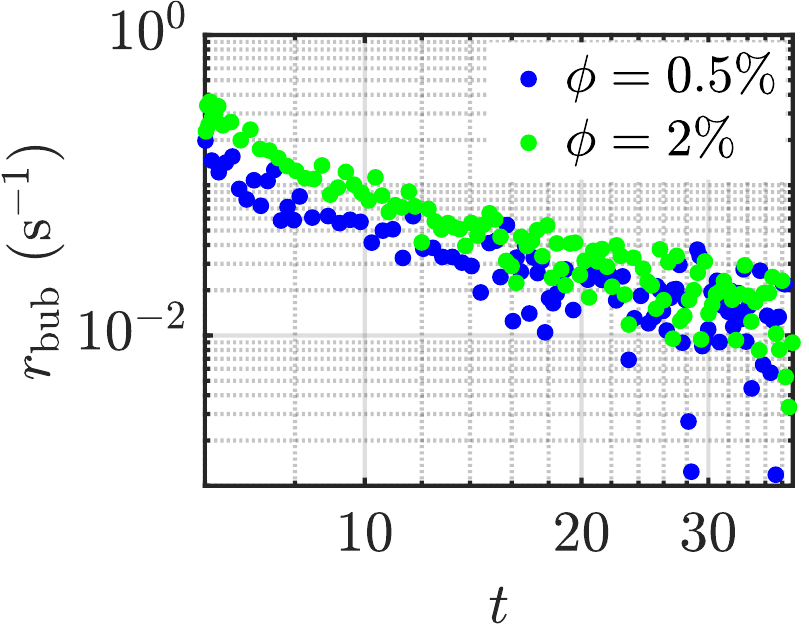}
        \caption{$r$, with slopes $-0.99$ (0.5\%) for $t>7$ and $-1.06$ (2\%) for $t>8$, and corresponding $\mathcal{R}^2$ values of $0.81$ and $0.86$.}
        \label{fig:b}
    \end{subfigure}
    \begin{subfigure}[b]{0.312\textwidth}
        \includegraphics[width=\textwidth]{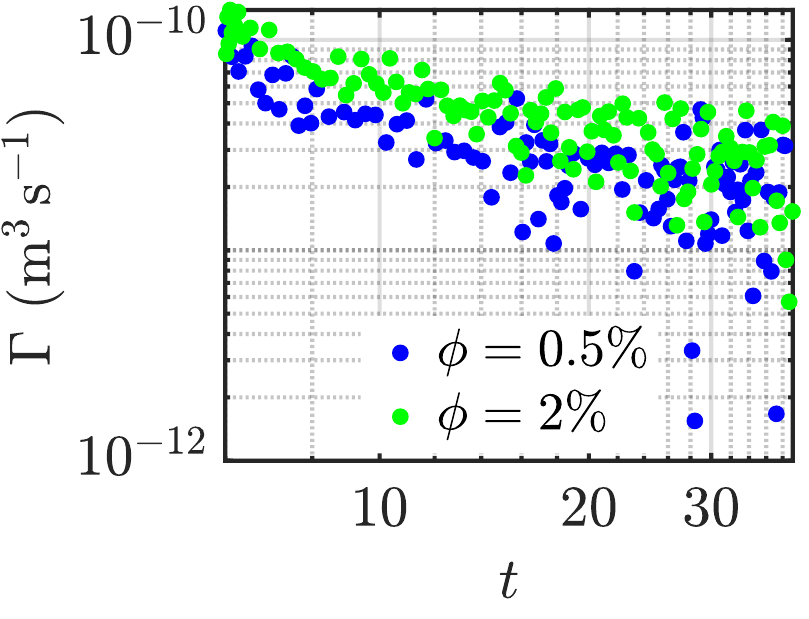}
        \caption{$\Gamma$, with slopes $-0.36$ (0.5\%) for $t>7$ and $-0.2$ (2\%) for $t>8$, and corresponding $\mathcal{R}^2$ values of $0.76$ and $0.79$.}
        \label{fig:c}
    \end{subfigure}    
    \caption{Temporal evolution of the volumetric coalescence rate $R(t)$, coalescence hazard $r(t)$, and coalescence kernel $\Gamma(t)$ in HIT DNS. The coalescence-rate data become increasingly scattered at later times because turbulence decay lengthens the coalescence timescale, making events progressively rarer. Thus, although individual coalescence events are tracked accurately, the fixed sampling interval contains too few events to yield statistically converged rate estimates.}
    \label{fig:coalescence}
\end{figure}

Figure~\ref{fig:coalescence} shows the temporal evolution of the volumetric coalescence rate $R(t)$, the coalescence hazard $r(t)$, and the coalescence kernel $\Gamma(t)$. Figure~\ref{fig:coalescence}(a) shows that the volumetric coalescence rate \(R(t)\) decreases monotonically after the short mixed adjustment and follows a power-law decay with slopes of \(-1.59\) and \(-1.90\) for \(\langle \phi \rangle=0.5\%\) and \(2\%\), respectively. In the pure-coalescence regime, the scaling in Table~\ref{tab:scalings_exponents_m_coal} gives
\begin{equation}
    R(t)\sim \Gamma n^2
    \sim \varepsilon^{1/3}d^{7/3}(d^{-3})^2
    \sim \varepsilon^{1/3}d^{-11/3}
    \sim t^{(3m-11)/2}.
    \label{eq:R_scaling_dns}
\end{equation}
Using the DNS dissipation exponents, gives predicted slopes of approximately \(-1.7\) and \(-2.1\), respectively, consistent with the measured DNS trends (small deviation is due to error in bubble tracking). The comparison is significant because \(R(t)\) combines two distinct effects: the pairwise encounter kernel and the number of available bubble pairs. Although bubble growth increases the geometric contribution to the encounter kernel, this effect is outweighed by the decay of turbulent relative velocity and, more importantly, by the rapid depletion of bubble pairs as \(n(t)\) decreases. Thus, the strong decay of \(R(t)\) reflects the loss of available encounters in a coarsening population, rather than a collapse of the effective pairwise coalescence kernel. The scatter in \(R(t)\) is expected because event counts are evaluated over finite time intervals chosen to resolve the characteristic coalescence time; nevertheless, ensemble averaging reveals a clear power-law trend.

Figure~\ref{fig:coalescence}(b) shows that the coalescence hazard \(r(t)=R(t)/n(t)\) follows an approximately universal decay, with fitted slopes of \(-0.99\) and \(-1.06\) for \(\langle \phi \rangle=0.5\%\) and \(2\%\), respectively. These values are close to the pure-coalescence prediction
\begin{equation}
    r(t)\sim t^{-1},
    \label{eq:r_scaling_dns}
\end{equation}
which is independent of the precise dissipation decay exponent \(m\). This collapse makes \(r(t)\) a particularly clean DNS test of the reduced theory: normalizing \(R(t)\) by \(n(t)\) removes the leading dependence on the number of bubbles remaining in the domain and isolates the removal probability per bubble. Although the total number of coalescence events depends on void fraction and on the available number of pairs, the per-bubble coalescence clock becomes nearly self-similar once the population has entered the sub-Hinze, coalescence-dominated regime. Thus, the near-\(t^{-1}\) decay reflects a normalized coalescence dynamics rather than the absolute event rate.

Figure~\ref{fig:coalescence}(c) shows that the effective pairwise coalescence kernel, \(\Gamma(t)=R(t)/n^2(t)\), varies much more weakly than the total coalescence rate \(R(t)\). The fitted slopes are \(-0.36\) and approximately \(-0.2\) for \(\langle \phi \rangle=0.5\%\) and \(2\%\), respectively. In the pure-coalescence regime, the theoretical scaling gives
\begin{equation}
    \Gamma(t)
    \sim
    \varepsilon^{1/3}d^{7/3}
    \sim
    t^{(7-3m)/2}.
    \label{eq:Gamma_scaling_dns}
\end{equation}
Using the DNS dissipation exponents, gives predicted exponents of approximately \(-0.2\) and \(-0.1\), respectively, consistent with the weak temporal variation observed in the DNS. This weak dependence reflects a compensation between two competing effects: turbulence decay reduces the inertial relative velocity between bubbles, whereas coalescence-driven bubble growth increases the geometric contribution to the encounter kernel. This interpretation is consistent with the standard kernel decomposition \(\Gamma=h\lambda\)~\citep{coulaloglou1977description,prince1990bubble,chesters1991modelling,lehr2002bubble,liao2010review}. The closure implication is that \(R(t)\) and \(\Gamma(t)\) need not have the same temporal dependence. In the DNS, the rapid decay of \(R(t)\) is caused primarily by depletion of available bubble pairs, whereas the effective pairwise kernel remains comparatively slow varying.

The rate of bubble coalescence is a key quantity governing the evolution of bubbly flows and has been examined previously using DNS. In a Lagrangian framework, the coalescence rate can be obtained directly by tracking individual bubbles in space and time. In our experiments, however, the measurements are inherently Eulerian, and tracking several thousand bubbles simultaneously over extended spatial and temporal scales is not feasible. Instead, we exploit the fact that the flow downstream of the inlet rapidly enters a purely coalescence-dominated regime, in which bubble breakup is negligible and the total number of bubbles decreases monotonically. In this regime, the volumetric coalescence rate $R$ can be inferred from the temporal decrease in the bubble number density. Specifically, $R$ is defined as the rate of change of the bubble count between two successive axial measurement locations, normalised by the corresponding convection time $\Delta t$ between those locations. This rate-based analysis is new: the same duct-flow database was previously used to characterize turbulence decay, BSD/PDF evolution and characteristic diameters, whereas here it is reprocessed to infer closure-level coalescence diagnostics. The first station is excluded not merely as a fitting choice, but because the mixed breakup--coalescence regime can still be active. The inference of $R(t)$ from the decrease of $n(t)$ is valid only after the super-Hinze probability mass has become negligible and the population is predominantly sub-Hinze.

\begin{figure}[htbp]
\centering

\begin{minipage}[t]{0.48\textwidth}
    \vspace{0pt}
    \centering
    \includegraphics[width=\textwidth]{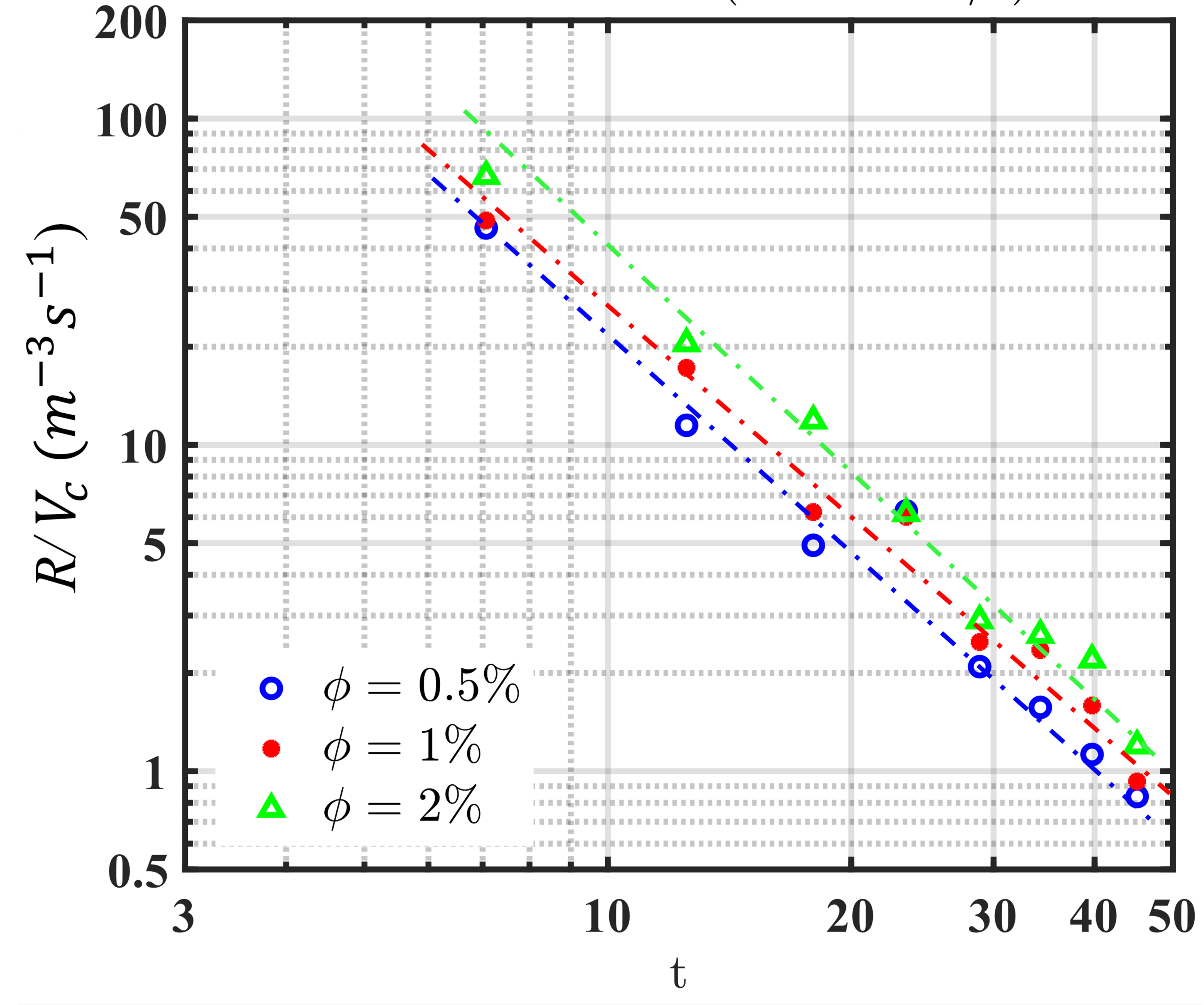}
    \subcaption{$V = 6.1\,\mathrm{m\,s^{-1}}$}
    \label{fig:coal-rate-v61}
\end{minipage}
\hfill
\begin{minipage}[t]{0.48\textwidth}
    \vspace{0pt}
    \centering
    \includegraphics[width=\textwidth]{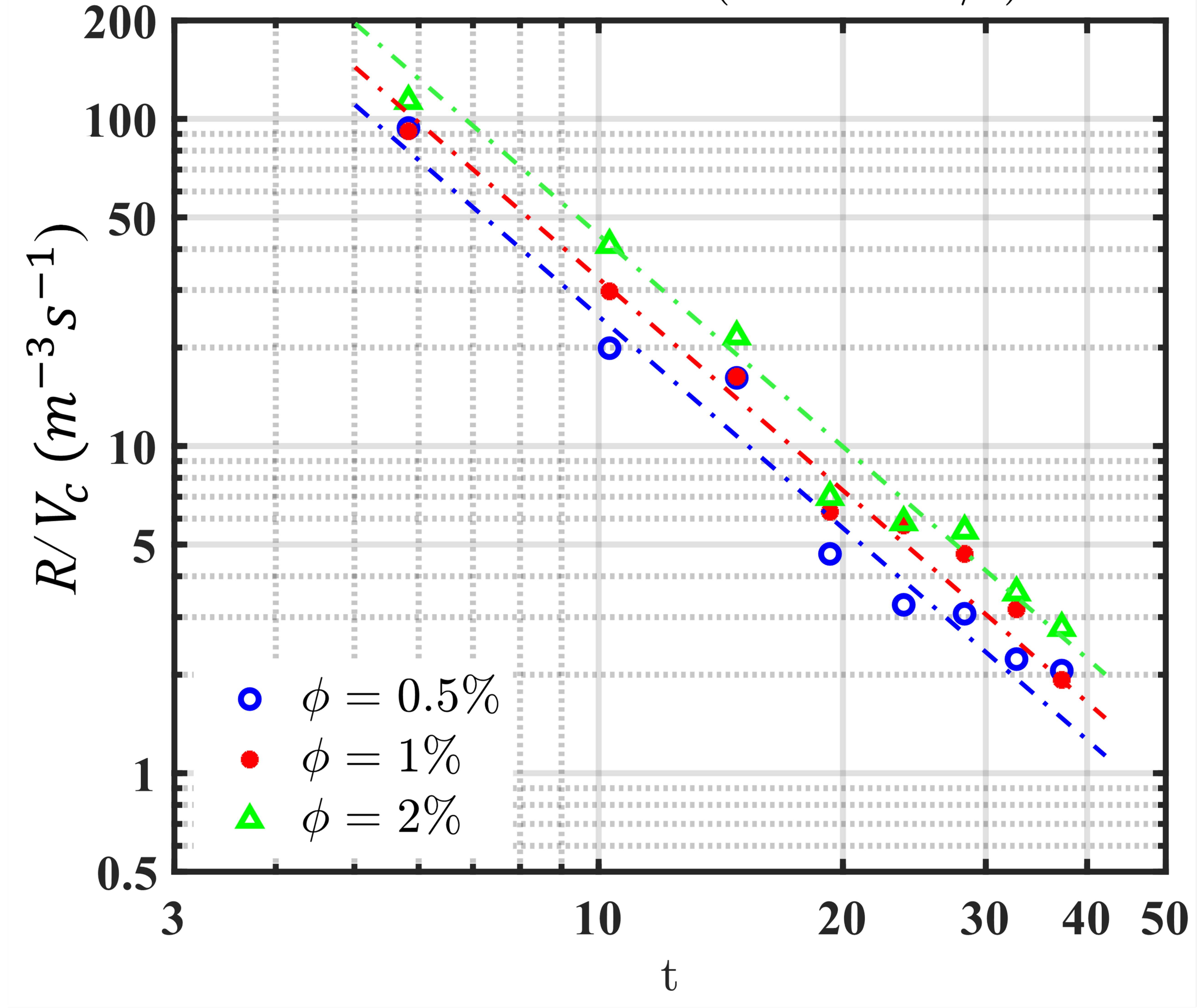}
    \subcaption{$V = 7.4\,\mathrm{m\,s^{-1}}$}
    \label{fig:coal-rate-v74}
\end{minipage}

\vspace{1em}

\begin{minipage}[t]{0.48\textwidth}
    \vspace{0pt}
    \centering
    \includegraphics[width=\textwidth]{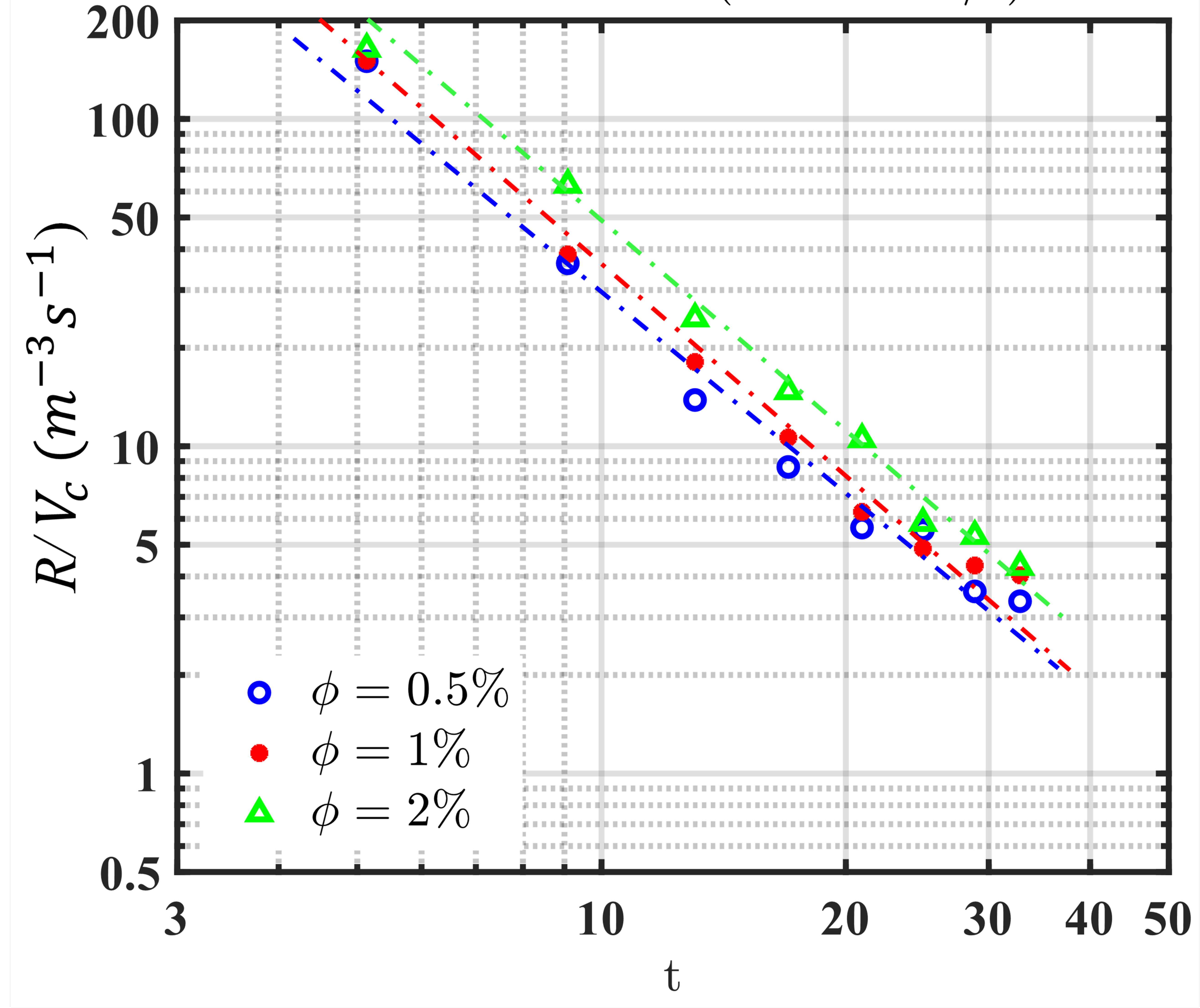}
    \subcaption{$V = 8.4\,\mathrm{m\,s^{-1}}$}
    \label{fig:coal-rate-v84}
\end{minipage}
\hfill
\begin{minipage}[t]{0.48\textwidth}
    \vspace{0pt}
    \centering

    \subcaption{Fitting parameters for $R(t)$, excluding the first data point.}
    \label{fig:coal-rate-fit-table}
    \vspace{0.5em}

    \renewcommand{\arraystretch}{1.25}
    \begin{tabular}{c c c c}
        \hline
        & $V_1$ & $V_2$ & $V_3$ \\ 
        $\phi$ (\%) & $\theta_R$ ($\mathcal{R}^2$) & $\theta_R$ ($\mathcal{R}^2$) & $\theta_R$ ($\mathcal{R}^2$) \\
        \hline
        $\phi_1$ & -2.37 (0.98) & -2.35 (0.96) & -2.18 (0.99) \\
        $\phi_2$ & -2.35 (0.97) & -2.35 (0.98) & -2.33 (0.97) \\ 
        $\phi_3$ & -2.48 (0.97) & -2.36 (0.96) & -2.34 (0.97) \\
        \hline
    \end{tabular}
\end{minipage}

\caption{Temporal evolution of the volumetric coalescence rate $R(t)$ for three bulk velocities and three void fractions. Panels (a--c) show $R(t)$ for $V=6.1$, $7.4$, and $8.4\,\mathrm{m\,s^{-1}}$, respectively, each shown for $\langle \phi \rangle=0.5\%$, $1\%$, and $2\%$. Panel (d) summarizes the fitted power-law exponents $\theta_R$ and coefficients of determination $\mathcal{R}^2$, obtained from log--log fits after excluding the first data point. }
\label{fig:R_exp}
\end{figure}


In interpreting the experimental trends it is important to distinguish temporal scaling from parametric offsets. For a fixed case, $R(t)$, $r(t)$ and $\Gamma(t)$ evolve downstream as turbulence decays. Changing $V$ or $\langle \phi \rangle$ mainly shifts their magnitudes at comparable downstream positions because it alters the upstream-conditioned population, gas loading and turbulent encounter environment. The fitted exponents remain similar because once the flow is sub-Hinze, the same decaying-turbulence coalescence balance controls the downstream evolution.

\begin{figure}[htbp]
\centering

\begin{minipage}[t]{0.48\textwidth}
    \vspace{0pt}
    \centering
    \includegraphics[width=\textwidth]{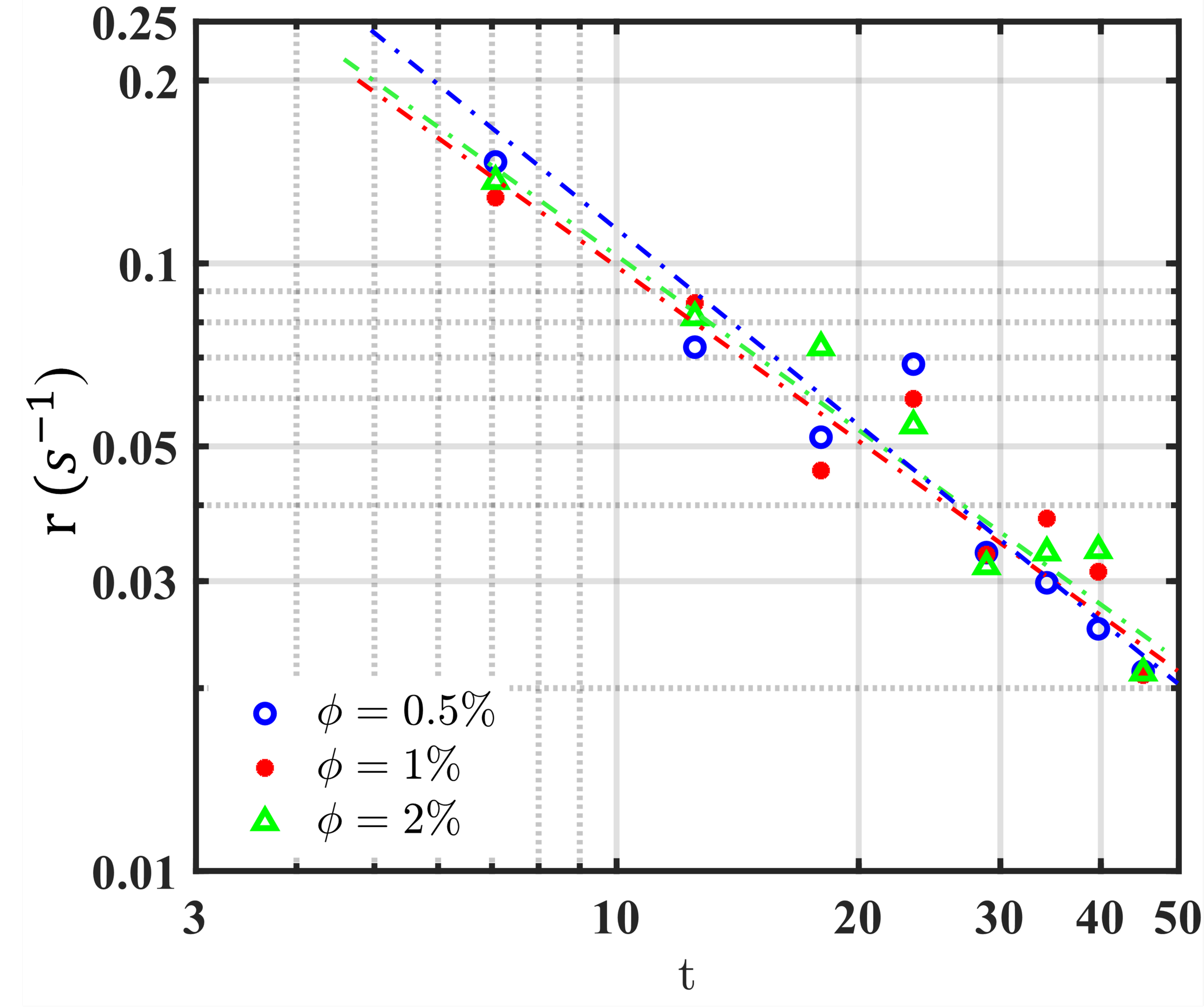}
    \subcaption{$V = 6.1\,\mathrm{m\,s^{-1}}$}
    \label{fig:hazard-v61}
\end{minipage}
\hfill
\begin{minipage}[t]{0.48\textwidth}
    \vspace{0pt}
    \centering
    \includegraphics[width=\textwidth]{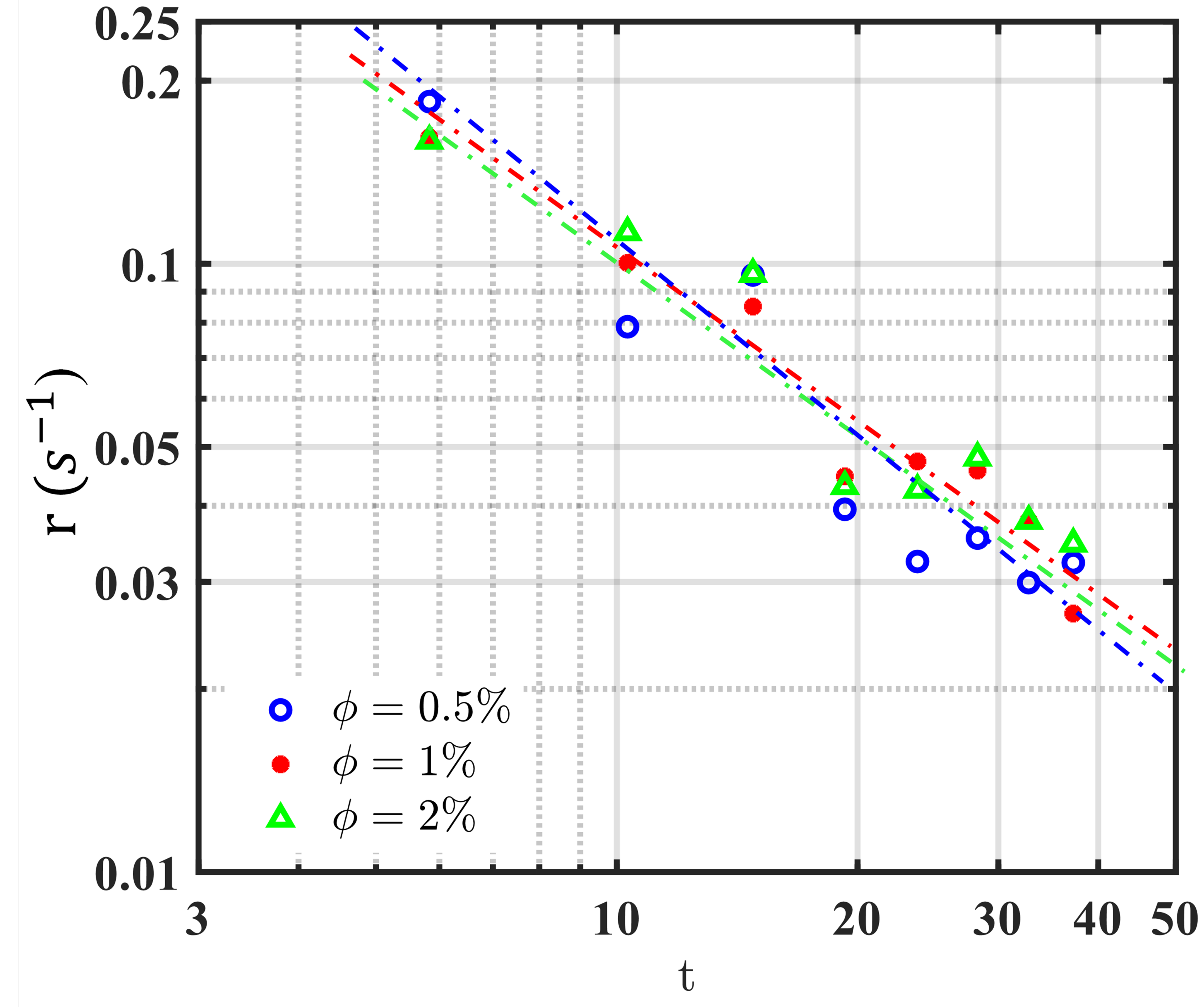}
    \subcaption{$V = 7.4\,\mathrm{m\,s^{-1}}$}
    \label{fig:hazard-v74}
\end{minipage}

\vspace{1em}

\begin{minipage}[t]{0.48\textwidth}
    \vspace{0pt}
    \centering
    \includegraphics[width=\textwidth]{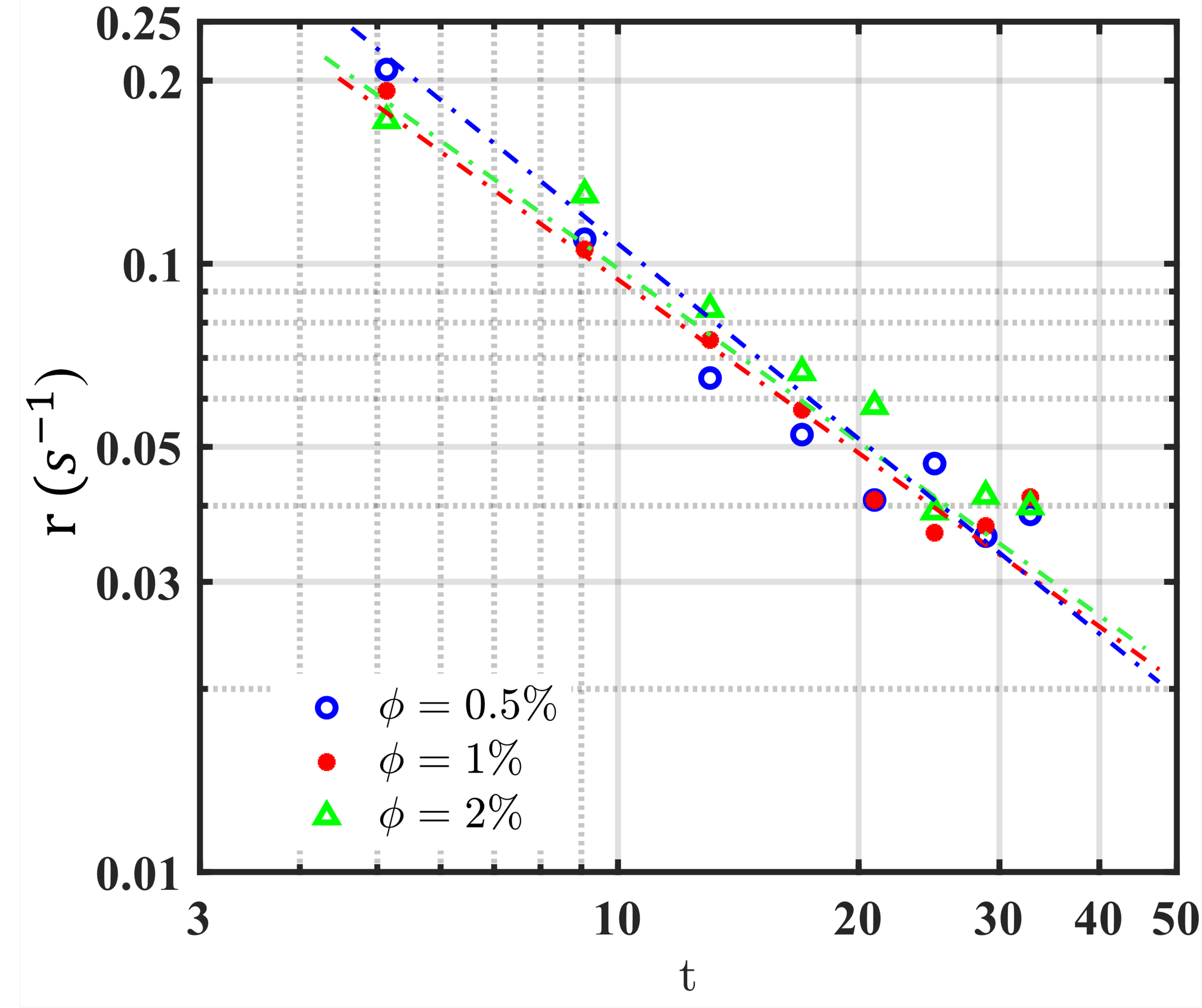}
    \subcaption{$V = 8.4\,\mathrm{m\,s^{-1}}$}
    \label{fig:hazard-v84}
\end{minipage}
\hfill
\begin{minipage}[t]{0.48\textwidth}
    \vspace{0pt}
    \centering

    \subcaption{Fitting parameters for $r(t)$, excluding the first data point.}
    \label{fig:hazard-fit-table}
    \vspace{0.5em}

    \renewcommand{\arraystretch}{1.25}
    \begin{tabular}{c c c c}
        \hline
        & $V_1$ & $V_2$ & $V_3$ \\ 
        $\phi$ (\%) & $\theta_r$ ($\mathcal{R}^2$) & $\theta_r$ ($\mathcal{R}^2$) & $\theta_r$ ($\mathcal{R}^2$) \\
        \hline
        $\phi_1$ & -1.04 (0.93) & -0.98 (0.90) & -1.03 (0.92) \\
        $\phi_2$ & -0.93 (0.92) & -0.99 (0.91) & -0.98 (0.93) \\ 
        $\phi_3$ & -1.02 (0.92) & -0.94 (0.92) & -0.93 (0.91) \\
        \hline
    \end{tabular}
\end{minipage}

\caption{Temporal evolution of the coalescence hazard $r(t)=R(t)/n(t)$ for three bulk velocities and three void fractions. Panels (a--c) show the power-law decay of $r(t)$ for $V=6.1$, $7.4$, and $8.4\,\mathrm{m\,s^{-1}}$, respectively, each shown for $\langle \phi \rangle=0.5\%$, $1\%$, and $2\%$. Panel (d) summarizes the fitted power-law exponents $\theta_r$ and coefficients of determination $\mathcal{R}^2$, obtained from log--log fits after excluding the first data point.}
\label{fig:r_exp}
\end{figure}

Figure~\ref{fig:R_exp} shows that the inferred experimental $R(t)$ decays as a power law for all bulk velocities and void fractions. The fitted exponents lie between $-2.18$ and $-2.48$, with $\mathcal{R}^2>0.95$. For the experimental dissipation-decay range, $m\simeq1.85$--$2.10$, the pure-coalescence prediction $R(t)\sim t^{(3m-11)/2}$ gives exponents of approximately $-2.73$ to $-2.35$, consistent with the measurements. At fixed $V$, increasing $\phi$ raises the magnitude of $R$ because a larger gas loading produces more bubbles and therefore more possible pairs. In the effective monodisperse balance, $R\sim \Gamma n^2$, so the absolute number of coalescence events can increase through the quadratic dependence on $n$ even if the pairwise kernel changes weakly. At fixed $\langle \phi \rangle$, increasing $V$ raises the near-inlet turbulence intensity, lowers the initial Hinze scale and can generate more small bubbles upstream or near the inlet before the pure-coalescence region is reached. The resulting upstream-conditioned population enters the sub-Hinze region with a larger available pair count and stronger encounter environment. The fitted exponents remain within a narrow range because the downstream decay of $R$ is governed primarily by the common coalescence-controlled balance, not by the amplitude shift.

\begin{figure}[htbp]
\centering

\begin{minipage}[t]{0.48\textwidth}
    \vspace{0pt}
    \centering
    \includegraphics[width=\textwidth]{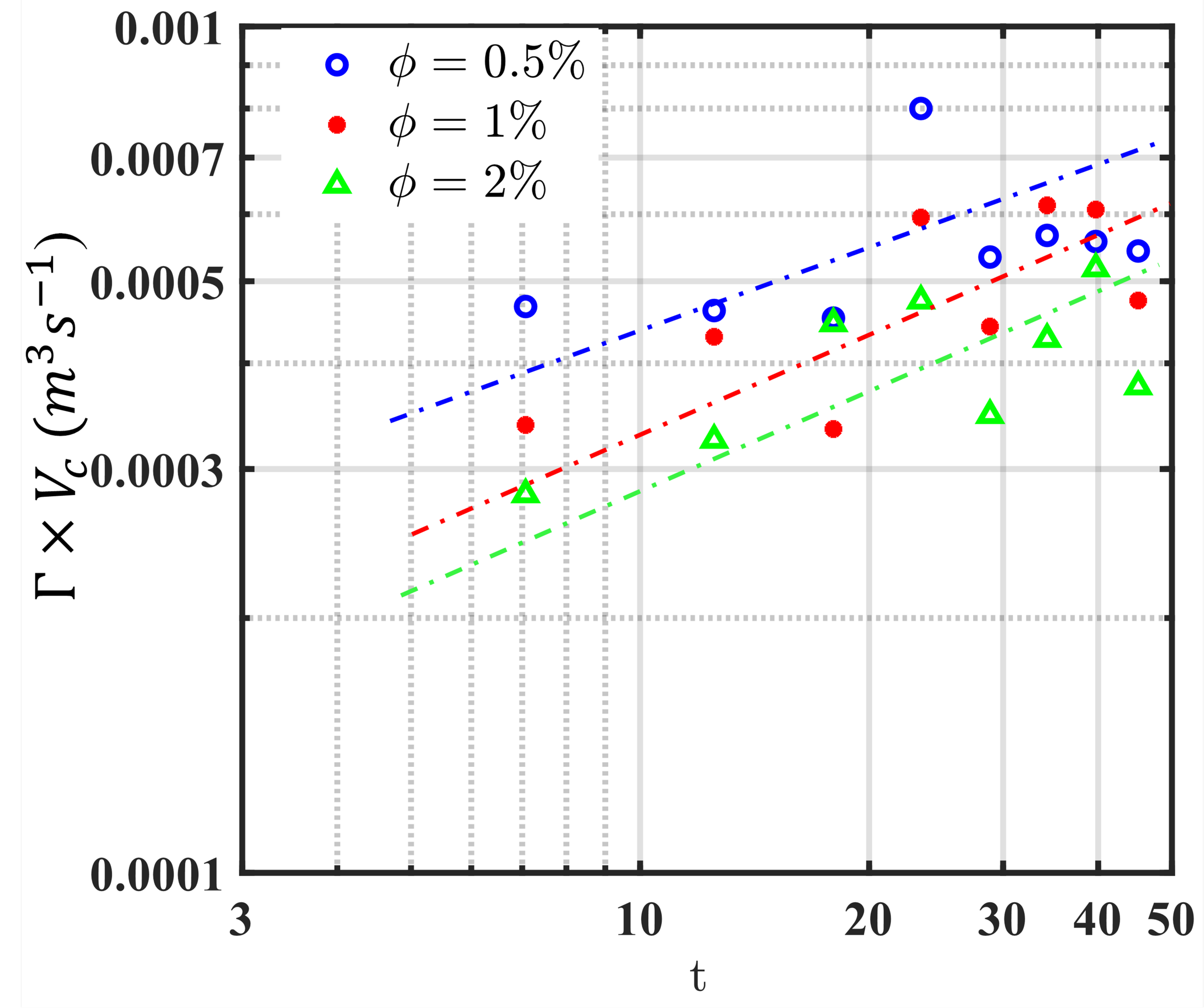}
    \subcaption{$V = 6.1\,\mathrm{m\,s^{-1}}$}
    \label{fig:gamma-v61}
\end{minipage}
\hfill
\begin{minipage}[t]{0.48\textwidth}
    \vspace{0pt}
    \centering
    \includegraphics[width=\textwidth]{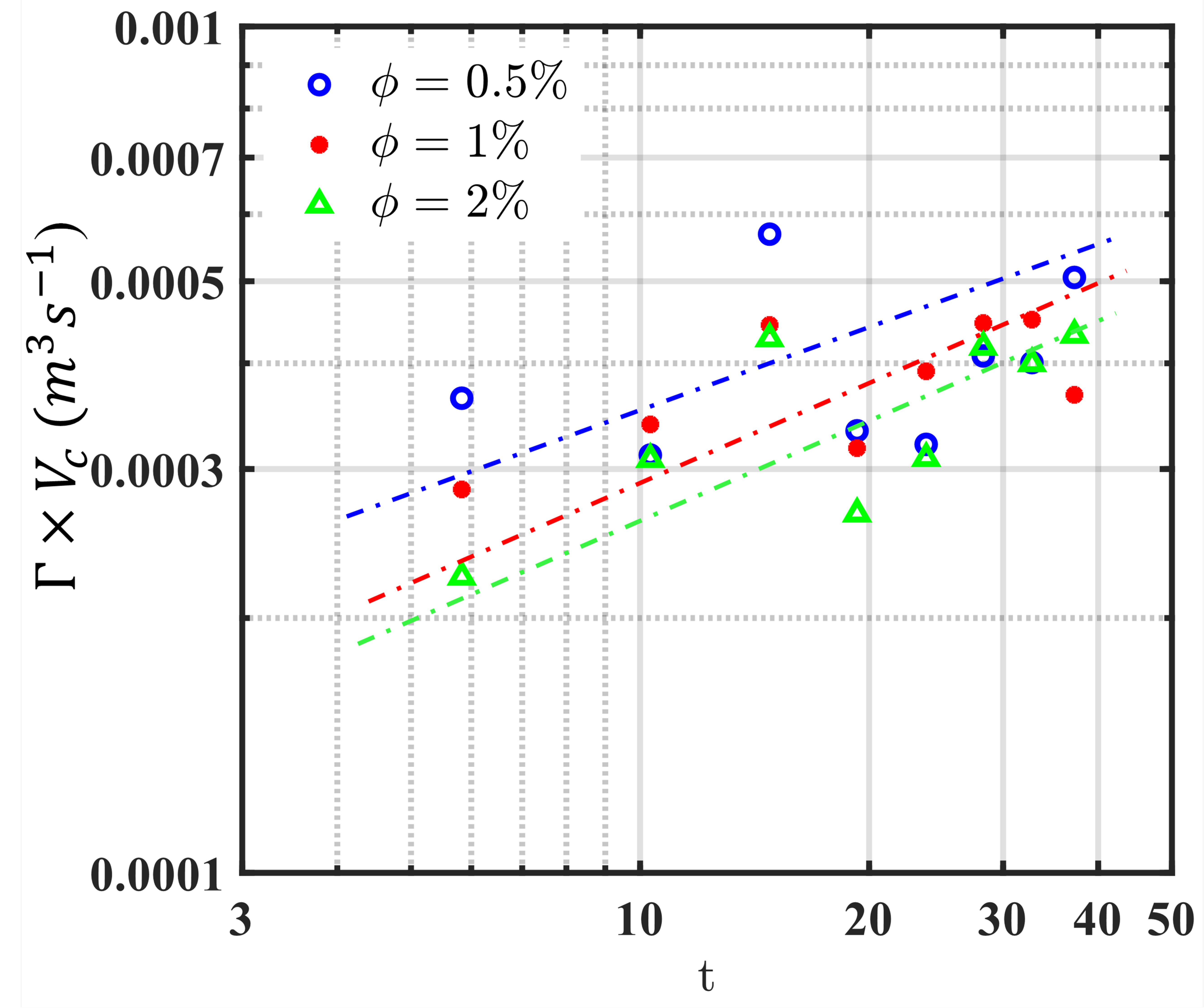}
    \subcaption{$V = 7.4\,\mathrm{m\,s^{-1}}$}
    \label{fig:gamma-v74}
\end{minipage}

\vspace{1em}

\begin{minipage}[t]{0.48\textwidth}
    \vspace{0pt}
    \centering
    \includegraphics[width=\textwidth]{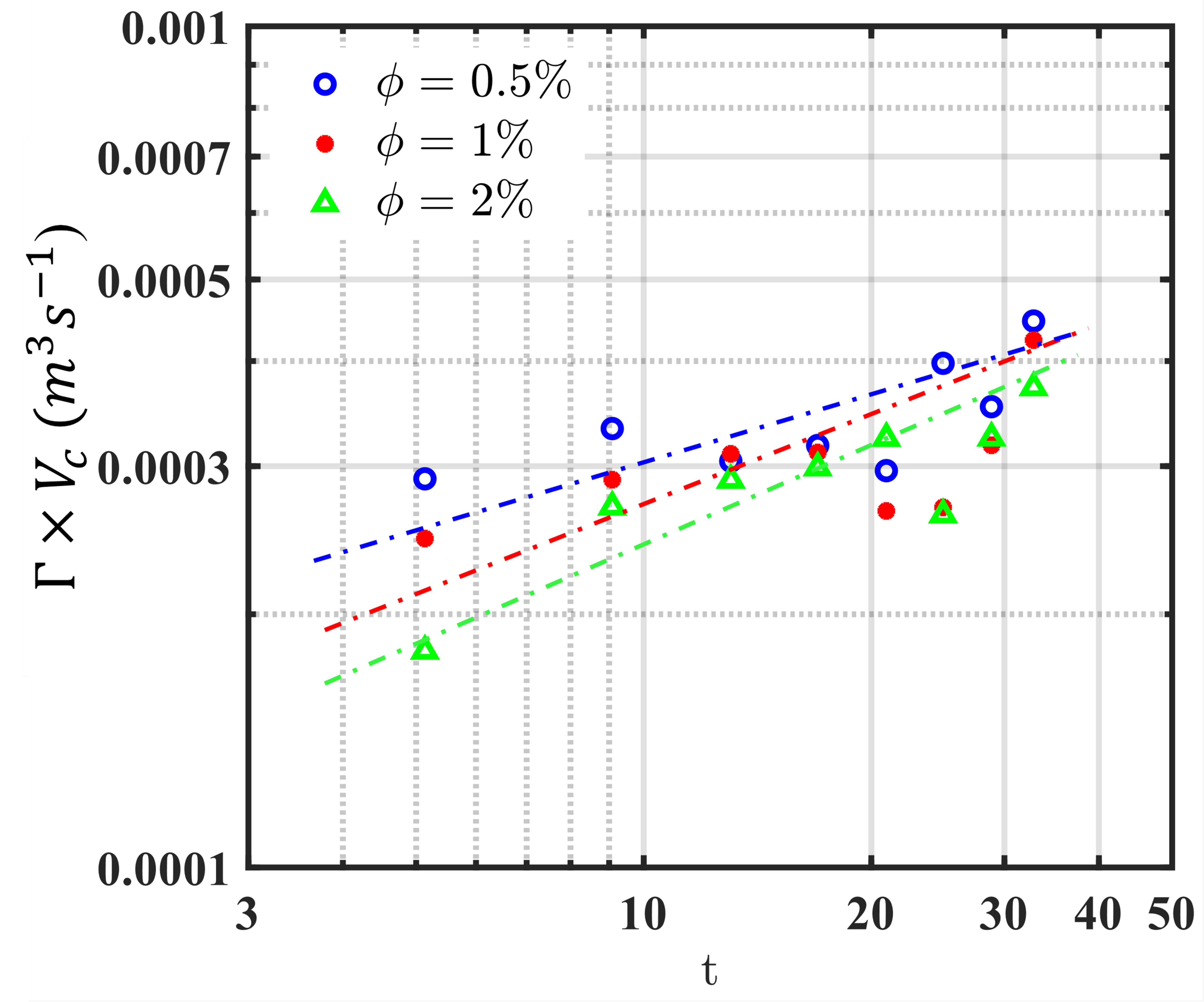}
    \subcaption{$V = 8.4\,\mathrm{m\,s^{-1}}$}
    \label{fig:gamma-v84}
\end{minipage}
\hfill
\begin{minipage}[t]{0.43\textwidth}
    \vspace{0pt}
    \centering

    \subcaption{Fitting parameters for $\Gamma(t)$, excluding the first data point.}
    \label{fig:gamma-fit-table}
    \vspace{0.5em}

    \renewcommand{\arraystretch}{1.25}
    \begin{tabular}{c c c c}
        \hline
        & $V_1$ & $V_2$ & $V_3$ \\ 
        $\phi$ (\%) & $\theta_\Gamma$ ($\mathcal{R}^2$) & $\theta_\Gamma$ ($\mathcal{R}^2$) & $\theta_\Gamma$ ($\mathcal{R}^2$) \\
        \hline
        0.5 & 0.32 (0.79) & 0.33 (0.71) & 0.27 (0.70) \\
        1   & 0.38 (0.83) & 0.38 (0.75) & 0.35 (0.77) \\ 
        2   & 0.38 (0.73) & 0.38 (0.79) & 0.38 (0.79) \\
        \hline
    \end{tabular}
\end{minipage}

\caption{Temporal evolution of the effective pairwise coalescence kernel, $\Gamma(t)=R(t)/n^2(t)$, for three bulk velocities and three void fractions. Panels (a--c) show $\Gamma(t)$ for $V=6.1$, $7.4$, and $8.4\,\mathrm{m\,s^{-1}}$, respectively, each shown for $\langle \phi \rangle=0.5\%$, $1\%$, and $2\%$. Panel (d) summarizes the fitted power-law exponents $\theta_\Gamma$ and coefficients of determination $\mathcal{R}^2$, obtained from log--log fits after excluding the first data point.}
\label{fig:gamma_exp}
\end{figure}

The hazard $r(t)$ provides the strongest practical validation because it removes much of the population-size effect retained by $R(t)$. Since $r(t)=\frac{R(t)}{n(t)}=\Gamma(t)n(t)$, changes in $n$ and $\Gamma$ can partially compensate. Figure~\ref{fig:r_exp} shows that the experimental exponents for $r(t)$ cluster near $-1$, ranging from about $-0.93$ to $-1.04$, with $\mathcal{R}^2\simeq0.90$--$0.93$. Small offsets in the magnitude of $r$ with $V$ or $\phi$ may remain because operating conditions alter both the local neighbour density and the effective pairwise kernel. Nevertheless, the temporal exponent is nearly unchanged. The per-bubble coalescence clock therefore follows the same law as the DNS and the pure-coalescence theory, making $r$ a robust diagnostic of the sub-Hinze regime.

The experimentally obtained coalescence kernel $\Gamma$ in Figure~\ref{fig:gamma_exp} increases downstream, with fitted exponents approximately between $0.27$ and $0.38$. The coefficients of determination are lower than those for $R$ and $r$, as expected for a quantity inferred from $R/n^2$ and therefore sensitive to uncertainty in $n$, finite sampling, polydispersity and centreline measurement-volume effects. The DNS establishes the pairwise-kernel scaling directly from lineages; the experiment provides supportive evidence that the effective pairwise kernel can increase downstream as turbulence decays and bubble size grows. Here $\Gamma$ should be interpreted as an effective pairwise kernel, not as a pure coalescence efficiency. Thus variations in $\Gamma$ reflect the net outcome of collision frequency, contact time, film drainage, deformation, rebound and successful merger, and cannot by themselves identify which component changed.

This distinction also clarifies the dependence on operating conditions. The temporal derivative of $\Gamma$ along a fixed case and the parametric offset of $\Gamma$ between cases need not have the same sign. At comparable downstream locations, lower values of $\Gamma$ at larger $\langle \phi \rangle$, if observed, are consistent with crowding and many-body hydrodynamic interactions that can hinder liquid-film drainage or shorten effective contact times, even while the total rate $R$ is larger because $n^2$ is larger. At fixed $\langle \phi \rangle$, lower values of $\Gamma$ at larger $V$, if observed, are consistent with larger turbulent relative velocities, stronger deformation and a greater tendency for rebound or failed coalescence. Since $\Gamma=R/n^2$ does not separate $h$ from $\lambda$, these interpretations are mechanistic but not unique. They are not inconsistent with the downstream increase of $\Gamma(t)$ for each fixed case: along the duct, turbulence decays, violent short-contact encounters weaken and dynamically populated bubble sizes grow, so the effective pairwise kernel can increase even while $R(t)$ decreases through pair depletion.

These distinctions matter in flows where turbulence is generated intensely in one region and then decays while the dispersed phase continues to evolve. Examples include pump discharge flows, injectors and spargers, stirred-tank circulation loops, bubble columns or reactors with spatially varying turbulence and breaking-wave bubble clouds~\citep{kantarci2005bubble,kulkarni2007mass,garcia2009bioreactor,deane2002scale}. In such systems, closures based only on instantaneous local quantities can misinterpret a decrease in the total event rate as a decrease in pairwise coalescence strength, even when the dominant effect is pair depletion. The present decomposition separates these effects through $ R(t)\sim \Gamma(t)n^2(t)$. A high value of $R$ does not necessarily imply a strong pairwise kernel; it may simply reflect a large number of available pairs. Conversely, a lower or decreasing $\Gamma$ does not preclude a larger total coalescence rate if $n^2$ is sufficiently large. The total sink, normalized hazard and pairwise kernel therefore carry different physical information and should not be collapsed into a single empirical coalescence-rate correction. The agreement between the measured coalescence rates and the theoretical scalings can be attributed to weak bubble-induced agitation relative to the background turbulence. In the theory, we assumed that bubbles are dynamically passive at the bulk turbulence scale; that is, the carrier-phase turbulence sets the dominant bubble--bubble relative velocity. This condition is supported in both the HIT DNS and the duct-flow experiment. In the duct-flow measurements, the bubblance parameter, $b=\phi u_d^2 / \mathcal{U}^2$, is \(b\sim10^{-5}\)--\(10^{-3}\) ~\citep{Almeras2017,vandenberg2020bubbles,kumar2026bubble}. Thus, bubble-induced agitation is much weaker than the carrier-phase turbulent kinetic energy, so random bubble--bubble encounters are governed primarily by turbulent fluctuations of the carrier phase. Mean shear, buoyancy-driven overtaking, lift-induced migration, and wake capture may modify prefactors in the duct flow, but they do not set the leading-order temporal exponent of the coalescence-rate closures.

The closure-level implication is that non-stationary PBM and IATE source terms should include the drift of the bubble distribution relative to the moving Hinze scale. Breakup source terms should be weighted by the breakable population, represented by $\mathcal{P}_{d>d_H}(t)$ or an equivalent drift variable such as $d_c(t)/d_H(t)$, where $d_c$ is a representative diameter such as $d_{32}$. This weighting determines when the model should transition from mixed breakup--coalescence to the pure-coalescence source terms measured here. Once $\mathcal{P}_{d>d_H}\ll1$, the model should recover
\begin{equation}
    R(t)\sim t^{(3m-11)/2},\qquad
    r(t)\sim t^{-1},\qquad
    \Gamma(t)\sim t^{(7-3m)/2}.
\end{equation}
These relations provide asymptotic constraints for non-stationary PBM/IATE source terms in decaying turbulence. The issue is not merely the calibration of empirical constants: stationary kernels evaluate rates at a fixed local state, whereas in the present flow, $\varepsilon(t)$, $d_H(t)$, $n(t)$ and $d(t)$ move the system through regime space. The measured and inferred $R(t)$, $r(t)$ and $\Gamma(t)$ therefore convert Hinze-scale drift from a geometric observation into a source-term constraint for non-stationary bubbly-flow closures.\\

\section{Conclusion} \label{sec:conclusion}

This combined theoretical, numerical and experimental study examines bubble interactions in decaying turbulent bubbly flow. The central mechanism is governed by Hinze-scale ($d_H$) drift: if dissipation $\varepsilon(t)\sim t^{-m}$, then the breakup threshold moves as $d_H(t)\sim \varepsilon^{-2/5}\sim t^{2m/5}$. The bubble-size distribution therefore evolves relative to a moving threshold, not a fixed Hinze scale. Its trajectory depends on both the initial distribution relative to $d_H$ and the decay exponent $m$. For a conditioned distribution near the Hinze scale, as expected downstream of pumps, agitators, spargers or injectors, the later drift is controlled mainly by $m$. For $m>5/3$, $d_H$ grows faster than the coalescence-driven characteristic bubble size, so $d/d_H$ decreases, the super-Hinze probability mass is depleted and breakup is dynamically suppressed. For $m<5/3$, an initially super-Hinze fraction need not be swept into the sub-Hinze range. Thus the sign of the relative drift provides a non-stationary regime criterion absent from closures based on a locally quasi-stationary breakup threshold.

Our theoretical analysis clarifies the early mixed regime. It is not simply breakup-dominated, but coalescence-dominated and breakup-assisted. While the growing Hinze scale removes bubbles from the breakable class, residual breakup supplies smaller bubbles and increases the number of collision partners. Once the super-Hinze probability mass becomes negligible, the reduced balance predicts power-law growth of characteristic diameter and corresponding decay of bubble number density and interfacial area, with exponents set by $m$. The source-term quantities obey $R(t)\sim t^{(3m-11)/2}$, $r(t)\sim t^{-1}$ and $\Gamma(t)\sim t^{(7-3m)/2}$, where \(R\) is the total coalescence-removal rate, \(r=R/n\) is the per-bubble hazard, and \(\Gamma=R/n^2\) is the effective pairwise kernel. The inverse-time law for $r$ is a compact result: after breakup becomes negligible, the normalized coalescence clock slows self-similarly.

The DNS provides an event-resolved realization of this mechanism in decaying homogeneous isotropic turbulence. In this controlled setting, without walls, mean shear or spatial development, dilute bubbly HIT retains a decay close to canonical single-phase HIT, so $m$ remains a useful organizing exponent. The simulations show the full population reorganization: the PDF drifts toward smaller $d/d_H$, the super-Hinze tail is depleted, $d_{32}$ and $d_{99.8}$ grow, and both $A(t)$ and $n(t)$ decrease. The mixed regime also shows the predicted breakup-assisted coalescence, with residual breakup increasing the supply of collision partners and contributing to faster early growth. Event-resolved lineage tracking then exposes the source terms directly: $R(t)$ decays strongly, $r(t)$ follows the near-$t^{-1}$ law and $\Gamma(t)$ varies much more weakly than $R(t)$. The decreasing total event rate therefore mainly reflects depletion of available bubble pairs, not failure of pairwise coalescence.

The duct experiments demonstrate the application relevance of the same mechanism in a realistic high-Reynolds-number flow. The recent study of \citet{kumar2026bubble} established the pump-driven duct-flow database, including turbulence decay, BSD/PDF evolution, $d_{32}$, $d_{99.8}$, $d_H$ and void-fraction redistribution. The present work uses that database differently: it connects the duct observations to the scaling framework, interprets the PDF evolution through Hinze-scale drift and extracts population-level and source-term behaviour relevant to closures. The duct flow is spatially developing, confined, wall-bounded and inhomogeneous, yet the mechanism persists when the local dissipation decay exponent is used. Downstream of the pump, the distribution shifts toward smaller $d/d_H$, the super-Hinze population diminishes and coalescence-dominated evolution emerges. Changes in bulk velocity $V$ and void fraction $\phi$ mainly alter amplitudes and near-inlet conditioning, whereas downstream exponents remain organized by the sub-Hinze balance.

These results provide closure-level constraints for population-balance modelling. Classical PBM formulations represent coalescence through a kernel $\Gamma(d,d')=h\lambda$, where \(h\) is the collision frequency and \(\lambda\) is the coalescence efficiency \citep{coulaloglou1977description,prince1990bubble,chesters1991modelling,luo1996breakup,lehr2002bubble,liao2010review,ramkrishna2000population}. The present work supplies a non-stationary effective-kernel constraint through $\Gamma(t)$, while the total coalescence sink follows $R(t)\sim \Gamma(t)n^2(t)$. This separation matters because stationary fitted kernels can conflate pair depletion with pairwise coalescence strength. In a decaying flow, $R(t)$ may decrease rapidly even when $\Gamma(t)$ varies weakly or increases, so a closure based only on the total sink can misidentify the controlling mechanism.

The same argument constrains interfacial-area-transport modelling. IATE formulations require source and sink terms for interfacial area generation by breakup and loss by coalescence \citep{wu1998one,hibiki1999experimental,ishii2010thermo,Chen2021}. In decaying turbulence, the breakup source should be weighted by the breakable population, equivalently the probability mass above the moving Hinze scale, $\mathcal{P}_{d>d_H}(t)=\int_{d_H(t)}^\infty P(d,t)\,{\rm d}d$. Once $\mathcal{P}_{d>d_H}$ becomes negligible, breakup no longer supplies new bubbles or interfacial area at leading order, and the model should reduce to coalescence-controlled decay of number density and interfacial area. Hinze-scale drift therefore determines when breakup source terms become negligible and when coalescence-only source terms should control topology and available interface.

The results presented here apply most directly to dilute, decay-dominated bubbly turbulence and to flows in which turbulence is generated intensely in one region and then decays while the bubble population continues to evolve, including pump discharge flows, injectors, spargers, stirred-tank circulation loops, spatially varying bubble columns, aeration devices and breaking-wave bubble clouds. Natural extensions include mean shear, buoyancy-driven slip, surfactant-modified film drainage, stronger polydispersity and higher void fractions where bubble feedback modifies the turbulence. Within its range of validity, the measured and inferred $R(t)$, $r(t)$ and $\Gamma(t)$ convert Hinze-scale drift from a geometric observation into a source-term constraint for non-stationary turbulent bubbly-flow closures.

\section*{Acknowledgements}

The authors thank Jason Rom for assistance with the experiments. 

\section*{Funding}
The information, data, or work presented herein was funded in part by the Advanced Research Projects Agency--Energy, U.S. Department of Energy (DOE) under Award Number DE-AR0001587, and Office of Critical Minerals and Energy Innovation (Former EERE), US DOE under Award Number DE-EE0009396. The views and opinions of authors expressed herein do not necessarily state or reflect those of the United States Government or any agency thereof. \\

S.~S.~J acknowledges support from the donors of the ACS Petroleum Research Fund through Doctoral New Investigator Grant 69196-DNI9. The authors also acknowledge computing resources provided through the U.S. Department of Energy 2024 and 2025 ALCC awards (TUR147 and BubbleLaden). The Argonne Leadership Computing Facility at Argonne National Laboratory is supported by the Office of Science of the U.S. Department of Energy under Contract No. DE-AC02-06CH11357. The Oak Ridge Leadership Computing Facility at Oak Ridge National Laboratory is supported by the Office of Science of the U.S. Department of Energy under Contract No. DE-AC05-00OR22725.

\section*{Author contributions}
\noindent \textbf{Vivek Kumar}: Conceptualization, Methodology, Theoretical Modeling, Software, Investigation, Experimental Setup Design, Numerical and Experimental Data Curation, Formal Analysis, Validation, Visualization, and Writing – Original Draft; Writing – Review and Editing. \textbf{Prasoon Suchandra}: Data Curation and Analysis, Methodology, Writing – Review \& Editing. \textbf{Shivam Prajapati}: Investigation and  Writing – Review \& Editing. \textbf{Suhas S. Jain}: Conceptualization, Supervision, Methodology, Writing – Review \& Editing. \textbf{Cyrus Aidun}: Conceptualization, Methodology, Supervision, Funding Acquisition, Project Administration, Writing – Review \& Editing.

\section*{Declaration of Interests}
The authors report no conflict of interest.

\section*{Data Availability}
Data will be made available upon reasonable request to the corresponding author. The supplementary video is provided as \textit{S.mp4}.

\section*{Author ORCID}
V. Kumar, https://orcid.org/0000-0001-5481-725X; \\P. Suchandra, https://orcid.org/0000-0003-3710-4615; \\S. Prajapati, https://orcid.org/0000-0002-8284-8771; \\S. S. Jain, https://orcid.org/0000-0002-9306-7903; \\
C. K. Aidun, https://orcid.org/0000-0003-0517-6040

\section*{Appendix} 
\subsection{Simplifying coalescence efficiency ($\lambda$)}
The coalescence efficiency can be described by the widely known film–drainage (FD) model,
\begin{equation}
\lambda_\mathrm{ij}=\exp\!\left(-\frac{t_{\mathrm{drain}}}{t_{\mathrm{contact}}}\right),
\label{eq:lambda_fd}
\end{equation}
where $t_{\mathrm{drain}}$ is the time for the intervening liquid film to thin to rupture thickness and $t_{\mathrm{contact}}$ is the hydrodynamic contact time during a collision~\citep{lehr2002bubble}. In bubbly turbulence with deformable, mobile interfaces, FD formulations distinguish regimes by interfacial mobility; for gas bubbles in clean liquids, the relevant asymptote is the inertia-controlled drainage limit~\citep{chesters1991modelling,chan2011film,liao2010review}.

\paragraph{Drainage time (deformable, inertia–controlled).}
Using~\citet{chesters1991modelling} parallel–film model in the inertia limit, the drainage time reduces to
\begin{equation}
t_{\mathrm{drain}} \;=\; \frac{1}{2}\,\frac{\rho_c\,u_\mathrm{\!rel}\,\mathcal{R}^2}{\gamma},
\qquad r=\tfrac{d}{2},
\label{eq:tdrain_inertia}
\end{equation}
with continuous–phase density $\rho_c$, surface tension $\gamma$, bubble radius $r$, and approach speed $u_\mathrm{\!rel}$~\citep{liao2010review}.\footnote{For unequal sizes see the~\citep{luo1995coalescence} generalization; here we restrict to monodisperse.}

\paragraph{Contact time and approach velocity.}
A standard bubble-buble contact time in turbulent flow is
\begin{equation}
t_{\mathrm{contact}} \;\sim\; \frac{d^{2/3}}{\varepsilon^{1/3}},
\label{eq:tcontact_turb}
\end{equation}
and an inertial–range estimate for the approach velocity is
\begin{equation}
u_\mathrm{\!rel} \;\approx\; \sqrt{2}\,\big[\varepsilon\,(2d)\big]^{1/3}.
\label{eq:urel_turb}
\end{equation}

With monodisperse simplification ($d_i=d_j=d$), and combining Eq. \eqref{eq:tdrain_inertia}–Eq. \eqref{eq:urel_turb} gives
\begin{align}
\frac{t_{\mathrm{drain}}}{t_{\mathrm{contact}}}
&= \frac{\rho_c}{\sigma}\,2^{-13/6}\,\varepsilon^{2/3}\,d^{5/3},
\\[3pt]
\Rightarrow\quad
\lambda(d)
&=\exp\!\Big[-A\,\tfrac{\rho_c}{\sigma}\,\varepsilon^{2/3}\,d^{5/3}\Big],
\qquad A=2^{-13/6}\approx 0.223.
\label{eq:lambda_mono_inertia}
\end{align}

\paragraph{Numerical constants for water (20$^\circ$C).}
Using $\rho_c=998\,\mathrm{kg\,m^{-3}}$ and $\gamma=0.072\,\mathrm{N\,m^{-1}}$ (air–water), Eq. \eqref{eq:lambda_mono_inertia} is fully specified.

\paragraph{Examples (equal sizes).}
\begin{itemize}
\item \textbf{Case A (72 L/min and 0.5\%) at $\mathcal{L}=0$:} $\varepsilon=200~\mathrm{m^2\,s^{-3}}$, $d=200~\mu$m.  
From Eq. \eqref{eq:urel_turb} $u_{\!rel}\!\approx\!0.609~\mathrm{m\,s^{-1}}$; using Eq. \eqref{eq:tdrain_inertia}–Eq. \eqref{eq:tcontact_turb},
$t_{\mathrm{drain}}\!\approx\!4.22\times10^{-5}$ s, $t_{\mathrm{contact}}\!\approx\!5.85\times10^{-4}$ s, so
\[
\lambda \;=\;\exp\!\left(-\frac{t_{\mathrm{drain}}}{t_{\mathrm{contact}}}\right)\approx \exp(-0.072)\approx \mathbf{0.93}.
\]

\item \textbf{Case B (72 L/min and 0.5\%) at $\mathcal{L}=40$:} $\varepsilon=10~\mathrm{m^2\,s^{-3}}$, $d=500~\mu$m.  
$u_{\!rel}\!\approx\!0.305~\mathrm{m\,s^{-1}}$, $t_{\mathrm{drain}}\!\approx\!1.32\times10^{-4}$ s, $t_{\mathrm{contact}}\!\approx\!2.92\times10^{-3}$ s, hence
\[
\lambda \;\approx\; \exp(-0.045)\approx \mathbf{0.95}.
\]
\end{itemize}

Hence, it can be concluded that $\lambda$ remains high and constant for our scaling regime. \vk{The high coalescence efficiency results from the film-drainage time, $t_{\mathrm{drain}}$, being significantly shorter than the contact time, $t_{\mathrm{contact}}$, associated with the eddy turnover time.}

\begin{equation}
  \lambda \sim \text{constant}, \quad \Rightarrow \quad K_o \sim  \text{constant}
  \tag{2}
\end{equation}

\bibliographystyle{jfm}
\bibliography{jfm}

@article{kumar2024hydrostatic,
title = {On the hydrostatic limit for thin film flow with applications to thermosyphons},
journal = {Applied Thermal Engineering},
volume = {245},
pages = {122869},
year = {2024},
issn = {1359-4311},
doi = {https://doi.org/10.1016/j.applthermaleng.2024.122869},
url = {https://www.sciencedirect.com/science/article/pii/S1359431124005374},
author = {Vivek Kumar and Muhammad Rizwanur Rahman and M.R. Flynn}
}

@article{kumar2023particle,
  title={Particle separation using modified Taylor’s flow},
  author={Kumar, Vivek and Jain, Palak and Upadhyay, Ravi Kant and Bharath, KS and Waghmare, Prashant R},
  journal={Microfluidics and Nanofluidics},
  volume={27},
  number={10},
  pages={66},
  year={2023},
  publisher={Springer}
}

@article{javadi2026large,
  title={Large eddy simulations of side channel pump in different operating conditions},
  author={Javadi, Ardalan and Kumar, Vivek and Aidun, Cyrus K},
  journal={Engineering Applications of Computational Fluid Mechanics},
  volume={20},
  number={1},
  pages={2587723},
  year={2026},
  publisher={Taylor \& Francis}
}

@article{kumar2026viscosity,
  title={Viscosity and dynamic surface tension measurement: A guideline for appropriate measurement.},
  author={Kumar, Vivek and Quintero, Jsm and Baldygin, Aleksey and Molina, Paul and Willers, Thomas and Waghmare, Prashant R},
  journal={Journal of Colloid and Interface Science},
  pages={139929},
  year={2026},
  publisher={Elsevier}
}

@article{jain2025stationary,
  title={Stationary states of forced two-phase turbulence},
  author={Jain, Suhas S and Elnahhas, Ahmed},
  journal={arXiv preprint arXiv:2501.02417},
  year={2025}
}

@article{kantarci2005bubble,
  title={Bubble column reactors},
  author={Kantarci, Nigar and Borak, Fahir and Ulgen, Kutlu O},
  journal={Process biochemistry},
  volume={40},
  number={7},
  pages={2263--2283},
  year={2005},
  publisher={Elsevier}
}

@article{kulkarni2007mass,
  title={Mass transfer in bubble column reactors: effect of bubble size distribution},
  author={Kulkarni, Amol A},
  journal={Industrial \& engineering chemistry research},
  volume={46},
  number={7},
  pages={2205--2211},
  year={2007},
  publisher={ACS Publications}
}

@article{garcia2009bioreactor,
  title={Bioreactor scale-up and oxygen transfer rate in microbial processes: an overview},
  author={Garcia-Ochoa, Felix and Gomez, Emilio},
  journal={Biotechnology advances},
  volume={27},
  number={2},
  pages={153--176},
  year={2009},
  publisher={Elsevier}
}

@book{batchelor1953theory,
  title={The theory of homogeneous turbulence},
  author={Batchelor, George Keith},
  year={1953},
  publisher={Cambridge university press}
}

@article{saffman1956collision,
  title={On the collision of drops in turbulent clouds},
  author={Saffman, PGF and Turner, JS},
  journal={Journal of Fluid Mechanics},
  volume={1},
  number={1},
  pages={16--30},
  year={1956},
  publisher={Cambridge University Press}
}

@misc{moninm,
  title={Statistical Fluid Mechanics},
  author={Monin, AS and Yaglom, A},
  year={1975},
  publisher={MIT Press}
}

@article{luo1995coalescence,
  title={Coalescence, breakup and liquid circulation in bubble column reactors.},
  author={Luo, Hean},
  year={1995}
}

@article{saffman1967large,
  title   = {The large-scale structure of homogeneous turbulence},
  author  = {Saffman, P. G.},
  journal = {Journal of Fluid Mechanics},
  volume  = {27},
  number  = {3},
  pages   = {581--593},
  year    = {1967},
  doi     = {10.1017/S0022112067000552}
}

@article{loitsyanskii1939some,
  title   = {Some basic laws for isotropic turbulent flow},
  author  = {Loitsyanskii, L. G.},
  journal = {Trudy Tsentral'nogo Aero-Gidrodinamicheskogo Instituta},
  volume  = {440},
  pages   = {3--23},
  year    = {1939}
}

@article{batchelor1956large,
  title   = {The large-scale structure of homogeneous turbulence},
  author  = {Batchelor, G. K. and Proudman, I.},
  journal = {Philosophical Transactions of the Royal Society of London. Series A, Mathematical and Physical Sciences},
  volume  = {248},
  number  = {949},
  pages   = {369--405},
  year    = {1956},
  doi     = {10.1098/rsta.1956.0002}
}

@article{chesters1991modelling,
  title={Modelling of coalescence processes in fluid-liquid dispersions: a review of current understanding},
  author={Chesters, A\_K},
  journal={Chemical engineering research and design},
  volume={69},
  number={A4},
  pages={259--270},
  year={1991}
}

@article{chan2011film,
  title={Film drainage and coalescence between deformable drops and bubbles},
  author={Chan, Derek YC and Klaseboer, Evert and Manica, Rogerio},
  journal={Soft Matter},
  volume={7},
  number={6},
  pages={2235--2264},
  year={2011},
  publisher={Royal Society of Chemistry}
}

@article{Kolmogorov1949,
  author = {Kolmogorov, A. N.},
  title = {On the breakage of droplets in a turbulent flow},
  journal = {Dokl. Akad. Nauk SSSR},
  year = {1949},
  volume = {66},
  pages = {825-828}
}

@article{Hinze1955,
  author = {J. O. Hinze},
  title = {Fundamentals of the hydrodynamic mechanism of splitting in dispersion processes},
  journal = {AIChE Journal},
  volume = {1},
  number = {3},
  pages = {289--295},
  year = {1955},
  doi = {10.1002/aic.690010303}
}

@article{MartinezBazan1999,
  author = {Martínez-Bazán, C. and Montañés, J. L. and Lasheras, J. C.},
  title = {Breakup of an Air Bubble Injected into a Fully Developed Turbulent Flow. Part 1: Breakup Frequency},
  journal = {Journal of Fluid Mechanics},
  year = {1999},
  volume = {401},
  pages = {157-182}
}

@article{Deane2002,
  author = {Deane, G. B. and Stokes, M. D.},
  title = {Scale dependence of bubble creation mechanisms in breaking waves},
  journal = {Nature},
  year = {2002},
  volume = {418},
  pages = {839-844}
}

@article{martinez1999breakup,
  title={On the breakup of an air bubble injected into a fully developed turbulent flow. Part 1. Breakup frequency},
  author={Mart{\'\i}nez-Baz{\'a}n, Carlos and Monta{\~n}es, Jose L and Lasheras, Juan C},
  journal={Journal of Fluid Mechanics},
  volume={401},
  pages={157--182},
  year={1999},
  publisher={Cambridge University Press}
}

@book{clift1978bubbles,
  title={Bubbles, drops, and particles},
  author={Clift, Roland and Grace, John R and Weber, Martin E},
  year={1978},
  publisher={Academic Press}
}

@article{delnoij1997computational,
  title={Computational fluid dynamics applied to gas--liquid contactors},
  author={Delnoij, E and Kuipers, JAM and van Swaaij, WPM and Akker, HEA},
  journal={Chemical Engineering Science},
  volume={52},
  number={21-22},
  pages={3623--3638},
  year={1997},
  publisher={Elsevier}
}

@article{magnaudet2000motion,
  title={The motion of high-Reynolds-number bubbles in inhomogeneous flows},
  author={Magnaudet, Jacques and Eames, Ian},
  journal={Annual Review of Fluid Mechanics},
  volume={32},
  number={1},
  pages={659--708},
  year={2000},
  publisher={Annual Reviews 4139 El Camino Way, PO Box 10139, Palo Alto, CA 94303-0139, USA}
}

@article{MoinMahesh1998,
  author  = {Moin, Parviz and Mahesh, Krishnan},
  title   = {Direct numerical simulation: A tool in turbulence research},
  journal = {Annual Review of Fluid Mechanics},
  volume  = {30},
  pages   = {539--578},
  year    = {1998}
}

@book{SagautCambon2008,
  title     = {Homogeneous Turbulence Dynamics},
  author    = {Sagaut, Pierre and Cambon, Claude},
  year      = {2008},
  publisher = {Springer}
}

@article{LaizetLamballais2009,
  author  = {Laizet, Sylvain and Lamballais, Eric},
  title   = {High-order compact schemes for incompressible flows},
  journal = {Journal of Computational Physics},
  volume  = {228},
  pages   = {5989--6015},
  year    = {2009}
}

@article{Moureau2011,
  author  = {Moureau, Vincent and Domingo, Pedro and Vervisch, Luc},
  title   = {From large-eddy simulation to direct numerical simulation of turbulent reacting flows},
  journal = {Combustion and Flame},
  volume  = {158},
  pages   = {1340--1357},
  year    = {2011}
}

@article{salibindla2020lift,
  author  = {Salibindla, Ashwanth K. R. and Masuk, Ashik Ullah Mohammad and Tan, Shiyong and Ni, Rui},
  title   = {Lift and drag coefficients of deformable bubbles in intense turbulence determined from bubble rise velocity},
  journal = {Journal of Fluid Mechanics},
  volume  = {894},
  pages   = {A20},
  year    = {2020},
  doi     = {10.1017/jfm.2020.238}
}

@book{pope2000turbulent,
  author    = {Pope, Stephen B.},
  title     = {Turbulent Flows},
  publisher = {Cambridge University Press},
  address   = {Cambridge},
  year      = {2000},
  isbn      = {9780521598866}
}

@article{coulaloglou1977description,
  title={Description of interaction processes in agitated liquid-liquid dispersions},
  author={Coulaloglou, CA and Tavlarides, Lawrence L},
  journal={Chemical Engineering Science},
  volume={32},
  number={11},
  pages={1289--1297},
  year={1977},
  publisher={Elsevier}
}

@article{prince1990bubble,
  title={Bubble coalescence and break-up in air-sparged bubble columns},
  author={Prince, Michael J. and Blanch, Harvey W.},
  journal={AIChE Journal},
  volume={36},
  number={10},
  pages={1485--1499},
  year={1990},
  publisher={Wiley},
  doi={10.1002/aic.690361003}
}

@article{wu1998one,
  title={One-group interfacial area transport in vertical bubbly flow},
  author={Wu, Q and Kim, S and Ishii, M and Beus, SG},
  journal={International Journal of Heat and Mass Transfer},
  volume={41},
  number={8-9},
  pages={1103--1112},
  year={1998},
  publisher={Elsevier}
}

@article{Tan_Zhong_Qi_Xu_Ni_2025, 
title={Dynamics of bubble collision and coalescence in three-dimensional turbulent flows}, 
volume={1020}, 
DOI={10.1017/jfm.2025.10603}, 
journal={Journal of Fluid Mechanics}, 
author={Tan, Shiyong and Zhong, Shijie and Qi, Yinghe and Xu, Xu and Ni, Rui}, 
year={2025}, 
pages={A15}}

@article{zeng2021investigation,
  title={Investigation on bubble diameter distribution in upward flow by the two-fluid and multi-fluid models},
  author={Zeng, Yongzhong and Xu, Weilin},
  journal={Energies},
  volume={14},
  number={18},
  pages={5776},
  year={2021},
  publisher={MDPI}
}

@article{jain:2025,
    title={Stationary states of forced two-phase turbulence}, 
    author={S. S. Jain and A. Elnahhas},
    journal = {Chemical Engineering Journal},
  volume = {524},
  pages = {169077},
  year = {2025},
  publisher = {Elsevier}}

@article{clift2005bubbles,
  title={Bubbles, drops, and particles},
  author={Clift, Roland and Grace, John R and Weber, Martin E},
  year={2005},
  publisher={Courier Corporation}
}

@book{ishii2010thermo,
  title={Thermo-fluid dynamics of two-phase flow},
  author={Ishii, Mamoru and Hibiki, Takashi},
  year={2010},
  publisher={Springer Science \& Business Media}
}

@article{zhang2018investigation,
  title={An investigation of the flow characteristics of multistage multiphase pumps},
  author={Zhang, Jinya and Li, Yongjiang and Vafai, K and Zhang, Yongxue},
  journal={International Journal of Numerical Methods for Heat \& Fluid Flow},
  volume={28},
  number={3},
  pages={763--784},
  year={2018},
  publisher={Emerald Publishing Limited}
}

@article{ni2024deformation,
  title={Deformation and breakup of bubbles and drops in turbulence},
  author={Ni, Rui},
  journal={Annual Review of Fluid Mechanics},
  volume={56},
  number={1},
  pages={319--347},
  year={2024},
  publisher={Annual Reviews}
}

@article{estevadeordal2005,
  title={PIV with LED: Particle Shadow Velocimetry (PSV)},
  author={Estevadeordal, Jordi and Goss, Larry},
  journal={43rd AIAA Aerospace Sciences Meeting and Exhibit},
  year={2005}
}

@article{khodaparast2013,
  title={A micro particle shadow velocimetry ($\rm{\mu}$PSV) technique to measure flows in microchannels},
  author={Khodaparast, Sepideh and Borhani, Navid and Tagliabue, Giulia and Thome, John Richard},
  journal={Experiments in Fluids},
  volume={54},
  number={1474},
  year={2013}
}

@article{hessenkemper2018,
  title={Particle Shadow Velocimetry (PSV) in bubbly flows},
  author={Hessenkemper, H. and Ziegenhein, T.},
  journal={International Journal of Multiphase Flow},
  volume={106},
  pages={268--279},
  year={2018}
}

@article{vela2022memoryless,
  title={Memoryless drop breakup in turbulence},
  author={Vela-Mart{\'\i}n, Alberto and Avila, Marc},
  journal={Science Advances},
  volume={8},
  number={50},
  pages={eabp9561},
  year={2022},
  publisher={American Association for the Advancement of Science}
}

@article{calado2024dynamics,
  title={Dynamics of bubble deformation and breakup in decaying isotropic turbulence},
  author={Calado, Andre and Balaras, Elias},
  journal={Physical Review Fluids},
  volume={9},
  number={12},
  pages={123604},
  year={2024},
  publisher={APS}
}

@article{lehr2002bubble,
  title={Bubble-size distributions and flow fields in bubble columns},
  author={Lehr, F and Millies, M and Mewes, D},
  journal={AIChE journal},
  volume={48},
  number={11},
  pages={2426--2443},
  year={2002},
  publisher={Wiley Online Library}
}

@article{serizawa1975turbulence,
  title   = {Turbulence structure of air-water bubbly flow--II. Local properties},
  author  = {Serizawa, Akimi and Kataoka, Isao and Michiyoshi, Itaru},
  journal = {International Journal of Multiphase Flow},
  volume  = {2},
  number  = {3},
  pages   = {235--246},
  year    = {1975},
  doi     = {10.1016/0301-9322(75)90012-9}
}

@article{Lance1991,
  title={Turbulence in the liquid phase of a uniform bubbly air--water flow},
  author={Lance, M. and Bataille, J.},
  journal={J. Fluid Mech.},
  volume={222},
  pages={95--118},
  year={1991}
}

@article{Almeras2017,
  title={Experimental investigation of the turbulence induced by a bubble swarm rising within incident turbulence},
  author={Alm{\'e}ras, E. and Mathai, V. and Lohse, D. and Sun, C.},
  journal={J. Fluid Mech.},
  volume={825},
  pages={1091--1112},
  year={2017}
}

@article{Risso2018,
  author    = {Fr\'ed\'eric Risso},
  title     = {Agitation, Mixing, and Transfers Induced by Bubbles},
  journal   = {Annual Review of Fluid Mechanics},
  year      = {2018},
  volume    = {50},
  pages     = {25--48},
  doi       = {10.1146/annurev-fluid-122316-045606}
}

@article{hesketh1987bubble,
  title={Bubble breakage in pipeline flow},
  author={Hesketh, RP and Etchells, AW and Russell, TB},
  journal={Chemical Engineering Science},
  volume={42},
  number={3},
  pages={649--653},
  year={1987},
  publisher={Elsevier},
  doi={10.1016/0009-2509(87)85016-6}
}

@article{lance1991turbulence,
  title={Turbulence in the liquid phase of a uniform bubbly air--water flow},
  author={Lance, Michel and Bataille, Jean},
  journal={Journal of Fluid Mechanics},
  volume={222},
  pages={95--118},
  year={1991},
  publisher={Cambridge University Press},
  doi={10.1017/S0022112091001019}
}

@article{chan2018bubble,
  title={Formation and dynamics of bubbles in breaking waves: Part II. The evolution of the bubble size distribution and breakup/coalescence statistics},
  author={Chan, WHR and Dodd, MS and Johnson, PL and Urzay, J and Moin, P},
  journal={Annu. Res. Briefs},
  volume={2018},
  pages={21--34},
  year={2018}
}

@article{deane2002scale,
  title={Scale dependence of bubble creation mechanisms in breaking waves},
  author={Deane, Grant B. and Stokes, M. Dale},
  journal={Nature},
  volume={418},
  pages={839--844},
  year={2002},
  doi={10.1038/nature00967}
}

@article{balachandar2010turbulent,
  title={Turbulent dispersed multiphase flow},
  author={Balachandar, S and Eaton, John K},
  journal={Annual review of fluid mechanics},
  volume={42},
  number={1},
  pages={111--133},
  year={2010},
  publisher={Annual Reviews}
}

@article{garrett2000connection,
  title={Connection between bubble size spectra and energy dissipation rates in the upper ocean},
  author={Garrett, Chris and Li, Ming},
  journal={Journal of Physical Oceanography},
  volume={30},
  number={9},
  pages={2163--2171},
  year={2000},
  doi={10.1175/1520-0485(2000)030<2163:CBSSAE>2.0.CO;2}
}

@article{hinze1955fundamentals,
  title   = {Fundamentals of the hydrodynamic mechanism of splitting in dispersion processes},
  author  = {Hinze, J. O.},
  journal = {AIChE Journal},
  volume  = {1},
  number  = {3},
  pages   = {289--295},
  year    = {1955},
  doi     = {10.1002/aic.690010303}
}

@article{jassal2025particle,
  title={Particle shadow velocimetry and its potential applications, limitations and advantages vis-{\`a}-vis particle image velocimetry},
  author={Jassal, Gauresh Raj and Song, Maxwell and Schmidt, Bryan E},
  journal={Experiments in Fluids},
  volume={66},
  number={1},
  pages={21},
  year={2025},
  publisher={Springer}
}

@article{soligo2019breakage,
  title={Breakage, coalescence and size distribution of surfactant-laden droplets in turbulent flow},
  author={Soligo, Giovanni and Roccon, Alessio and Soldati, Alfredo},
  journal={Journal of Fluid Mechanics},
  volume={881},
  pages={244--282},
  year={2019},
  publisher={Cambridge University Press}
}

@book{lumley1972,
   author = {Tennekes, H. and Lumley, J. L.},
   title = {A first course in turbulence},
   publisher = {The MIT Press},
   year = {1972}
}

@article{vandenberg2020bubbles,
  title={Bubbles increase turbulent energy dissipation in dilute bubbly flows},
  author={van den Berg, Thomas H. and Rensen, Jeroen M. and Luther, Stefan and Lohse, Detlef},
  journal={Journal of Fluid Mechanics},
  volume={859},
  pages={R5},
  year={2020},
  publisher={Cambridge University Press},
  doi={10.1017/jfm.2018.840}
}

@article{jain2020,
  title={A conservative diffuse-interface method for compressible two-phase flows},
  author={Jain, Suhas S and Mani, Ali and Moin, Parviz},
  journal={Journal of Computational Physics},
  volume={418},
  pages={109606},
  year={2020},
  publisher={Elsevier}
}

@article{jain2022a,
title = {Accurate conservative phase-field method for simulation of two-phase flows},
journal = {Journal of Computational Physics},
volume = {469},
pages = {111529},
year = {2022},
issn = {0021-9991},
author = {Suhas S. Jain},
}

@article{Li2024,
  author = {Shiwang Li and Yixiang Liao},
  title = {CFD investigation of bubble breakup and coalescence in a rectangular pool-scrubbing column},
  journal = {Nuclear Engineering and Design},
  volume = {425},
  year = {2024},
  pages = {113342},
  doi = {10.1016/j.nucengdes.2024.113342}
}

@article{Chen2021,
  author = {Huiting Chen and Shiyu Wei and Weitian Ding and Han Wei and Liang Li and Henrik Saxén and Hongming Long and Yaowei Yu},
  title = {Interfacial Area Transport Equation for Bubble Coalescence and Breakup: Developments and Comparisons},
  journal = {Entropy},
  volume = {23},
  year = {2021},
  pages = {1106},
  doi = {10.3390/e23091106}
}

@article{Yu2024,
  author = {Yu Li and Zhongqiu Liu and Guodong Xu and Baokuan Li},
  title = {Effect of breakup and coalescence kernels on polydispersed bubbly flow in continuous casting mold},
  journal = {International Journal of Multiphase Flow},
  volume = {177},
  year = {2024},
  pages = {104872},
  doi = {10.1016/j.ijmultiphaseflow.2024.104872}
}

@article{Nguyen2013,
  author = {Van Thai Nguyen and Chul-Hwa Song and Byoung-Uhn Bae and Dong-Jin Euh},
  title = {Modeling of bubble coalescence and break-up considering turbulent suppression phenomena in bubbly two-phase flow},
  journal = {International Journal of Multiphase Flow},
  volume = {54},
  year = {2013},
  pages = {31-42},
  doi = {10.1016/j.ijmultiphaseflow.2013.03.001}
}

@article{Sajjadi2013,
  author = {Baharak Sajjadi and Abdul Aziz Abdul Raman and Raja Shazrin Shah Raja Ehsan Shah and Shaliza Ibrahim},
  title = {Review on applicable breakup/coalescence models in turbulent liquid-liquid flows},
  journal = {Review of Chemical Engineering},
  volume = {29},
  year = {2013},
  pages = {131-158},
  doi = {10.1515/revce-2012-0014}
}

@article{ruth2022experimental,
  title={Experimental observations and modelling of sub-Hinze bubble production by turbulent bubble break-up},
  author={Ruth, Daniel J and Aiyer, Aditya K and Rivi{\`e}re, Ali{\'e}nor and Perrard, St{\'e}phane and Deike, Luc},
  journal={Journal of Fluid Mechanics},
  volume={951},
  pages={A32},
  year={2022},
  publisher={Cambridge University Press}
}

@article{mazzitelli2003relevance,
  title={On the relevance of the lift force in bubbly turbulence},
  author={Mazzitelli, Irene M and Lohse, Detlef and Toschi, Federico},
  journal={Journal of Fluid Mechanics},
  volume={488},
  pages={283--313},
  year={2003},
  publisher={Cambridge University Press}
}

@article{farsoiya2023role,
  title={Role of viscosity in turbulent drop break-up},
  author={Farsoiya, Palas Kumar and Liu, Zehua and Daiss, Andreas and Fox, Rodney O and Deike, Luc},
  journal={Journal of Fluid Mechanics},
  volume={972},
  pages={A11},
  year={2023},
  publisher={Cambridge University Press}
}

@article{crialesi2023interaction,
  title={The interaction of droplet dynamics and turbulence cascade},
  author={Crialesi-Esposito, Marco and Chibbaro, Sergio and Brandt, Luca},
  journal={Communications Physics},
  volume={6},
  number={1},
  pages={5},
  year={2023},
  publisher={Nature Publishing Group UK London}
}

@article{hibiki1999experimental,
  title={Experimental study on interfacial area transport in bubbly two-phase flows},
  author={Hibiki, Takashi and Ishii, Mamoru},
  journal={International Journal of Heat and Mass Transfer},
  volume={42},
  number={16},
  pages={3019--3035},
  year={1999},
  publisher={Elsevier}
}

@article{alameedy2025comprehensive,
  title={Comprehensive review of severe slugging phenomena and innovative mitigation techniques in oil and gas systems},
  author={Alameedy, Usama and Al-Sarkhi, Abdelsalam and Abdul-Majeed, Ghassan},
  journal={Chemical Engineering Research and Design},
  volume={213},
  pages={78--94},
  year={2025},
  publisher={Elsevier}
}

@article{fabre1992modeling,
  title={Modeling of two-phase slug flow},
  author={Fabre, J and Lin{\'e}, A},
  journal={Annual review of fluid mechanics},
  pages={21--46},
  year={1992}
}

@article{kolmogorov1991local,
  title={The local structure of turbulence in incompressible viscous fluid for very large Reynolds numbers},
  author={Kolmogorov, Andrei Nikolaevich},
  journal={Proceedings of the Royal Society of London. Series A: Mathematical and Physical Sciences},
  volume={434},
  number={1890},
  pages={9--13},
  year={1991},
  publisher={The Royal Society London}
}

@article{risso1998oscillations,
  title   = {Oscillations and breakup of a bubble immersed in a turbulent field},
  author  = {Risso, F. and Fabre, J.},
  journal = {Journal of Fluid Mechanics},
  volume  = {372},
  pages   = {323--355},
  year    = {1998},
  doi     = {10.1017/S0022112098002705}
}

@article{hatashita2025scalings,
  title={Scalings and simulation requirements in two-phase flows},
  author={Hatashita, Luis H and Nathan, Pranav and Jain, Suhas S},
  journal={arXiv preprint arXiv:2510.07727},
  year={2025}
}

@article{liao2009review,
  author  = {Liao, Y. and Lucas, D.},
  title   = {A literature review of theoretical models for drop and bubble breakup in turbulent dispersions},
  journal = {Chemical Engineering Science},
  year    = {2009},
  volume  = {64},
  number  = {15},
  pages   = {3389--3406}
}

@article{liao2010review,
  author  = {Liao, Y. and Lucas, D.},
  title   = {A literature review on mechanisms and models for the coalescence process of fluid particles},
  journal = {Chemical Engineering Science},
  year    = {2010},
  volume  = {65},
  number  = {10},
  pages   = {2851--2864}
}

@article{yao2023breakup,
  author  = {Yao, W. and others},
  title   = {Bubble breakup criteria for population balance modelling of gas--liquid flows},
  journal = {Chemical Engineering Science},
  year    = {2023}
}

@article{anas2020finite,
  title   = {Freely decaying turbulence in a finite domain at finite Reynolds number},
  author  = {Anas, Mohammad and Joshi, Pranav and Verma, Mahendra K.},
  journal = {Physics of Fluids},
  volume  = {32},
  number  = {9},
  pages   = {095109},
  year    = {2020},
  doi     = {10.1063/5.0015009}
}

@article{skrbek2000decay,
  title   = {On the decay of homogeneous isotropic turbulence},
  author  = {Skrbek, L. and Stalp, Steven R.},
  journal = {Physics of Fluids},
  volume  = {12},
  number  = {8},
  pages   = {1997--2019},
  year    = {2000},
  doi     = {10.1063/1.870447}
}

@article{thornber2016impact,
  title   = {Impact of domain size and statistical errors in simulations of homogeneous decaying turbulence and the Richtmyer--Meshkov instability},
  author  = {Thornber, B.},
  journal = {Physics of Fluids},
  volume  = {28},
  number  = {4},
  pages   = {045106},
  year    = {2016},
  doi     = {10.1063/1.4944877}
}

@article{meldi2017turbulence,
  title   = {Turbulence in a box: quantification of large-scale resolution effects in isotropic turbulence free decay},
  author  = {Meldi, M. and Sagaut, P.},
  journal = {Journal of Fluid Mechanics},
  volume  = {818},
  pages   = {697--715},
  year    = {2017},
  doi     = {10.1017/jfm.2017.158}
}

@article{ishihara2009study,
  title={Study of high--Reynolds number isotropic turbulence by direct numerical simulation},
  author={Ishihara, Takashi and Gotoh, Toshiyuki and Kaneda, Yukio},
  journal={Annual review of fluid mechanics},
  volume={41},
  number={1},
  pages={165--180},
  year={2009},
  publisher={Annual Reviews}
}

@article{yoffe2018onset,
  title={Onset criteria for freely decaying isotropic turbulence},
  author={Yoffe, SR and McComb, WD},
  journal={Physical Review Fluids},
  volume={3},
  number={10},
  pages={104605},
  year={2018},
  publisher={APS}
}

@article{luo1996breakup,
  author  = {Luo, H. and Svendsen, H. F.},
  title   = {Theoretical model for drop and bubble breakup in turbulent dispersions},
  journal = {AIChE Journal},
  year    = {1996},
  volume  = {42},
  number  = {5},
  pages   = {1225--1233}
}

@book{ramkrishna2000population,
  title={Population balances: Theory and applications to particulate systems in engineering},
  author={Ramkrishna, Doraiswami},
  year={2000},
  publisher={Elsevier}
}

@article{luo1998coalescence,
  author  = {Luo, H. and others},
  title   = {A model for bubble coalescence in turbulent dispersions},
  journal = {Chemical Engineering Science},
  year    = {1998}
}

@article{Hatashita,
  title = {Scalings and simulation requirements in two-phase flows},
  author = {Hatashita, Luis H. and Nathan, Pranav and Jain, Suhas S.},
  journal = {Phys. Rev. Fluids},
  pages = {},
  year = {2026},
  month = {Jun},
  publisher = {American Physical Society},
  doi = {10.1103/3t33-4k53},
  url = {https://link.aps.org/doi/10.1103/3t33-4k53}
}

@article{vivek_prl,
  title={Cascade-driven fragmentation–aggregation transitions in decaying turbulence},
  author={Kumar, Vivek and Suchandra, Prasoon and Prajapati, Shivam and Jain, Suhas S and Aidun, Cyrus K},
  journal={arxiv},
  volume={submitted},
  year={2026}
}

@article{perrard2021bubble,
  title   = {Bubble deformation by a turbulent flow},
  author  = {Perrard, St{\'e}phane and Rivi{\`e}re, Ali{\'e}nor and Mostert, Wouter and Deike, Luc},
  journal = {Journal of Fluid Mechanics},
  volume  = {920},
  pages   = {A15},
  year    = {2021},
  doi     = {10.1017/jfm.2021.379}
}

@article{masuk2021simultaneous,
  title   = {Simultaneous measurements of deforming Hinze-scale bubbles with surrounding turbulence},
  author  = {Masuk, Ashik Ullah Mohammad and Salibindla, Ashwanth K. R. and Ni, Rui},
  journal = {Journal of Fluid Mechanics},
  volume  = {910},
  pages   = {A21},
  year    = {2021},
  doi     = {10.1017/jfm.2020.964}
}

@article{riviere2021subhinze,
  title   = {Sub-Hinze scale bubble production in turbulent bubble break-up},
  author  = {Rivi{\`e}re, Ali{\'e}nor and Mostert, Wouter and Perrard, St{\'e}phane and Deike, Luc},
  journal = {Journal of Fluid Mechanics},
  volume  = {917},
  pages   = {A40},
  year    = {2021},
  doi     = {10.1017/jfm.2021.243}
}

@article{shawkat2007spectra,
  title   = {On the liquid turbulence energy spectra in two-phase bubbly flow in a large diameter vertical pipe},
  author  = {Shawkat, M. E. and Ching, C. Y. and Shoukri, M.},
  journal = {International Journal of Multiphase Flow},
  volume  = {33},
  number  = {9},
  pages   = {908--926},
  year    = {2007},
  doi     = {10.1016/j.ijmultiphaseflow.2006.11.001}
}

@article{prakash2016energy,
  title   = {Energy spectra in turbulent bubbly flows},
  author  = {Prakash, Vivek N. and Mercado, Julian Mart{\'i}nez and van Wijngaarden, Leen and Mancilla, Ernesto and Tagawa, Yoshiyuki and Lohse, Detlef and Sun, Chao},
  journal = {Journal of Fluid Mechanics},
  volume  = {791},
  pages   = {174--190},
  year    = {2016},
  doi     = {10.1017/jfm.2016.49}
}

@article{elghobashi1994predicting,
  title   = {On predicting particle-laden turbulent flows},
  author  = {Elghobashi, S.},
  journal = {Applied Scientific Research},
  volume  = {52},
  pages   = {309--329},
  year    = {1994},
  doi     = {10.1007/BF00936835}
}

@incollection{elghobashi2006overview,
  title     = {An updated classification map of particle-laden turbulent flows},
  author    = {Elgobashi, Said},
  booktitle = {IUTAM Symposium on Computational Approaches to Multiphase Flow},
  editor    = {Balachandar, S. and Prosperetti, A.},
  series    = {Fluid Mechanics and Its Applications},
  volume    = {81},
  pages     = {3--10},
  publisher = {Springer},
  address   = {Dordrecht},
  year      = {2006},
  doi       = {10.1007/1-4020-4977-3_1}
}

@article{kumar2026bubble,
  title={Bubble coalescence dynamics in a high-Reynolds number decaying turbulent flow},
  author={Kumar, Vivek and Suchandra, Prasoon and Javadi, Ardalan and Jain, Suhas S and Aidun, Cyrus K},
  journal={Journal of Fluid Mechanics},
  volume={1033},
  pages={A6},
  year={2026},
  publisher={Cambridge University Press}
}

@article{chan2021identifying,
  title={Identifying and tracking bubbles and drops in simulations: a toolbox for obtaining sizes, lineages, and breakup and coalescence statistics},
  author={Chan, Wai Hong Ronald and Dodd, Michael S and Johnson, Perry L and Moin, Parviz},
  journal={Journal of Computational Physics},
  volume={432},
  pages={110156},
  year={2021},
  publisher={Elsevier}
}

@article{vassilicos2015dissipation,
  title={Dissipation in turbulent flows},
  author={Vassilicos, J Christos},
  journal={Annual review of fluid mechanics},
  volume={47},
  number={1},
  pages={95--114},
  year={2015},
  publisher={Annual Reviews}
}

@article{bodenschatz2014vdtt,
  title   = {Variable density turbulence tunnel facility},
  author  = {Bodenschatz, Eberhard and Bewley, Gregory P. and Nobach, Holger and Sinhuber, Michael and Xu, Haitao},
  journal = {Review of Scientific Instruments},
  volume  = {85},
  number  = {9},
  pages   = {093908},
  year    = {2014},
  doi     = {10.1063/1.4896138}
}

@article{dhruva1997transverse,
  title   = {Transverse structure functions in high-Reynolds-number turbulence},
  author  = {Dhruva, B. and Tsuji, Y. and Sreenivasan, K. R.},
  journal = {Physical Review E},
  volume  = {56},
  number  = {5},
  pages   = {R4928--R4930},
  year    = {1997},
  doi     = {10.1103/PhysRevE.56.R4928}
}

@article{sreenivasan1997phenomenology,
  title   = {The phenomenology of small-scale turbulence},
  author  = {Sreenivasan, K. R. and Antonia, R. A.},
  journal = {Annual Review of Fluid Mechanics},
  volume  = {29},
  pages   = {435--472},
  year    = {1997},
  doi     = {10.1146/annurev.fluid.29.1.435}
}

@article{Rensen2005,
  author  = {Rensen, Judith and Luther, Stefan and Lohse, Detlef},
  title   = {The effect of bubbles on developed turbulence},
  journal = {Journal of Fluid Mechanics},
  year    = {2005},
  volume  = {538},
  pages   = {153--187},
  doi     = {10.1017/S0022112005005276}
}

@article{RibouxRissoLegendre2010,
  author  = {Riboux, Guillaume and Risso, Fr{\'e}d{\'e}ric and Legendre, Dominique},
  title   = {Experimental characterization of the agitation generated by bubbles rising at high Reynolds number},
  journal = {Journal of Fluid Mechanics},
  year    = {2010},
  volume  = {643},
  pages   = {509--539},
  doi     = {10.1017/S0022112009992084}
}

@article{nathan2025accurate,
  title={Accurate calculation of bubble and droplet properties in diffuse-interface two-phase simulations},
  author={Nathan, Pranav J and Jain, Suhas S},
  journal={Journal of Computational Physics},
  volume={538},
  pages={114190},
  year={2025},
  publisher={Elsevier}
}

\end{document}